\documentclass[12pt, a4paper]{report}
\usepackage[margin=2.56cm]{geometry}
\usepackage{layout}
\usepackage{epic}
\usepackage{graphicx}\graphicspath{{./images/}{./OpenFoam_matlab_images/}{./Matlab_images/}{./OpenFoam_matlab_T_images_without_time/}}
\usepackage{subfigure}
\usepackage[subfigure,titles]{tocloft}
\usepackage{titlesec}
\usepackage{calc}
\usepackage{gensymb}
\usepackage{amsmath}
\usepackage{pifont}
\usepackage{listings}
\usepackage{xcolor}
\usepackage{url}
\usepackage{tikz}\usetikzlibrary{decorations.markings}

\usepackage{booktabs,threeparttable}
\usepackage{rotating,booktabs}
\usepackage{slashbox}

\usepackage{eso-pic}

\title{\bf \large Numerical Investigation of Nonisothermal Reversed Stagnation-point Flow}
\author{\normalsize by\\[0.8cm]
		\bf \large Chio Chon Kit\\[2.8cm]
		\bf \large Master of Science in Electromechanical Engineering\\[3cm]
		\bf \large 2012\\\\
\includegraphics[width=2.86cm]{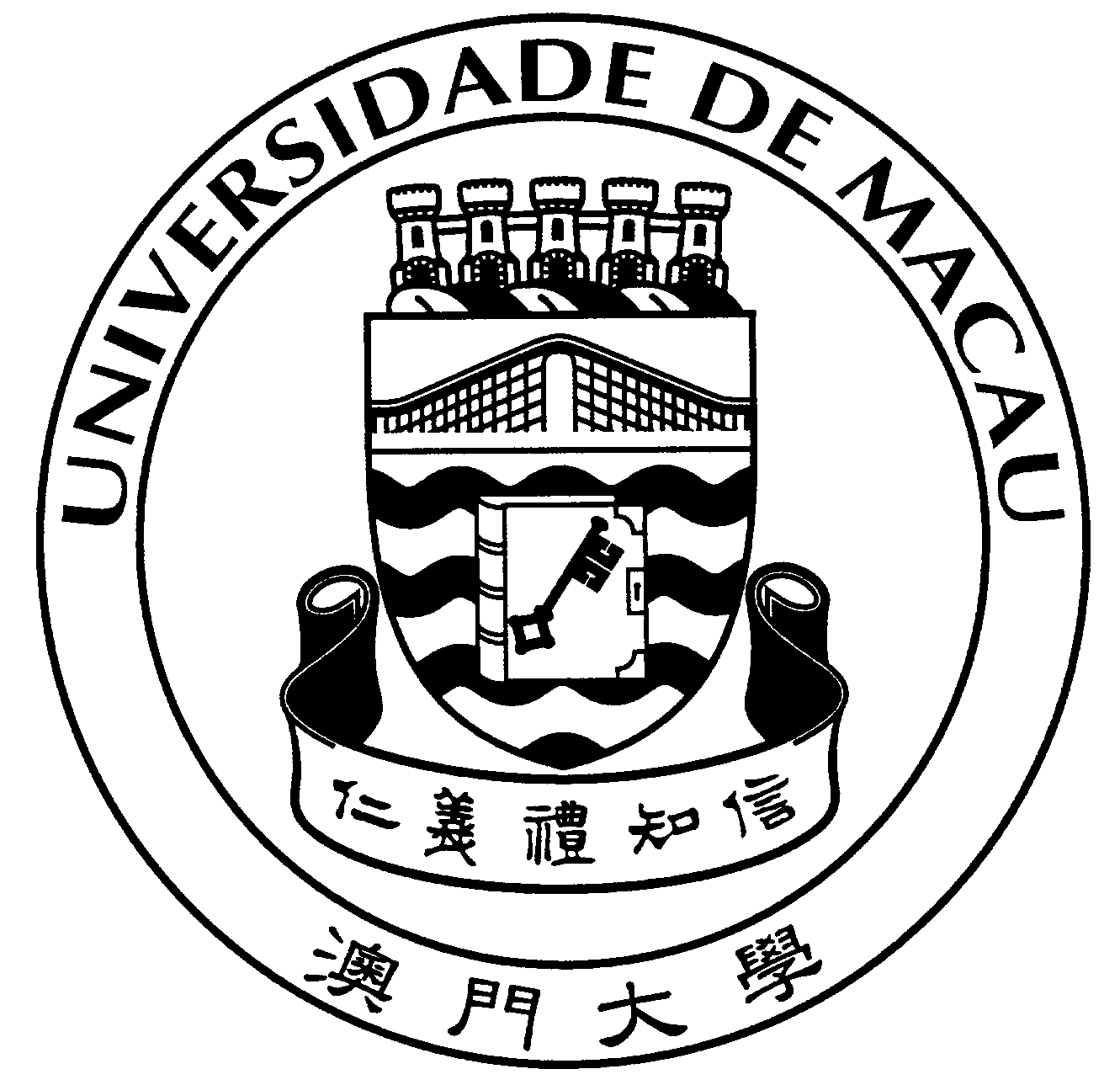}\\
		\bf \large Faculty of Science and Technology\\
		\bf \large University of Macau
}
\date{}

\titleformat{\chapter}{\large \bfseries \centering}{CHAPTER \thechapter:}{.5em}{}
\titleformat{\section}{\large \bfseries}{\thesection}{.5em}{}

\DeclareMathSizes{12}{13}{10}{9}

\begin{document}
\unitlength=0.42mm
\newtheorem{thm}{Theorem}
\newtheorem{cor}{Corollary}
\newtheorem{lem}{Lemma}
\maketitle
\newpage

\textheight=655pt

\thispagestyle{empty}
\begin{center}
\hspace*{1em}\\[4em]
Numerical Investigation of Nonisothermal Reversed
\\Stagnation-point Flow\\[1cm]
by\\[1cm]
Chio Chon Kit\\[1cm]
A thesis submitted in partial fulfillment of the\\
requirements of the degree of\\[1cm]
Master of Science in Electromechanical Engineering\\[1cm]
Faculty of Science and Technology\\
University of Macau\\[1cm]
2012\\[3cm]
\end{center}

Approved by \hrulefill{}\\
\hspace*{7cm} Supervisor\\[1cm]
Date \hrulefill{}
\newpage


\thispagestyle{empty}
\hspace*{1em}\\[4em]
In presenting this thesis in partial fulfillment of the requirements for a\linebreak Master's
degree at the University of Macau, I agree that the Library and the Faculty of Science and Technology
shall make its copies freely available for inspection. However, reproduction of this
thesis for any purposes or by any means shall not be allowed without my written
permission. Authorization is sought by contacting the author at\\\\
\hspace*{1cm}Address: \hspace{1.35em}CALCADO JANUARIO\\
\hspace*{8em}EDIFICIO WAI CHOI YUEN\\ \hspace*{8em}3 ANDAR D\\ \hspace*{8em}MACAU\\
\hspace*{1cm}Telephone: \hspace{0.5em}\\
\hspace*{1cm}E-mail: \hspace{2.05em}s9a9m92001@gmail.com\\\\\\

\hspace*{6cm}Signature \hrulefill{}\\\\
\hspace*{6cm}Date \hrulefill{}
\newpage

\pagestyle{empty}
\begin{center}
University of Macau\\
Abstract\\
Numerical Investigation of Nonisothermal Reversed
\\Stagnation-point Flow\\
by Chio Chon Kit\\
Thesis Supervisor:\\
Associate Professor Sin Vai Kuong\\
Electromechanical Engineering\\[1cm]
\end{center}
This thesis investigates the nature of the development of two-dimensional laminar nonisothermal flow of an incompressible fluid close to the reversed stagnation-point. Proudman and Johnson (1962) \cite{proudman1962boundary} first studied the flow and obtained an asymptotic solution by neglecting the viscous terms. This is not practice in neglecting the viscous terms within the total flow field.  Viscous terms in this analysis are now included, and two-dimensional nonisothermal reversed stagnation-point flow is investigated by solving the Navier-Stokes equations coupled to energy equation.

\newpage



\pagestyle{plain}
\pagenumbering{roman}
\thispagestyle{empty}
\addtocontents{toc}{\protect \thispagestyle{empty}}
\renewcommand{\contentsname}{TABLE OF CONTENTS}
\renewcommand{\cftchappresnum}{CHAPTER \space}
\settowidth{\cftchapnumwidth}{\widthof{\cftchappresnum+1ex}}
\tableofcontents
\newpage

\addcontentsline{toc}{chapter}{LIST OF FIGURES}
\renewcommand{\listfigurename}{LIST OF FIGURES}
\setlength{\cftfignumwidth}{3em}
\listoffigures
\newpage

\addcontentsline{toc}{chapter}{LIST OF TABLES}
\renewcommand{\listtablename}{LIST OF TABLES}
\setlength{\cfttabnumwidth}{3em}
\listoftables
\newpage


\chapter*{LIST OF ABBREVIATIONS}
{\bf CFD} \hspace{3.85em} {\bf C}omputational {\bf F}luid {\bf D}ynamics\\
{\bf FEM} \hspace{3.75em} {\bf F}inite {\bf E}lement {\bf M}ethod\\
{\bf FD} \hspace{4.7em} {\bf F}inite {\bf D}ifference\\
{\bf FV} \hspace{4.7em} {\bf F}inite {\bf V}olume\\
{\bf ODE} \hspace{3.88em} {\bf O}rdinary {\bf D}ifferential {\bf E}quation\\
{\bf PDE} \hspace{3.88em} {\bf P}artial {\bf D}ifferential {\bf E}quation\\
\\
$\rho$ \hspace{5.8em} density\\
$\mu$ \hspace{5.72em} dynamic viscosity\\
$\nu$ \hspace{5.8em}  kinematic viscosity\\
$c_p$ \hspace{5.5em}  heat capacity\\
$k$ \hspace{5.8em} thermal conductivity\\
\addcontentsline{toc}{chapter}{LIST OF ABBREVIATIONS}
\newpage


\chapter*{ACKNOWLEDGMENTS}
I would like to express my gratitude to all those who gave me the possibility to complete this thesis. The preparation of this thesis would not have been possible without their support. \\

First, my deep-felt gratitude to my supervisor, Professor Sin Vai Kuong, Ph.D., Department of Electromechanical Engineering, Faculty of Science and Technology, University of Macau, who has walked me through all the stages of the analysis and simulation. Really thank for his supervision and guidance, without his consistent and illuminating instructions, this thesis could not have reached its present form.
\\

Professor Vong Seak Weng, Department of Mathematics, Faculty of Science and Technology, University of Macau, for the information in mathematical analysis.
\\

Professor U Lei, Institute of Applied Mechanics, National Taiwan University, for his expertise. Despite the distance, he has e-mailed some suggestions that I needed.
\\

I also give my sincere gratitude to my friends and my fellow classmates who gave me their help and spent time in listening to me and helping me work out my problems when I had difficulties in the thesis. Their kind support and guidance have been of great value in this thesis. Meanwhile, I wish to thank Mr. Wong Ian Kai for providing the \LaTeX ~template in preparing the thesis.
\\

My thanks would go to my beloved family for their loving considerations and great confidence in supporting me all through these years.  And I also want to express my gratitude towards my dearest parents who brought me to this wonderful world.
\\

In conclusion, I recognize that this research would not have been possible without the financial assistance of Research Committee of University of Macau (Graduate Research Scholarships), Department of Electromechanical Engineering at the University of Macau (Teaching Assistantships), Science and Technology Development Fund (FDCT) of Macao SAR and acknowledge to those agencies.

\newpage



\voffset=36pt

\pagenumbering{arabic}
\pagestyle{myheadings}

\chapter*{PUBLICATIONS ARISING FROM THIS THESIS}
\thispagestyle{empty}
V.~K. Sin and C.~K. Chio, {\em Computation of NonIsothermal Reversed
  Stagnation-Point Flow over a Flat Plate}, ch.~Computational Simulations and
  Applications, pp.~159--174.
\newblock InTech, 2011.
\newblock ISBN: 978-953-307-430-6  (Chapter 2 and part of Chapter 3)
\\\\
V.~Sin and C.~Chio, ``Reversed stagnation-point flow: Numerical simulation and
  asymptotic solution,'' in {\em System Science and Engineering (ICSSE), 2011
  International Conference on}, pp.~17--22, IEEE, 2011 (Part of Chapter 3)
\\\\
V.~K. Sin and C.~K. Chio., ``{Another Approach of Similarity Solution in
  Reversed Stagnation-point Flow},'' in {\em World Academy of Science,
  Engineering and Technology}, vol.~59, 2011 (Part of Chapter 3)
\\\\
V.~K. Sin and C.~K. Chio., ``{Unsteady Reversed Stagnation-Point Flow over a
  Flat Plate},'' {\em International Journal of Computational and Mathematical
  Sciences}, vol.~6, pp.~153--158, 2012 (Chapter 7)

\newpage

\chapter{INTRODUCTION}

\thispagestyle{empty}
The Navier-Stokes equations describe the motion of fluid substances by applying Newton's second law to fluid motion. It is wonder that given their wide range of practical uses, mathematicians are difficult or impossible to obtain an exact solution in almost every real situation because of the analytic difficulties associated with the nonlinearity due to convective acceleration. The existence of exact solutions is fundamental not only in their own right as solutions of particular flows, but also are useful as accuracy checks for numerical solutions.
\\\\
Computational fluid dynamic modeling has been a very active area of research in recent years as evidenced by numerous papers in the literature. Advances in computer capacity concurrent with the maturation of flow and heat transfer modeling have made feasible these coupled simulations. The goal of research in this area is to make simulations simple in the design and analysis environment for real-world applications. Such capability would be very beneficial to those industries, including enhanced oil recovery, which is a technique for increasing the amount of crude oil that can be extracted from an oil field. 

\section{PREVIOUS AND RELATED WORK}
\begin{figure}[htb]
\begin{center}
\includegraphics[width = 9cm]{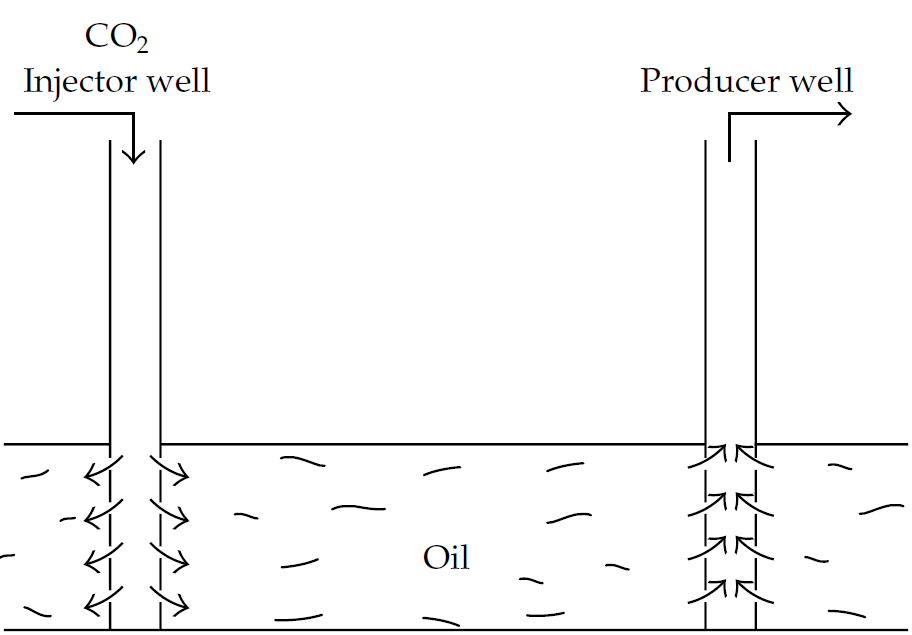}
\caption{Oil recovery industry}
\label{goil}
\end{center}
\end{figure}

Several application of such have also appeared in the recent literature, for example, in some simplified cases a fluid travels through a rigid body (e.g., missile, sports ball, automobile, spaceflight vehicle), or in oil recovery industry crude oil that can be extracted from an oil field is achieved by hot water injection, as shown in Fig. (\ref{goil}), or equivalently, an external flow impinges on a stationary point called stagnation-point that is on the surface of a submerged body in a flow, of which the velocity at the surface of the submerged object is zero. Moreover, the streamline is perpendicular to the surface of the rigid body. The study of the flow motion at stagnation point is of importance in oil recovery industry that develops techniques to efficiently recover oil, gas, and other minerals while reducing environmental impacts using various pollution remediation and greenhouse gas reduction techniques.

The classic problems of two-dimensional stagnation-point flows can be analyzed exactly by Hiemenz \cite{hiemenz1911grenzschicht}. The result is an exact solution for flow directed perpendicular to an infinite flat plate. Howarth \cite{howarth1951cxliv} and Davey \cite{davey1961boundary} extended the two-dimensional and axisymmetric flows to three dimensions, and Wang \cite{wang1985unsteady} studied the case for obliquely-impacting jets. The similarity solutions for the temperature field were studied by Eckert \cite{eckert1942berechnung}. Case corresponding a step change in wall temperature or in wall heat flux in laminar steady flows at a stagnation point has been also investigated by several authors (see Chao et al. \cite{chao1965unsteady}, Sano \cite{sano1981unsteady} and Gorla \cite{gorla1988final}). Further, Lok et al. \cite{lok2006mixed} investigated the mixed convection near non-orthogonal stagnation point flow on a vertical plate with uniform surface heat flux, where the results published are very good with present value of the normalized temperature at the wall for the constant wall temperature boundary condition.
\\

On the contrary, when the external flow is extracted away from the stagnation-point shown in Fig. (\ref{goil}), the flow in the vicinity of this "reversed stagnation point" is governed by boundary-layer separation and vorticity generation and the reversed stagnation-point flow develops.  Reversed stagnation-point flow is a flow in which the component of velocity normal to a wall is outward the wall everywhere in the region concerned. Reversed stagnation-point flows against an infinite flat wall do not have analytic solution in two dimensions, but certain reverse flows have solution in three dimensions \cite{davey1961boundary}.
\\

Proudman and Johnson \cite{proudman1962boundary} first suggested that the convection terms dominate in considering the inviscid equation in the body of the fluid. By introducing a very simple function of a particular similarity variable and neglecting the viscous forces in their analytic result for region sufficient far from the wall, they obtained an asymptotic solution in reversed stagnation-point flow, describing the development of the region of separated flow for large time $t$. In their solution, the phenomenon of separation is described near a plane that represents the rear-stagnation point of a cylinder is set in motion impulsively with a constant velocity normal to the surface of the plane. Robins and Howarth \cite{robins1972boundary} have recently extended the asymptotic solution, finding higher order terms by singular perturbation methods. They indicated that the viscous forces cannot be ignored in the governing equation because of a consistent asymptotic expansion in both this outer inviscid region and also in the inner region near the plane. Smith \cite{smith1977development} generalized the solution of Proudman and Johnson with both viscous and convection terms in balance by considering the monotonic potential flow when the time is relatively large. Shapiro \cite{shapiro2006analytical} obtained a solution for unsteady reversed stagnation-point flow with injection or suction. These unsteady flows fit within a class of similarity transformations originally identified by Birkhoff using a group-theoretic approach.

\section{THESIS OBJECTIVE}
This thesis focuses on the challenging problem of numerical modeling for a nonisothermal reversed stagnation-point flow. The primary objective of the present study is to determine the main characteristics of the flow at reversed stagnation-point. This includes the flow profile, the separation zone, the dividing streamline and the nonisothermal temperature profile. The other objective is to briefly verify the simulation data for the reversed stagnation point.

\section{THESIS OUTLINE}
Chapter 2 is dedicated to developing the governing equation of flow and heat transfer modeling used in this research. Chapter 3 investigates the nature of the development of two-dimensional laminar flow of an incompressible fluid near the reversed stagnation-point. Similarity solutions of two-dimensional reversed stagnation-point flow are investigated by simplifying the full Navier-Stokes equations coupled to the energy equation, describing the motion of nonisothermal fluid substances. The model is valid if the fluid velocity is small compared with the speed of sound and the fluid is treated as Newtonian. Chapter 4 describes the implementation of the algorithm used to achieve the objectives outlined above. Specific information is provided about the individual solvers and details of the interface of simulation. Details of the mesh generation and the efficacy of the CFD solver are provided. Chapter 5 focuses on numerical simulations on the reversed stagnation-point flow. This chapter provides result and discussion of the model provided. The final chapter summarizes and provides conclusions of the research, and recommends future work.

\chapter{GOVERNING EQUATIONS}
\thispagestyle{empty}
The nonlinear behavior of fluid flow is emphasized in this chapter. The Navier-Stokes equations describe the motion of a fluid in two- or three-dimensional space. These equations are to be solved for an unknown velocity vector and pressure. We restrict attention here to the reversed stagnation-point flow in an incompressible fluids domain. To construct an effective method for handling the Navier-Stokes equations, which are systems of partial differential equations, a similarity transformation is applied and a simplified similarity equation is considered.

\section{INCOMPRESSIBILITY}
In an incompressible fluid, the density of an element of fluid is not affected by any changes in pressure. If the relative speeds within a flow are low enough (typically Mach number less than $0.3$), thermodynamic effects and density changes due to changes in pressure become negligible. If density is constant and mass is conserved so is volume. This condition is called the equation of continuity and expressed mathematically as the divergence of flow velocity $\boldsymbol{\vec{V}}$ is zero
\begin{equation}
\nabla \cdot \boldsymbol{\vec{V}}=0
\label{e0}
\end{equation}

Essentially what goes into a differential volume must exit it simultaneously. Coupling this equation with conservation of momentum makes the system fully determined, without need of the energy equation or an equation of state, and yields extremely efficient simulations. The flows and solution methods  can be greatly different; however, they all start with the same underlying defined as the differential element of the continuous Navier-Stokes equations.

\section{MOMENTUM EQUATIONS}
Models for Newtonian fluids undergoing incompressible flow make use of the approximation that dynamic viscosity $\mu$ is a constant. Performing a force balance and making use of the continuity equation leads to the Navier-Stokes equations \cite{zikanov2011essential}:
\begin{equation}
\rho \frac{D\boldsymbol{\vec{V}}}{Dt}=-\nabla p+\mu \nabla^2 \boldsymbol{\vec{V}}+\boldsymbol{\vec{f}}
  \label{e1}
\end{equation}
${D\boldsymbol{\vec{V}}}/{Dt}$ is the material derivative of flow velocity
\begin{equation}
\frac{D\boldsymbol{\vec{V}}}{Dt}=\frac{\partial\boldsymbol{\vec{V}}}{\partial t}+(\boldsymbol{\vec{V}} \cdot \nabla )\boldsymbol{\vec{V}}
\end{equation}
representing the convective acceleration in the fluid motion. The physical principle of momentum transfer is Newton's second law. Equation (\ref{e1}) is just Newton's second law $\boldsymbol{\vec{F}} = m\boldsymbol{\vec{a}}$ for a fluid element subject to the external force $\boldsymbol{\vec{f}}$ and to the forces arising from pressure gradient $-\nabla p$ and viscosity $\mu \nabla^2 \boldsymbol{\vec{V}}$.

\section{ANALYTICAL ANALYSIS}
We begin with writing the governing equation in conservative velocity form in the Cartesian coordinates  \cite{white1991viscous} and neglecting the external force $\boldsymbol{\vec{f}}$:
\begin{subequations}
 \begin{gather}
\frac{\partial u}{\partial x}+\frac{\partial v}{\partial y}=0
\label{eq:e0}
  \end{gather}
 \begin{gather}
    \frac{\partial u}{\partial t}+ u \frac{\partial u}{\partial x}+v\frac{\partial u}{\partial y}=  -\frac{1}{
     \rho}\frac{\partial p}{\partial x}+ \nu \left(\frac{\partial^2 u}{\partial
     x^2}+\frac{\partial ^2u}{\partial y^2}\right)
     \label{e3}\\\notag\\
     \frac{\partial v}{\partial t}+ u \frac{\partial v}{\partial x}+v\frac{\partial v}{\partial y}=  -\frac{1}{
     \rho}\frac{\partial p}{\partial y}+ \nu \left(\frac{\partial^2 v}{\partial
     x^2}+\frac{\partial ^2v}{\partial y^2}\right)
     \label{e4}
  \end{gather}
  \label{e3_4}
\end{subequations}

\begin{figure}[!htb]
\centering
\begin{tikzpicture}[>=stealth]
\draw[
    decoration={markings,mark=at position 1 with {\arrow[scale=1.5]{>}}},
    postaction={decorate},
    shorten >=0.4pt
    ] (-4,-0.6) -- (4,-0.6);
\coordinate [label=-45:$x$] (a) at (4,-0.6);
\draw[
    decoration={markings,mark=at position 1 with {\arrow[scale=1.5]{>}}},
    postaction={decorate},
    shorten >=0.4pt
    ](0,-0.6) -- (0,4);
\coordinate [label=45:$y$] (a) at (0,4);
\fill[black] (0,-0.6) circle (2pt);
\coordinate [label=45:$O$] (a) at (0,-0.6);
\node[right=0] at (-1.3,-0.3) {$u=0$};
\node[right=0] at (3,2) {$u=-Ax$};
\node[right=0] at (1.3,-0.3) {$T_w$};
\node[right=0] at (2.8,1.2) {$T_{\infty}$};
\draw (0.6,3) .. controls (0.6,1) and (2,0) .. (3.5,0);
\draw[
    decoration={markings,mark=at position 1 with {\arrow[scale=2]{>}}},
    postaction={decorate},
    shorten >=0.4pt
    ] (1.15,1.15) -- (1.05,1.25);
\draw (-0.6,3) .. controls (-0.6,1) and (-2,0) .. (-3.5,0);
\draw[
    decoration={markings,mark=at position 1 with {\arrow[scale=2]{>}}},
    postaction={decorate},
    shorten >=0.4pt
    ] (-1.15,1.15) -- (-1.05,1.25);
\draw (1.3,3) .. controls (1.3,1) and (2.7,0.7) .. (3.5,0.7);
\draw[
    decoration={markings,mark=at position 1 with {\arrow[scale=2]{>}}},
    postaction={decorate},
    shorten >=0.4pt
    ] (2.02,1.15) -- (1.92,1.25);
\draw (-1.3,3) .. controls (-1.3,1) and (-2.7,0.7) .. (-3.5,0.7);
\draw[
    decoration={markings,mark=at position 1 with {\arrow[scale=2]{>}}},
    postaction={decorate},
    shorten >=0.4pt
    ] (-2.02,1.15) -- (-1.92,1.25);
\end{tikzpicture}
\caption{Coordinate system of nonisothermal reversed stagnation-point flow}
\label{csp}
\end{figure}
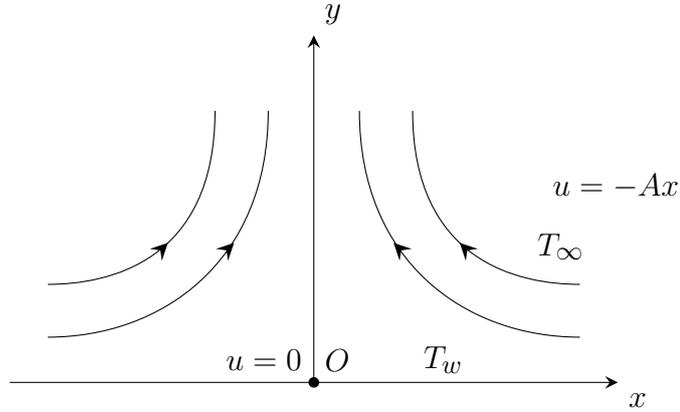

Here $u$ and $v$ are the components of flow velocity $\boldsymbol{\vec{V}}(u,v)$, $\rho$ is the fluid density, $p$ is the fluid pressure, $\nu=\mu/\rho$ is the kinematic viscosity. The viscous fluid flows in a rectangular Cartesian coordinates $(x,y,z)$, Fig.~\ref{csp}, illustrates the motion of external flow directly moving perpendicular out of an infinite flat plane wall. The origin is the so-called stagnation point and $z$ is the normal to the plane.

What we are concerned about is the two-dimensional reversed stagnation-point flow in unsteady state. The total fluid domain is bounded by an infinite plane $y=0$, the fluid remains at rest when time $t<0$.  At $t=0$, it starts impulsively in motion which is determined by the stream function
\begin{equation}
\psi= -\alpha x y
\end{equation}

where $\alpha$ is a positive rate of strain \cite{white1991viscous}. At large distances far above the planar boundary, the existence of the potential flow implies an inviscid boundary condition. Far away from the wall the flow is of a constant $V_0$ along the $y$-axis. Because of the axisymmetric configuration, the flow field is considered in right-hand side region only. For such a flow the components of velocity are easily given from the relationships
\begin{subequations}
\begin{gather}
u= -\alpha x\\
v=V_0
\end{gather}
\end{subequations}

Here $\alpha$ is a constant proportional to $V_0/L$, $V_0$ is the external flow velocity removing from the plane and $L$ is the characteristic length. We have $u=0$ at $x=0$ and $v=0$ at $y=0$, but the no-slip boundary at wall $(y=0)$ cannot be satisfied.
\\

The equation of continuity (\ref{eq:e0}) is integrated by introducing the stream function $\psi$:
  \begin{equation}
   u=   \displaystyle \frac{\partial \psi}{\partial y}\qquad\mathrm{and}\qquad v= \displaystyle- \frac{\partial \psi}{\partial x}
   \label {stream}
  \end {equation}
For reversed stagnation flow without friction (ideal fluid flow), the stream function may be written as
  \begin{equation}
\psi =\psi_i = -A_ixy
   \label {eq:e00}
\end {equation}
where $A_i$ is a constant and from which
  \begin{equation}
u_i=-A_ix \qquad\mathrm{and}\qquad v_i = A_iy.
   \label {eq:e01}
 \end {equation}
We have $u_i=0$ at $x=0$ and $v_i=0$ at $y=0$, but the no-slip boundary at wall $(y=0)$ cannot be satisfied.
\\

Since for a (real) viscous fluid the flow motion is determined by only two factors,  the kinematic viscosity $\nu$  and $\alpha$, consistent with the initial and boundary conditions that $\psi$ is proportional to $x$ for all value of $y$ and $t$. Provided that the surface is an infinite plane wall, a following modified stream function is introduced, see Proudman and Johnson \cite{proudman1962boundary}:
\begin{subequations}
\begin{gather}
\psi = -\sqrt{A\nu}xf(\eta, \tau)\label{psi1}\\
\eta = \sqrt{\frac{A}{\nu}}y\\
\tau = At
\end{gather}
\end{subequations}

where $\eta$ is the non-dimensional distance from wall and $\tau$ is the non-dimensional time. Noting that the stream function automatically satisfies equation of continuity (\ref{eq:e0}). Substituting $u$ and $v$ into the governing equations results a simplified partial differential equation. From the definition of the stream function, we have
\begin{subequations}
  \begin{gather}
u= \frac{\partial \psi}{\partial y}  =  -Axf_{\eta}\\
v=-\frac{\partial \psi}{\partial x} =  \sqrt{A\nu}f
  \end{gather}
\label{se}
\end{subequations}
Note that $A$ has the dimension as "1/time". The governing equations can be simplified by a similarity transformation when several independent variables appear in specific combinations, in flow geometries involving infinite or semi-infinite surfaces.  By introducing coordinate variable transformation, the number of independent variables is reduced by one or more. The original system of partial differential equations can be simplified into the following pair of partial differential equations
\begin{subequations}
  \begin{gather}
    -A^2xf_{\eta\tau}+A^2x(f_{\eta})^2-A^2xff_{\eta\eta}=   -\frac{1}{ \rho}
     \frac{\partial p}{\partial x}- A^2xf_{\eta\eta\eta}
    \label{e5}\\
     A\sqrt{A\nu}f_{\tau}+A\sqrt{A\nu}ff_{\eta}=   -\frac{1}{ \rho}
     \frac{\partial p}{\partial y}+ A\sqrt{A\nu} f_{\eta\eta}
    \label{e6}
  \end{gather}
\end{subequations}

The pressure gradient can be again reduced by a further differentiation equation~(\ref{e6}) with respect to $x$. That is
  \begin{equation}
\frac{\partial^2 p}{\partial x \partial y}=0
  \end{equation}
and equation (\ref{e5}) reduces to
  \begin{equation}
    [f_{\eta\tau}-(f_{\eta})^2+ff_{\eta\eta}-f_{\eta\eta\eta}]_{\eta}=0.
     \label{e7}
  \end{equation}

or the equation becomes a differential equation for $f$ \cite{schlichting2000boundary}
  \begin{equation}
    f_{\eta\tau}-(f_{\eta})^2+ff_{\eta\eta}-f_{\eta\eta\eta}=\mathrm{function~of~}\tau\mathrm{~only}.
     \label{e7a}
  \end{equation}

The initial and boundary conditions are
  \begin{subequations}
     \begin{gather}
        f(\eta, 0) \equiv \eta ~~~~~~~~~~~~~~~(\eta \neq 0)\\
        f(0,\tau)= f_{\eta}(0,\tau)=0~~(t\neq 0)\\
        f(\infty,\tau)\sim \eta~~~~~~~~~~~~~~~~~~~~~~
    \end{gather}
 \label{e8}
  \end{subequations}
These follow from the impermeability condition of the wall (from $v(x,0,\tau)=0$ it follows that $ f(0,\tau)=0$) and from the no-slip condition (from $u(x,0,\tau)=0$ it follows that $f_{\eta}(0,\tau)=0$). \\

Proudman and Johnson suggested that at large distances from the wall ($\eta \rightarrow \infty$) the velocity $v(x,\eta,\tau)$ should pass over smoothly into that for inviscid $V_0$. Here they have employed $f_{\eta\eta}(\infty) = f_{\eta\eta\eta}(\infty) = 0,$ which implies that the flow matches smoothly with the inviscid flow as $\eta \rightarrow \infty$. This leads to the condition $f(\infty,\tau)\sim \eta,$ and thus, the last condition reduces the differential equation~(\ref{e7}) for $f$ \cite{schlichting2000boundary}
  \begin{equation}
    f_{\eta\tau}-(f_{\eta})^2+ff_{\eta\eta}-f_{\eta\eta\eta}=-1,
     \label{e9}
  \end{equation}
with the boundary conditions
  \begin{subequations}
     \begin{gather}
        f(0,\tau)= f_{\eta}(0,\tau)=0\\
        f_{\eta}(\infty,\tau) = 1.~~~~~~~~~~
    \end{gather}
 \label{e10}
  \end{subequations}
Here $f_{\eta\eta\eta}$ is proportional to the viscous stress, $f_{\eta\eta}$ is proportional to the shear stress, $f_{\eta}$ is proportional to the x-component of velocity in boundary layer and $f$ is s proportional to the stream function. \\

It should be noted that the dimensionless velocity distribution $f_{\eta}$ is, from (\ref{se}), independent of the length $x$, and thus equation~(\ref{e9}) is a similarity equation of the full Navier-Stokes equations at two-dimension reversed stagnation-point. The coordinates $x$ and $y$ are replaced by a dimensionless variable $\eta$. Under the boundary conditions $f_{\eta}(\infty,\tau) = 1$, when the flow is in steady state such that $f_{\eta\tau}\equiv 0$, the differential equation has no solution.

\section{ENERGY TRANSPORT}
\thispagestyle{empty}
In this section our considerations of reversed stagnation-point flow until now have referred only to velocity field. Now we shall extend to include the temperature field in the nonisothermal flow which is at a temperature $T$ different from that of the wall $T_w$. It will be assumed that heat energy is transferred to the flow through the wall. Once the velocities are known from the flow analysis, the temperature distributions can be determined by solving the energy equation in the reversed stagnation-point flow. \\

To include the temperature $T$ in our analysis we must now turn to the thermodynamic properties of fluids. The principle of conservation of energy yields the thermal energy equation \cite{louis1993convective} for constant-property fluid:
  \begin{equation}
     \rho c_p \left(\frac{\partial T}{\partial t}+u \frac{\partial T}{\partial x}+v\frac{\partial T}{\partial y}\right)=
     k \left(\frac{\partial^2 T}{\partial x^2}+\frac{\partial ^2 T}{\partial y^2}\right)
    +\mu \Phi
     \label {q:e13}
  \end{equation}
where $k$ is the thermal conductivity, $c_p$ is the heat capacity, and $\Phi$ is defined as
\begin{equation}
\Phi = 2 \left[ \left(\frac{\partial u}{\partial x}\right)^2 +\left(\frac{\partial v}{\partial y}\right)^2\right] + \left(\frac{\partial u}{\partial y}+\frac{\partial v}{\partial x} \right)^2 -\frac {2}{3} \left(\frac{\partial u}{\partial x} +\frac{\partial v}{\partial y} \right)^2
\end{equation}
and is called the viscous dissipation since it represents the irreversible conservation of mechanical forms of energy to a thermal form. \\

If both velocity field and temperature field exist, there is generally also a coupling between these two fields. Since the velocity components $u$ and $v$ appear in the energy equation, a simplification of the energy equation requires to know the actual value of the velocity components. This velocity field would be identical to the velocity components in the reversed stagnation-point flow
\begin{subequations}
\begin{gather}
u=-Axf_{\eta} \\
v=\sqrt{A\nu}f
\end{gather}
\end{subequations}

To transform equation (\ref{q:e13}) into a nondimensional form, it is convenient to work with a dimensionless temperature $\theta$ \cite{white1991viscous}:
  \begin{equation}
\theta(\eta,\tau)= \frac {T-T_w}{T_\infty-T_w}
     \label {q:e17}
  \end{equation}
where $T_w$ and $T_\infty$ are the wall temperature and ambient temperature. Considering the case that both $T_ w$ and $T_\infty$ are constant, the required boundary conditions are
  \begin{equation}
T(0,t)=T_w, \qquad
T(\infty,t)=T_\infty.
  \end{equation}
The fluid temperature $T$ can be treated as a function of $\eta$ and $\tau$ only. Under the assumption that the viscous dissipation is negligible compared to conduction at the wall, we may write the energy equation in the form
  \begin{equation}
\theta_{\eta\eta} - \frac{\rho c_p\nu}{k} f\theta_{\eta}= \frac{\rho c_p\nu}{k} \theta_{\tau}
     \label {q:e14}
  \end{equation}
subject to the boundary conditions
  \begin{equation}
   \theta(0,\tau)=0 \qquad
   \theta(\infty,\tau)=1
\label {q:e15}
\end{equation}

Equation (\ref{q:e14}) is a second-order partial differential equation with variable coefficient $f(\eta,\tau)$ and the Prandtl number $Pr= \rho c_p\nu / k$ is assumed to be constant. Consider the fluid of which $Pr=1$, the thermal boundary layer and the velocity boundary layer collapse, and thus, substituting $\theta=f'$, equation (\ref{e9}) and (\ref{q:e14}) represent the same equation. It is noticed that in these nonisothermal flows the velocity field is decoupled from the temperature field if the kinematic viscosity $\nu$ is constant and is assumed to be independent of the temperature and pressure. This assumption is valid when the temperature and pressure differences are small within the boundary layer.

\chapter{FLOW ANALYSIS}
\thispagestyle{empty}
We complete the governing equations of viscous reversed stagnation-point flow by discussing similar flow. Our objective is to obtain a similarity solution of the governing equation. Generally speaking, a similarity solution is one in which the number of variables can be reduced by a coordinate transformation. Let us discuss various laminar similarity solutions.

\section{INVISCID SOLUTION}
\begin{figure}[!htb]\centering
\begin{tikzpicture}[>=stealth]
\draw[
    decoration={markings,mark=at position 1 with {\arrow[scale=1.5]{>}}},
    postaction={decorate},
    shorten >=0.4pt
    ] (-4,-1.2) -- (4,-1.2);
\coordinate [label=-45:$x$] (a) at (4,-1.2);
\draw[
    decoration={markings,mark=at position 1 with {\arrow[scale=1.5]{>}}},
    postaction={decorate},
    shorten >=0.4pt
    ](0,-1.2) -- (0,4);
\coordinate [label=45:$y$] (a) at (0,4);
\fill[black] (0,-1.2) circle (2pt);

\draw (0.6,3) .. controls (0.6,1) and (2,0) .. (3.5,0);
\draw[
    decoration={markings,mark=at position 1 with {\arrow[scale=2]{>}}},
    postaction={decorate},
    shorten >=0.4pt
    ] (1.15,1.15) -- (1.05,1.25);
\draw (-0.6,3) .. controls (-0.6,1) and (-2,0) .. (-3.5,0);
\draw[
    decoration={markings,mark=at position 1 with {\arrow[scale=2]{>}}},
    postaction={decorate},
    shorten >=0.4pt
    ] (-1.15,1.15) -- (-1.05,1.25);
\draw (1.3,3) .. controls (1.3,1) and (2.7,0.7) .. (3.5,0.7);
\draw[
    decoration={markings,mark=at position 1 with {\arrow[scale=2]{>}}},
    postaction={decorate},
    shorten >=0.4pt
    ] (2.02,1.15) -- (1.92,1.25);
\draw (-1.3,3) .. controls (-1.3,1) and (-2.7,0.7) .. (-3.5,0.7);
\draw[
    decoration={markings,mark=at position 1 with {\arrow[scale=2]{>}}},
    postaction={decorate},
    shorten >=0.4pt
    ] (-2.02,1.15) -- (-1.92,1.25);

\draw (3.5,-0.4) .. controls (0.6,-0.4) and (0.6,-1) .. (3.5,-1);
\draw[
    decoration={markings,mark=at position 1 with {\arrow[scale=2]{>}}},
    postaction={decorate},
    shorten >=0.4pt
    ] (2.9,-0.4) -- (2.8,-0.4);
\draw[
    decoration={markings,mark=at position 1 with {\arrow[scale=2]{>}}},
    postaction={decorate},
    shorten >=0.4pt
    ] (2.8,-1) -- (2.9,-1);
\draw (-3.5,-0.4) .. controls (-0.6,-0.4) and (-0.6,-1) .. (-3.5,-1);
\draw[
    decoration={markings,mark=at position 1 with {\arrow[scale=2]{>}}},
    postaction={decorate},
    shorten >=0.4pt
    ] (-2.9,-0.4) -- (-2.8,-0.4);
\draw[
    decoration={markings,mark=at position 1 with {\arrow[scale=2]{>}}},
    postaction={decorate},
    shorten >=0.4pt
    ] (-2.8,-1) -- (-2.9,-1);
\end{tikzpicture}
\caption{Streamlines of reversed stagnation-point flow}
\label{vort}
\end{figure}
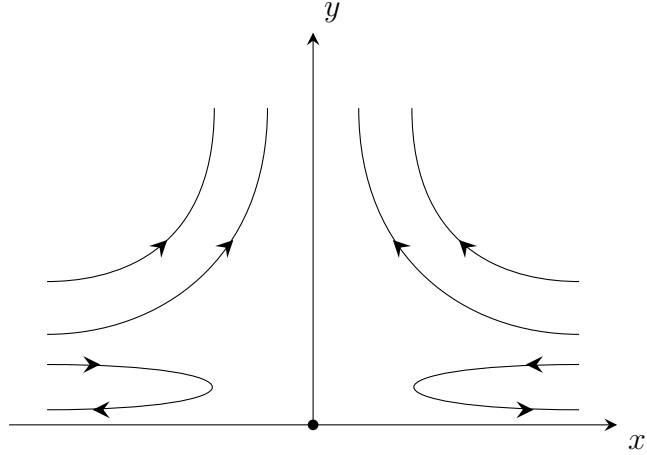

Proudman and Johnson \cite{proudman1962boundary} first thought over the early stages of the diffusion of the initial vortex sheet at $y=0$. The idea was to divide the flow into two regions: an outer flow region that is inviscid and can sometimes be approximated as potential flow, and an inner flow region where the viscous forces are of the same order as the inertial forces. The general feature of the predicted streamline pattern is sketched in Fig. (\ref{vort}).\\

Proudman and Johnson suggested that, when the flow is near the wall region, the viscous forces are dominant, and the viscous term in the governing Navier-Stokes equations is important only near the boundary. On the contrary, the viscous forces were neglected far away from the wall. The convection terms dominate the motion of external flow in considering the inviscid equation in the fluid. Ignoring the viscous stress in equation (\ref{e9}) yields an inviscid equation
  \begin{equation}
    f_{\eta\tau}-(f_{\eta})^2+ff_{\eta\eta}+1=0.
     \label{e11}
  \end{equation}
They contemplated the similarity of the inviscid equation in the form
\begin{equation}
f(\eta,\tau)=\lambda({\tau}) F(\gamma), ~~~\gamma=\eta/\lambda({\tau})
\label{eq2_250}
  \end{equation}

Substituting equation (\ref{eq2_250}) in (\ref{e11}) results in
\begin{equation}
\frac{\dot{\lambda}}{\lambda}\gamma F''-F'^2+FF''=-1,
\label{eq2_251}
  \end{equation}
so that 
\begin{equation}
\frac{\dot{\lambda}}{\lambda}=\mathrm{constant} =k, ~~\mathrm{or~~} \lambda=e^{k\tau}
\label{eq2_252}
  \end{equation}
A solution to this equation that satisfies $F=1$ with exponential error as $\eta \rightarrow \infty$ is only possible when $k= 1$; Proudman and Johnson finally obtained an asymptotic similarity solution of
\begin{equation}
f(\eta,\tau)={\eta} -\frac{2e^{\tau}}{c}(1-e^{-c{\eta}e^{-\tau}} )
\label{e14}
\end{equation}
where $c$ is a constant of integration, representing the uncertainty in the precise position of the time origin. In the asymptotic solution (\ref{e14}),  the constant $c$ always appears multiplying the similarity variable, i.e. $c\eta e^{-\tau}$. A change from $\tau$ to $\tau+\Delta \tau$ can be included in the constant $c$. The improved numerical evaluations of Robins and Howarth \cite{robins1972boundary} estimated the value of $c$ to be $3.51$. This solution describes an exponential decay of vorticity in the outer region, moving away from the plane with a constant velocity.

\section{VISCOUS SOLUTION}
\begin{figure}[!htb]
\begin{center}
\includegraphics[width = 14cm]{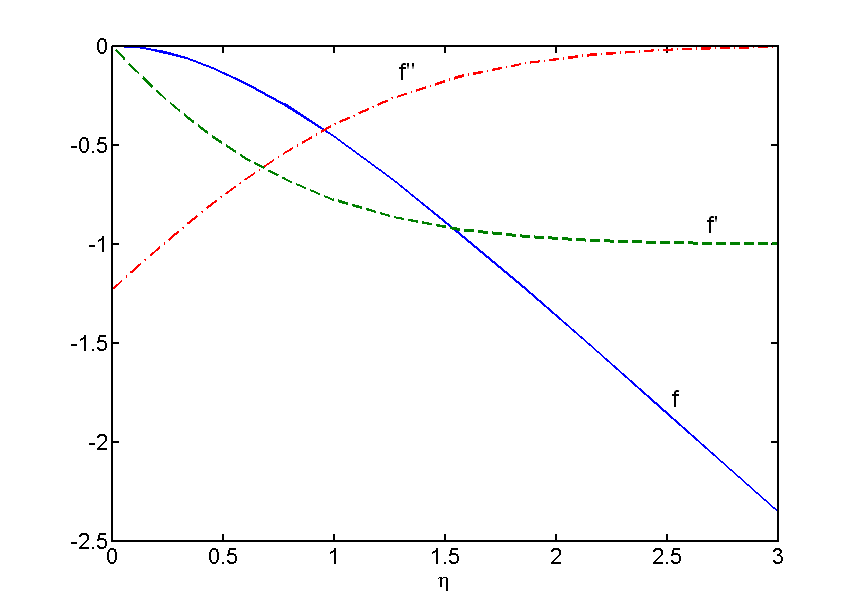}
\caption{Numerical solution of classic stagnation-point problem}
\end{center}
\label{fsf}
\end{figure}

The viscous layer develops as a consequence of the no-slip boundary condition at the wall. In the inner region the viscous term cannot be neglected and a further solution must be found which satisfies the no-slip condition on the wall. 
When $\tau \rightarrow \infty$, the solution (\ref{e14}) yields the steady flow
\begin{equation}
f \sim -\eta ~~~~\mathrm{and} ~~~~~f'\sim-1
\end{equation}
which becomes the outer boundary condition for the viscous flow near the boundary. Substituting in equation (\ref{e9}) yields
\begin{subequations}
    \begin{gather}
     f'''- ff''+(f')^2-1= 0 \\
     f(0)= f'(0)=0~~~~~~~~~~~~~\\
     f'(\infty) = -1~~~~~~~~~~~~~~~~~~~
   \end{gather}
  \label{q:e12a}
\end{subequations}
This is exactly the classic stagnation-point problem (Hiemenz \cite{hiemenz1911grenzschicht}) by changing the sign in $f$. It is a third-order nonlinear ordinary differential equation and does not have an analytic solution, and thus it is necessary to solve it numerically. The numerical solution of classic stagnation-point problem is shown in Figure (\ref{fsf}).\\

Although an asymptotic solution was obtained, it can easily been observed that this is not valid when the viscous term $f_{\eta\eta\eta}$ is neglected within the total flow field. Robins and Howarth \cite{robins1972boundary} studied higher order terms by singular perturbation methods and indicated that a consistent asymptotic expansion occurs in both outer inviscid region and also in the inner region that must exist close to the wall where the viscous forces need to be included. It is not quite appropriate to say Proudman and Johnson are wrong because of neglecting the viscous term in their analytic result for region sufficient far from the wall, but in such a case neglecting the viscous terms within the total flow field can be improved. The next section will discuss the nonexistence of the exact solutions with the boundary condition $f_{\eta}(\infty,\tau) = 1$ for steady case.

\section{INSOLUBILITY IN STEADY STATE}
In this section, a mathematical proof indicates that all of the steady solutions, however, do not satisfy the boundary condition $f_{\eta}(\infty,\tau) = 1$.\\

When $f_{\eta\tau}\equiv 0$, Eq.~(\ref{e9}) reduces to
\begin{subequations}
    \begin{gather}
       f'''- ff''+(f')^2-1= 0    \\
          f(0)=f'(0)=0 ~~~~~~~~~~~~~\\
          f'(\infty) = 1~~~~~~~~~~~~~~~~~~~~~
   \end{gather}
 \label{eq:e10}
\end{subequations}
where the prime denotes the derivative with respect to $\eta$.
\begin{lem}
No solution  $f'(\eta)$ exists which has stationary value of 1 for finite $\eta$.
 \label{l1}
\end{lem}
Proof. Rearrange equation~(\ref{eq:e10}) yields
 \begin{equation}
 f'''=1-(f')^2+ff''
 \label{eq:e13}
   \end{equation}
Suppose for $\eta = \eta_0$, we have
$f'(\eta_0)=1$ and $f''(\eta_0)=0$. Afterwards, it follows from the derivatives of equation~(\ref{eq:e13}) that $f'''$ and all higher derivatives are zero when $\eta = \eta_0$.  Consider a variable transformation
\begin{gather}
\varpi(\eta)=f'(\eta)\notag\\
\varpi(\eta_0)=1
\end{gather}
Expand the function into Taylor's series near $\eta_0$, we have
\begin{eqnarray*}
f'(\eta) = \varpi(\eta)
& = &\sum_{n=0}^{\infty} \frac{\varpi^{(n)}(\eta_0)}{n!}(\eta - \eta_0)^n \\
& = &\varpi(\eta_0)+\sum_{n=1}^{\infty} \frac{\varpi^{(n)}(\eta_0)}{n!}(\eta - \eta_0)^n \\
& = & 1+\sum_{n=1}^{\infty} \frac{\varpi^{(n)}(\eta_0)}{n!}(\eta - \eta_0)^n \\
& \equiv & 1
\end{eqnarray*}
Hence, the boundary condition $f'(0)=0$ is thus not satisfied and the Lemma is proved.

\begin{lem}
When $f'$ has a stationary value, if $|f'|<1$ it is a minimum and if $|f'|>1$ it is a maximum.
 \label{l2}
\end{lem}
Proof: From equation~(\ref{eq:e13}), when $f'$ has a stationary value, it means $f''=0$ and equation~(\ref{eq:e13}) becomes
 \begin{equation}
  f'''=1-(f')^2
 \label{eq:e14}
   \end{equation}
If $|f'|<1$, $f'''>0$ and it is minima. Else if $|f'|>1$, $f'''<0$ and it is maxima. Eventually, the lemma is proved.

\begin{lem}
If $f''(\eta)$ vanishes for $\eta = \eta_1, \eta_2$ ... with $ \eta_1<\eta_2<...$, then the sequence $f'(\eta_i)$ does not tend to 1 as  $\eta_i\to \infty$.
 \label{l3}
\end{lem}
Proof: Consider a region where $ (\eta_1,\eta_2)$ is far away from the origin.
Multiply $f''$ to equation~(\ref{eq:e13}) and integrate it between $\eta_1$ and $\eta_2$  with respect to $\eta$.
$$       f''f'''=f''-(f')^2f''+ f(f'')^2 $$
$$    \int_{\eta_1}^{\eta_2} f''f'''d\eta= \int_{\eta_1}^{\eta_2} [f''-(f')^2f''+ f(f'')^2]d\eta $$
$$      \displaystyle \frac {1}{2}[(f'')^2]_{\eta_1}^{\eta_2}=  [f'-\frac{1}{3}(f')^3]_{\eta_1}^{\eta_2}+\int_{\eta_1}^{\eta_2} f(f'')^2d\eta $$
When $f'(\infty)  \to 1$, it is required that $f''(\eta_1)=f''(\eta_2)=0$ and thus
$$   [f'-\frac{1}{3}(f')^3]_{\eta_1}^{\eta_2}=-L $$
whereas $\displaystyle L =\int_{\eta_1}^{\eta_2} f(f'')^2d\eta$ is always positive and we can obtain
$$   [f'-\frac{1}{3}(f')^3]_{\eta_1}^{\eta_2}<0 $$
$$
   f'(\eta_2)-\frac{1}{3}[f'(\eta_2)]^3< f'(\eta_1)-\frac{1}{3}[f'(\eta_1)]^3
$$
$$
[f'(\eta_2)]^3-3f'(\eta_2)> [f'(\eta_1)]^3-3f'(\eta_1)
$$
\\
Consider $G=f'^3-3f'$ as a function of $f'$, then
$$G'=3f'^2-3$$
As $f'=1$, then $G'(1)=0$, which makes G a minimum. We do not have $f'(\eta_i)=1$ as $\eta_i\to \infty$

\begin{thm}
Given any $f'(\eta) \to 1$ as $\eta \to \infty$, no solution of equation~(\ref{eq:e10}) exists.
 \label{t1}
\end{thm}
Proof : When $|f'|<1$, since $f' \to 1$ as $\eta \to \infty$, then $f''$ must be greater than zero. Hence, recall from equation~(\ref{eq:e13}),
$$  f'''=1-(f')^2+ ff'' >0.$$
for all $\eta>\eta_0$. After integrating $f'''(\eta) > 0$ from $\eta_0$ to $\eta > \eta_0$, we have
$$f'' (\eta) > f'' (\eta_0 ) = K > 0.$$
Another integration from $\eta_0$ to $\eta > \eta_0$ yields
$$f' (\eta) > f' (\eta_0 )+ K (\eta - \eta_0 ).$$
By Lemma $2$, $f'(\eta)$ has at most one stationary value because one cannot have two consecutive stationary values which are both minima. Since $f''(\eta)>0$, when $\eta \to \infty$, $f'(\eta) \to \infty.$ It violates that $f'(\eta) \to 1$. A similar argument shows that a solution cannot approach to 1 when $|f'|>1$.
\\

The remaining option is that $f(\eta)$ oscillates about 1 as $f'(\infty)\to 1$. But this would imply an infinite number of changes in concavity of $f'$ as $\eta \to \infty$. Once $f'(\eta)$ vibrates from concave upward $(f'''> 0)$ to concave downward $( f''' < 0)$, as $f(\eta)$ becomes great enough, it cannot turn concave upward again \cite{paullet2005nonexistence}. It requires a point $\eta_1$ such that $$f^ {(4)} (\eta_1 ) = 0$$ and $$f^{(5)} (\eta_1 )\geq 0$$
Differentiating equation~(\ref{eq:e13}) gives
\begin{equation}
f^ {(4)}=ff'''-f'f''
 \label{eq:e17}
\end{equation}
and
\begin{equation}
 f^{(5)}= ff^ {(4)} -(f'')^2
 \label{eq:e18}
\end{equation}
Evaluating these at $\eta_1$ yields
 $$f^{(5)}(\eta_1 )= -[f''(\eta_1 )]^2 \leq 0$$
If $f''(\eta_1 ) \neq 0$ we get an immediate contradiction. On the other hand, if $f''(\eta_1 ) = 0$, equation~(\ref{eq:e17}) implies that $f'''=0$. But from equation~(\ref{eq:e13}) we see that $f ''(\eta_1 ) = f''' (\eta_1 ) = 0$ implies the desired contradiction that $$f'(\eta) \equiv 1.$$

Thus, since $f'(\eta)$ cannot ultimately approach to 1 from above or below, nor in an oscillatory manner, no solution to equation~(\ref{eq:e10}) exists. Reversed stagnation-point flow against an impermeable flat wall does not exist in two-dimensional steady case.

\section{FINITE-DIFFERENCE FORMULATIONS}
\begin{figure}[!htb]
\centering
\includegraphics[width=9cm]{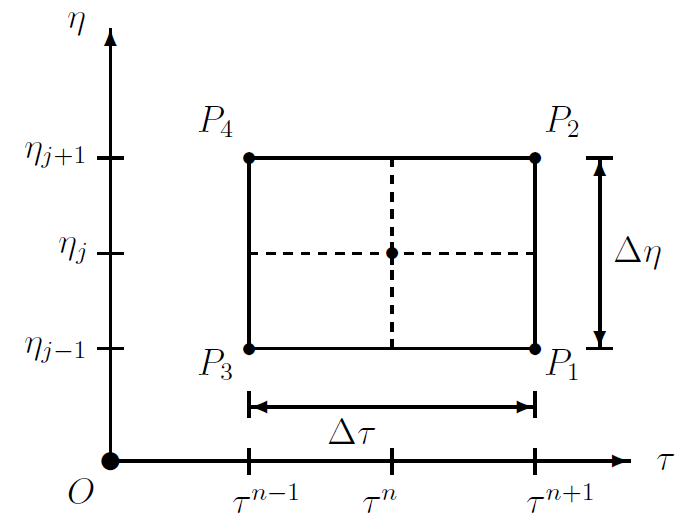}
\caption{Net rectangle for finite-difference method}
\label{kel}
\end{figure}

Similarity solutions of reversed stagnation-point flow with different boundary conditions have been published in \cite{sin2011reversed, sin2011another, chio2012unsteady}. Numerical simulation of reversed stagnation-point flow with full Navier-Stokes equations has been studied in \cite{sin2011computational}. According to the previous work, the governing equations in reversed stagnation-point flow are
  \begin{equation}
    f_{\eta\tau}-(f_{\eta})^2+ff_{\eta\eta}-f_{\eta\eta\eta}+1=0,
     \tag{\ref{e9}}
  \end{equation}
  \begin{equation}
\theta_{\eta\eta} - {Pr}~f\theta_{\eta}= {Pr}~\theta_{\tau}
     \tag{\ref{q:e14}}
  \end{equation}
The above equations subject to the boundary conditions (\ref{e9}) and (\ref{q:e14}) are nonlinear third-order partial differential equations. They do not admit similarity solution and numerical or perturbation methods are required to solve the problem.
\\\\
We shall, however, use here a numerical method. It is an implicit finite-difference method with second-order accuracy. The partial differential equations can be expressed as approximate expressions, so that it is easy to program the solution of large numbers of coupled equation.
\\\\
We start with rewriting the partial differential equations in the form:
\begin{subequations}
\begin{gather}
f_{\eta\tau}=f_{\eta\eta\eta}+(f_{\eta})^2-1+ff_{\eta\eta}\\
\theta_{\tau}=\frac{1}{Pr}\theta_{\eta\eta} -f\theta_{\eta}
\end{gather}
\end{subequations}
and introducing the new dependent variables
\begin{subequations}
\begin{gather}
h=1-f_{\eta}\\
g=\theta
\end{gather}
\end{subequations}
The equations can be rewritten as
\begin{subequations}
\begin{gather}
h_{\tau}=h_{\eta\eta}+2h-h^2+h_{\eta}\int (1-h)~d\eta\\
g_{\tau}=\frac{1}{Pr}g_{\eta\eta} -g_{\eta}\int (1-h)~d\eta
\end{gather}
\end{subequations}
We now contemplate the net rectangle in the $\tau-\eta$ plane shown in Fig. (\ref{kel}) and the net points defined as below:
$$\eta^0 =0,~~~~~~ \eta_j=\eta_{j-1}+\Delta \eta,~~~j=1,2,...J, ~\eta_J=\eta_{\infty}$$
$$\tau^0 =0,~~~~~~ \tau^n=\tau^{n-1}++\Delta \tau, ~~n=1,2,...J,~~~~~~~~~~~$$
Here $n$ and $j$ are just the sequence of numbers that indicate the coordinate location, not tensor indices or exponents.
The partial differential equations are easily discretized by central difference representations with second-order accuracy, for example the finite difference forms for any points are
\begin{equation}
h_{\eta}=\frac{h^n_{i+1}-h^n_{i-1}}{2\Delta \eta}
\end{equation}
and
\begin{equation}
h_{\eta\eta}=\frac{h^n_{i+1}-2h^n_{i}+h^n_{i-1}}{\Delta \eta}
\end{equation}
When $i=0$, since the value of $h^n_{i-1}$ is not logical, the derivative is replaced by the forward difference with second-order accuracy
\begin{equation}
h_{\eta}=\frac{-h^n_{i+2}+4h^n_{i+1}-3h^n_{i}}{2\Delta \eta}
\end{equation}

The finite-difference form of the ODE is written at the midpoint $(\tau^n,~\eta_{j})$, the discretized equation takes the form
\begin{subequations}
\begin{gather}
\frac{h^{n+1}_{i}-h^{n}_{i}}{\Delta \tau}=\frac{h^{n+1}_{i+1}-2h^{n+1}_{i}+h^{n+1}_{i-1}}{(\Delta \eta)^2}+2h^{n}_{i}~~~~~~~~~~~~~~~~~~~~~~~~~~~~~~~~~~~~~~~~~\notag\\
~~~~~~~~~~~~~~~~~~~~~~~~~~~~~~~~~~~-(h^{n}_{i})^2-\frac{h^{n}_{i+1}-h^{n}_{i-1}}{2\Delta \eta}\int^{i\Delta \eta}_{0}(1-h)~d\eta \\\notag\\
\frac{g^{n+1}_{i}-g^{n}_{i}}{\Delta \tau}=\frac{g^{n+1}_{i+1}-2g^{n+1}_{i}+g^{n+1}_{i-1}}{Pr(\Delta \eta)^2}-\frac{g^{n}_{i+1}-g^{n}_{i-1}}{2\Delta\eta}\int^{i\Delta \eta}_{0}(1-h)~d\eta
\end{gather}
\end{subequations}
This procedure yields the following linear tridiagonal system:
\begin{subequations}
\begin{gather}
-\beta h^{n+1}_{i+1}+(1+2\beta)h^{n+1}_{i}-\beta h^{n+1}_{i-1}=~~~~~~~~~~~~~~~~~~~~~~~~~~~~~~~~~~~~~~~~~~~~~~~~\notag\\
~~~~~~~~~~~~~~~~~~~~~~~~h^{n}_{i}+\Delta \tau\left[2h^{n}_{i}-(h^{n}_{i})^2-\frac{h^{n}_{i+1}-h^{n}_{i-1}}{2}\sum^{i}_{0}(1-h^{n}_{i})\right]\\
-\frac{\beta}{Pr} g^{n+1}_{i+1}+\left(1+\frac{2\beta}{Pr}\right)g^{n+1}_{i}-\frac{\beta}{Pr} g^{n+1}_{i-1}=~~~~~~~~~~~~~~~~~~~~~~~~~~~~~~~~~~~~~~~~~~\notag\\
~~~~~~~~~~~~~~~~~~~~~~~~~~~~~~~~~~~~~~~~~~~~g^{n}_{i}-\Delta \tau~\frac{g^{n}_{i+1}-g^{n}_{i-1}}{2}\sum^{i}_{0}(1-h^{n}_{i})
\end{gather}
\label{eq3_5}
\end{subequations}
where $\beta=\Delta \tau/ (\Delta \eta)^2 $.

\begin{figure}[htbp]
\centering
\subfigure{\includegraphics[width=13cm]{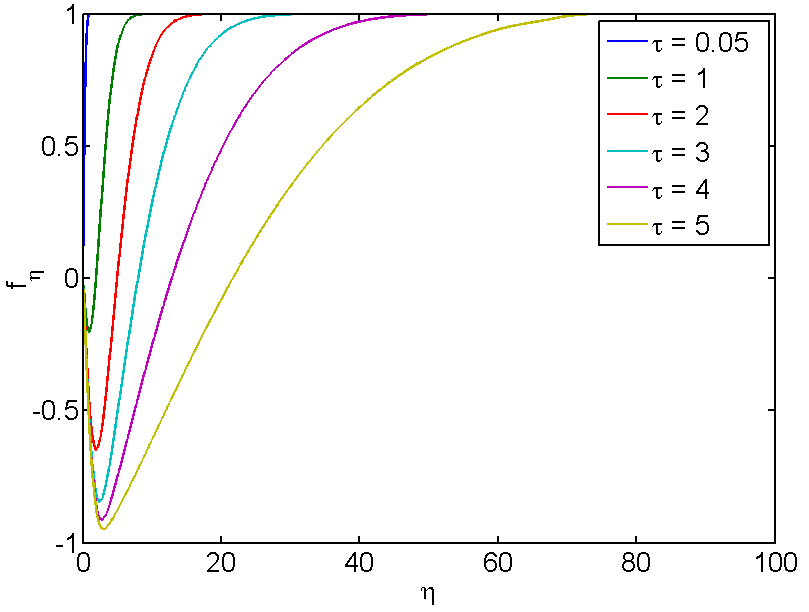}}
\subfigure{\includegraphics[width=13cm]{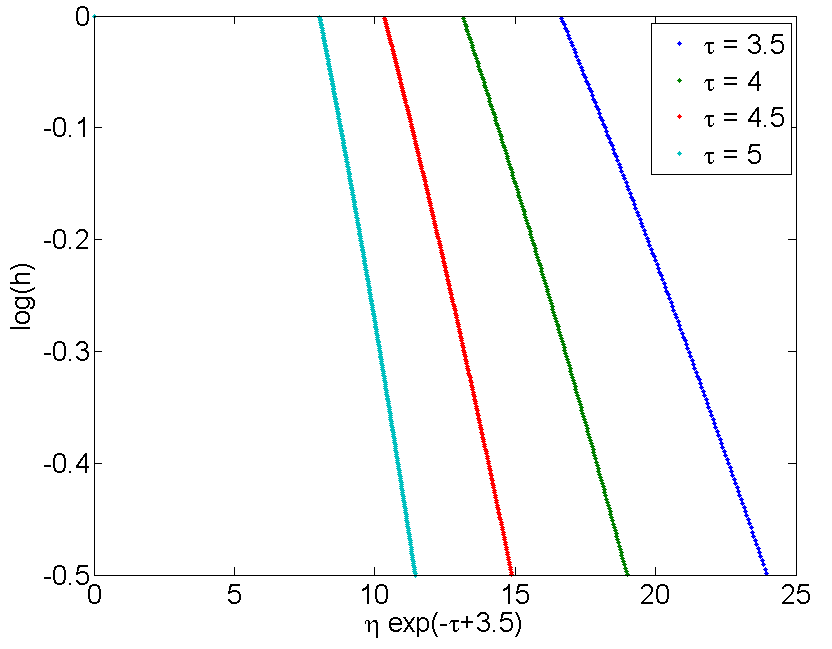}}
\caption{Numerical Solution of equation (\ref{e14}) against (a) $\tau$, (b) $\eta e^{-\tau+3.5}$}
\label{fig:Sub_0}
\end{figure}

\begin{figure}[htbp]
\centering
\subfigure[$Pr=0.7$]{\includegraphics[width=13cm]{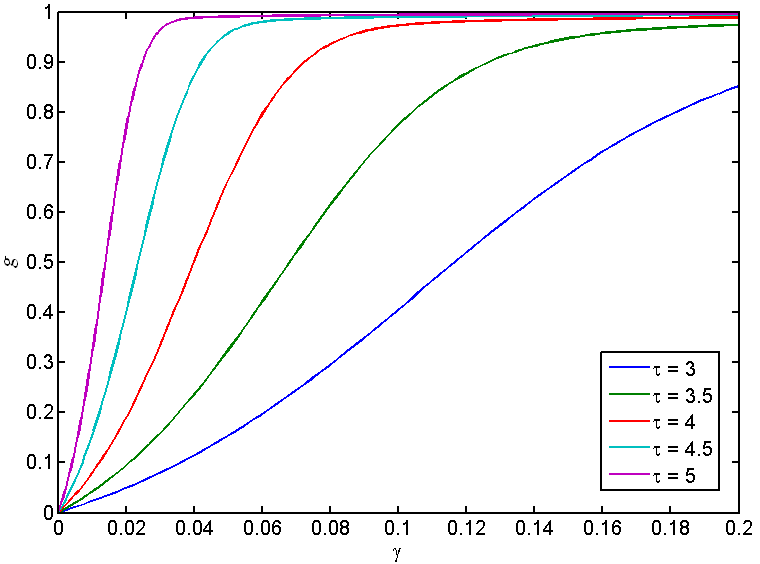}}
\subfigure[$Pr=1$]{\includegraphics[width=13cm]{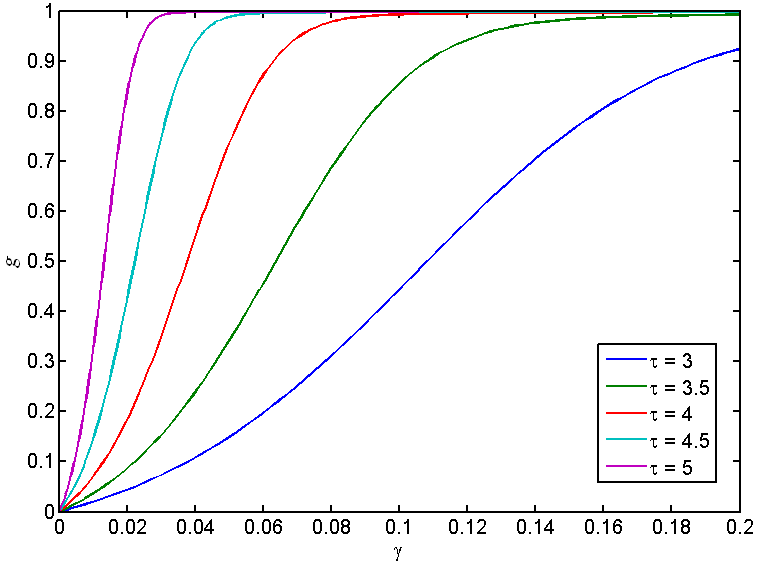}}
\caption{Asymptotic temperature solution $g$ for various value of $\gamma$}
\label{g5}
\end{figure}

\begin{figure}[htbp]
\centering
\subfigure[$Pr=3$]{\includegraphics[width=13cm]{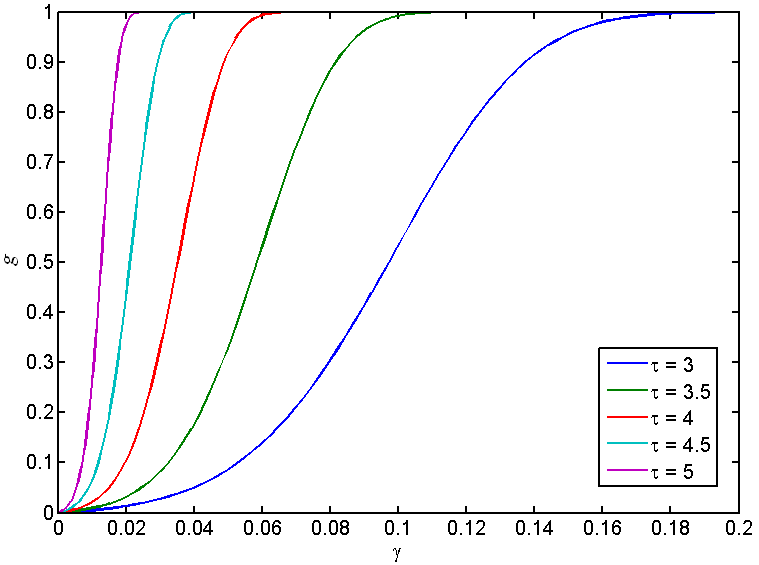}}
\subfigure[$Pr=10$]{\includegraphics[width=13cm]{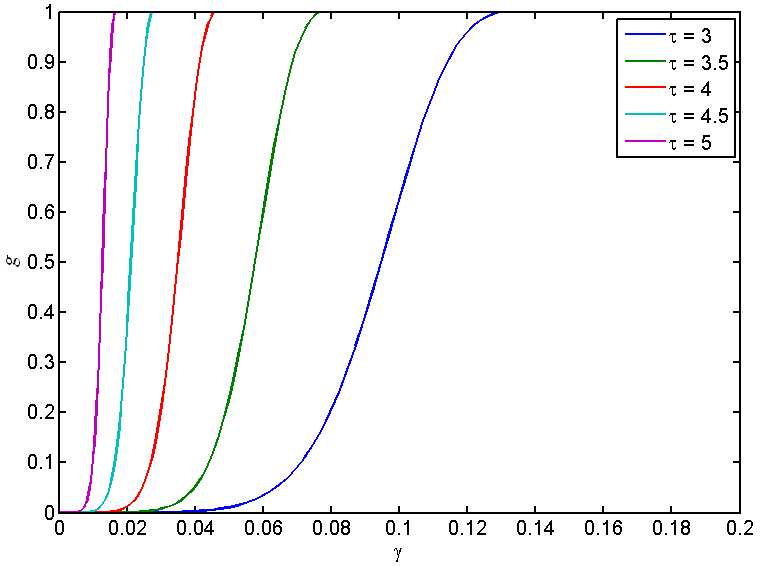}}
\caption{Asymptotic temperature solution $g$ for various value of $\gamma$}
\label{g5c}
\end{figure}

The initial conditions are the solutions of the following second-order linear parabolic differential equations
\begin{subequations}
\begin{gather}
h_{\tau}=h_{\eta\eta}\\
g_{\tau}=\frac{1}{Pr}g_{\eta\eta}
\end{gather}
\label{eq3_20}
\end{subequations}
As can be seen from the energy equation (\ref{q:e14}), equations (\ref{eq3_20}) are identical to the heat conduction equation for one-dimensional unsteady temperature field, and thus, there are many solutions to these differential equations in \cite{goldstein1936boundary}. The desired solutions of (\ref{eq3_20}) have the form
\begin{subequations}
\begin{gather}
h= 1-\mathrm{erf} \left(\frac{\eta}{2\sqrt{\tau}} \right)\\
g= \mathrm{erf} \left(\frac{\eta}{2\sqrt{\tau / Pr}} \right)
\end{gather}
\end{subequations}
where the error function $\mathrm{erf}(z)$ is defined as
\begin{equation}
\mathrm{erf} (z)=\frac{2}{\sqrt{\pi}}\int ^z_0 \exp(-\xi^2)~d\xi
\end{equation}
When $\tau\rightarrow 0,$ the boundary conditions are convenient to write in the form
\begin{equation}
h^{n}_{0}=g^{n}_{0}=0, ~~~~~~~h^{0}_{i}=\mathrm{erf} \left(\frac{\eta_i}{2\sqrt{\tau}} \right),~~~~~~~g^{0}_{i}=\mathrm{erf} \left(\frac{\eta_i}{2\sqrt{\tau / Pr}} \right)
\end{equation}
\\
Equations (\ref{eq3_5}) are defined as being implicit, as more than one unknown appears in the left hand side. They are unconditionally stable, however, set of linear algebraic equations is required to be solved by the tridiagonal matrix algorithm (TDMA), also known as the Thomas algorithm, which is a simplified form of Gaussian elimination that is applied to evaluate tridiagonal systems of equations.
\\\\
For the stability of the diffusion difference equation, the condition of $\beta\leq \frac{1}{2}$ must be satisfied. The procedure is straightforward, except for the algebra. The resulting algorithm of the finite-difference method is written in MATLAB, a numerical computing environment allowing matrix manipulations and plotting of functions and data. At our level of discretization, however, we are only able to resolve in small time range. The numerical results of (\ref{eq3_5}) are presented in Figures (\ref{fig:Sub_0}) to (\ref{g5c}).
\\

From the Proudman-Johnson solution (\ref{e14}),  we have
\begin{equation}
\log h=-c\eta e^{\tau}+\log 2,
\end{equation}
so the graph of $\log h$ against $\eta e^{\tau}$ should provide a straight line of gradient $-c$ if the Proudman-Johnson solution holds. In Figure (\ref{fig:Sub_0}b), the graph of $\log h$ against $e^{-\tau+3.5}$ is plotted for different values of $\tau$. As can be seen, the parallel straight lines for large values of $\tau$ agrees well for the Proudman-Johnson solution. The value of $c$ calculated from the gradient of the straight line in Figure (\ref{fig:Sub_0}b) is $c=3.5$,
which agrees to the estimation of Robins and Howarth.
\\

The following pages (Figures. (\ref{g5}) to (\ref{g5c})) show the numerical solution of temperature distributions with $\mathrm{Pr}$. It is noted that the dimensionless wall temperature gradient $g'(0)$ raises with increase of Prandtl number, but the temperature boundary layer thickness decrease with increase of Prandtl number. Prandtl number is the characteristic number for thermal boundary layers and heat transfer in forced convection. It can be explained by the definition of Prandtl number that inversely proportional to the thermal diffusivity $\alpha$. Prandtl number is a ratio of two quantities which characterize the momentum and heat transport of fluid. In heat transfer problems, the Prandtl number controls the relative thickness of the momentum and thermal boundary layers. When $\mathrm{Pr}$ is small, it means that the heat diffuses very quickly compared to the velocity field. This means that for liquid metals the thickness of the thermal boundary layer is much bigger than the velocity boundary layer.
\\

\begin{figure}[htbp]
\centering
\subfigure[$\tau =3$]{\includegraphics[width=10cm]{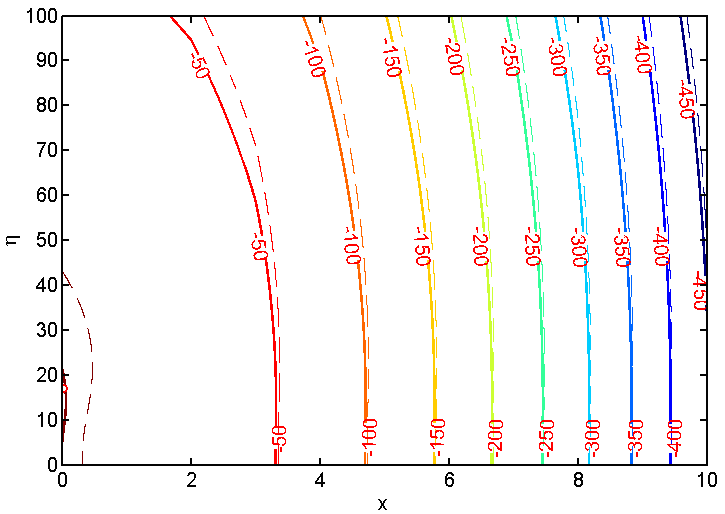}}
\subfigure[$\tau =4$]{\includegraphics[width=10cm]{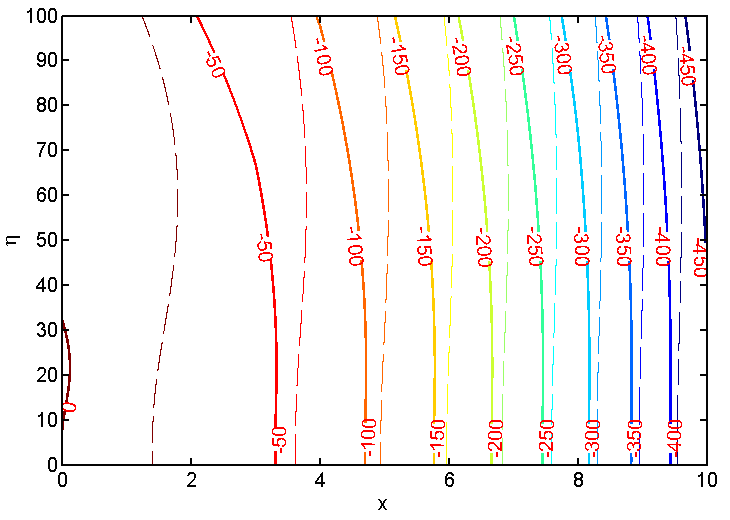}}
\subfigure[$\tau =5$]{\includegraphics[width=10cm]{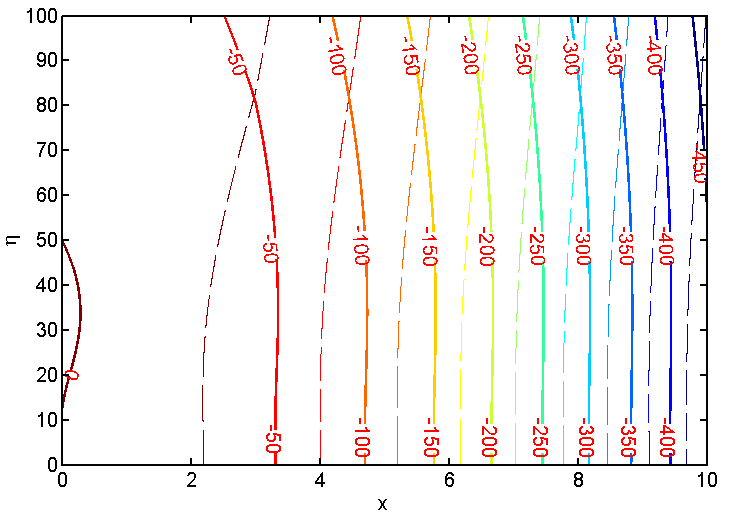}}
\caption{Numerical relations of pressure profiles}
\label{gp3}
\end{figure}

Figure (\ref{gp3}) shows the pressure distribution along the $x$- and $\eta$-direction at different values of $\tau = 3, 4$ and $5$. Lines without markers denote results obtained from numerical procedure and dotted lines are from asymptotic solution. Far away from the wall region, the solution agrees remarkably well for smaller values of $\tau$ with the known asymptotic solution, thus confirming the predictions of the analytical solution.  On the other hand, discrepancy occurs as a larger value of τ is applied in the numerical simulation. However, in the region near the stagnation point, a large difference is observed from the results obtained by these two methods in the region near the stagnation point. Numerical findings show that pressure profiles obtained from asymptotic solution and numerical simulation are in tremendously good agreement for smaller value of $\tau$.  Discrepancy of results in pressure profiles increases for larger value of $\tau$.

\begin{sidewaystable}
\caption{Numerical solution of the reversed stagnation-point flow in similarity variables at various time steps}
\centering
\begin{tabular}{r|rrrrrrr||r|rrrrrrrrrrrrr}
\hline
&&&&$f'(\eta)$&&&&&&&&$f'(\eta)$&\\
\hline
\backslashbox{$\eta$}{$\tau$}  &0.05&1&2&3&4 &5&&\backslashbox{$\eta$}{$\tau$}  &0.05&1&2&3&4&5\\
\hline
  0.05&  0.089& -0.0214& -0.0397& -0.0455& -0.00471 &-0.0476&~&2&  ~~~~~~  1&  0.2036& -0.5521& -0.8083& -0.8861&-0.9178\\
  0.1&  0.1769& -0.0409& -0.0779& -0.0896& -0.09290&-0.0940&& 2.5&  & 0.3895& -0.4683& -0.7849& -0.8887&-0.9350\\
  0.15& 0.2627& -0.0584& -0.1146& -0.1323& -0.1373&-0.1390&&  3&  &  0.5573& -0.3492& -0.7250& -0.8604&-0.9251\\
  0.2&  0.3453& -0.0741& -0.1498& -0.1737& -0.1804&-0.1826&&   4&  &  0.7989& -0.0697& -0.5543& -0.7659&-0.8766\\
  0.25&  0.4238& -0.0878& -0.1834& -0.2136& -0.2221&-0.2249&&  5&  &  0.9234&  0.2039& -0.3622& -0.6549&-0.8168\\
  0.3&  0.4977& -0.0998& -0.2155& -0.2521& -0.2623&-0.2658&&  6&  &  0.9751&  0.4374& -0.1715& -0.5390&-0.7523\\
  0.35&  0.5662& -0.1010& -0.2461& -0.2891& -0.3012&-0.3052&&   7&  &  0.9931&  0.6209&  0.0085& -0.4219&-0.6848\\
  0.4&  0.6289& -0.1184& -0.2752& -0.3247& -0.3386&-0.3432&&  8&  &  0.9983&  0.7564&  0.1731& -0.3059&-0.6151\\
  0.45& 0.6857& -0.1251& -0.3028& -0.3588& -0.3746&-0.3799&&  10&  &  0.9999&  0.9135&  0.4493& -0.8313&-0.4723\\
  0.5&  0.7364& -0.1303& -0.3289& -0.3916& -0.4092&-0.4151&&  12&  &  &  0.9756&  0.6546&  0.1211&-0.3249\\
  0.6&  0.82029& -0.1358& -0.3768& -0.4527& -0.4741&-0.4813&&  15&  &  &  0.9978&  0.8492&  0.3838&-0.175\\
  0.7&  0.8825& -0.1355& -0.4189& -0.5083& -0.5334&-0.5419&& 20&  &  &  1&  0.9765&  0.6997&0.2039\\
  0.8&  0.9264& -0.1296& -0.4555& -0.5582& -0.5872&-0.5971&&     25&  &  &  &  0.9987&  0.8817&0.4722\\
  0.9&  0.9558& -0.1188& -0.4869& -0.6029& -0.6356&-0.6469&& 30&  &  &  & 1&  0.9673&0.6804\\
  1&  0.9747& -0.1035& -0.5131& -0.6424& -0.6790&-0.6917&&    40&  &  &  &  &  0.9998&0.9266\\
  1.5&  0.9992&  0.0256& -0.5770& -0.7711& -0.8272&-0.8481&&  50&  &  &  &  &  1& 0.9977\\
\bottomrule
\end{tabular}
\end{sidewaystable}

\chapter{NUMERICAL SIMULATION}
As noted earlier, the simulation data at the reversed stagnation point is studied by solving the full Navier-Stokes and energy equations numerically, with the aid of a free, open source CFD software package of OpenFOAM. This chapter describes the fundamentals of the finite volume discretization. The technique has been described by many authors \cite{zikanov2011essential,wendt2009computational,openfoam,ferziger1996computational,tannehill1997computational} and is applied to solve the reversed stagnation-point flow.

\section{FINITE VOLUME METHOD}
\begin{figure}[!htb]
\centering
\includegraphics[height= 8cm]{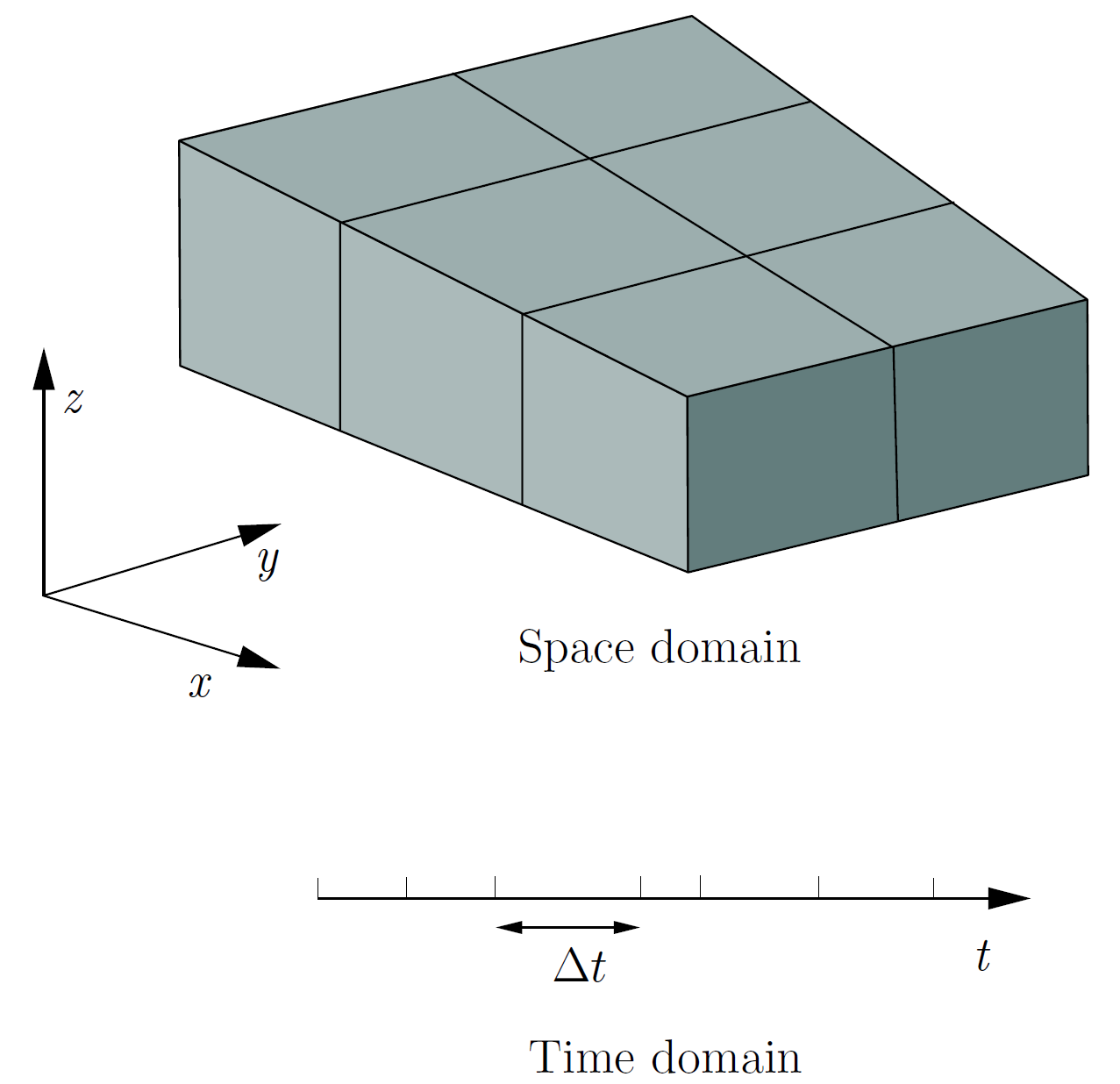}
\caption{Discretization of the solution domain from \cite{openfoam}}
\label{grid}
\end{figure}

The finite volume method (FVM) is a numerical technique that evaluates partial differential equations (PDEs) in the form of algebraic equations. Similar to the finite difference method, values are calculated at discrete places on a meshed geometry. An advantage of the finite volume method is that it is easy to develop to enable unstructured meshes. This feature gives convenience in processing complicated geometries. Once the mesh of the domain is formulated, those governing equations are able to be solved. The method is applied in many computational fluid dynamics packages, for instance, STAR-CD, FLUENT and OpenFOAM.
\\\\
 When solving the Navier-Stokes equations, discretization of the solution domain is shown in Figure (\ref{grid}). The solution domain consists of a space and a time domain. The space domain is subdivided into a set of very small but finite-sized cells or volumes covering the whole domain.  Each cell is stored a set of governing equations to describe the physical phenomenon. Discretization of time involves in subdividing the domain into a set of time steps $\Delta t$, which may alter during a numerical simulation depending on some condition calculated during the simulation.

\begin{figure}[!htb]
\centering
\includegraphics[height= 6.5cm]{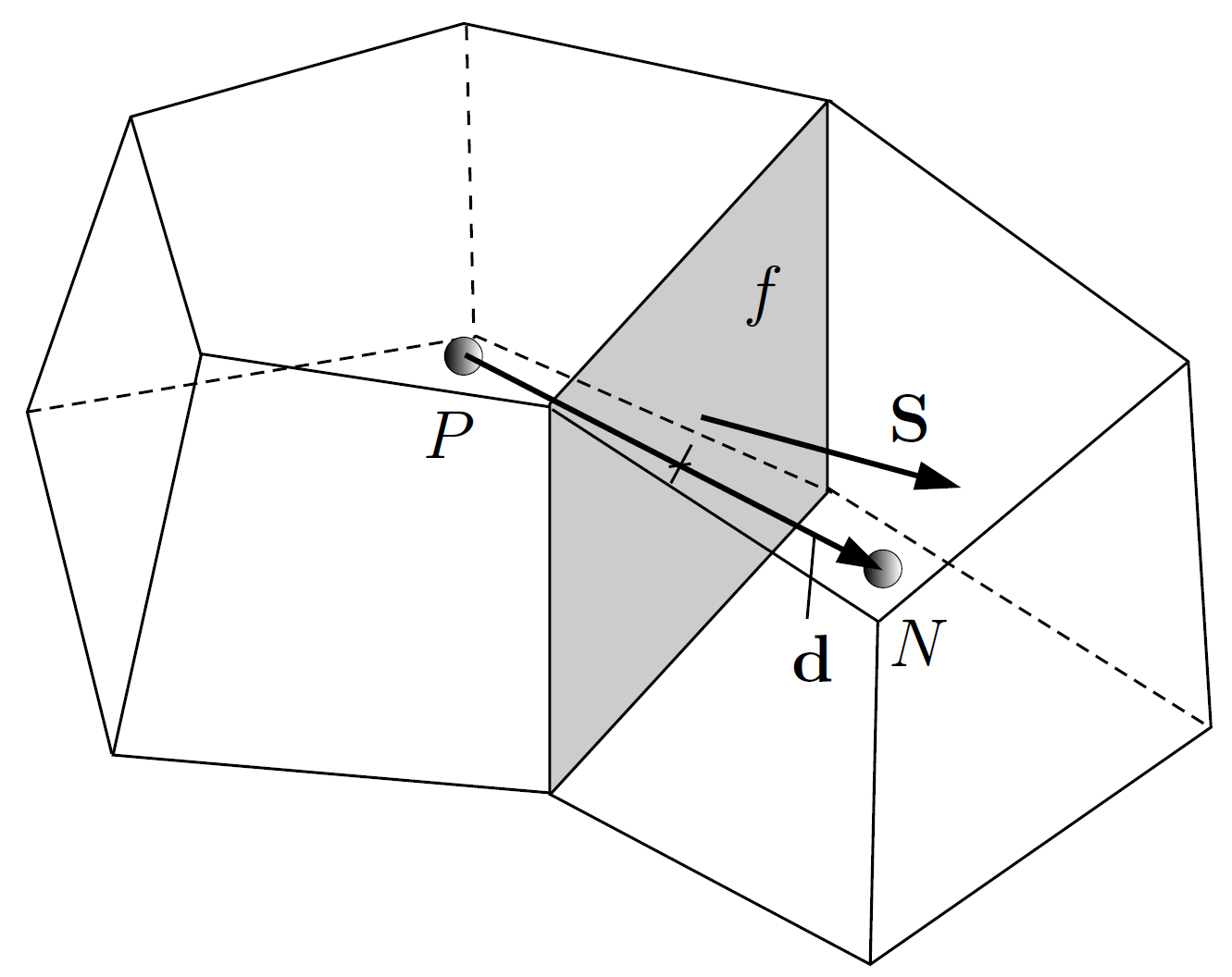}
\caption{Parameters in finite volume discretization from \cite{openfoam}}
\label{gridd}
\end{figure}

Discretization of space involves in subdividing the domain into a number of cells, or control volumes. A typical cell is shown in Figure (\ref{gridd}). As can be seen the cells fill the computational domain without overlap. Dependent variables and other properties are principally assigned at the cell centroid, but it is possible to assign them on faces or vertices. The cell is bounded by a set of flat faces, given the generic label. \\

\section{FUNDAMENTAL EQUATIONS}
The purpose of equation discretization is to transform one or more governing equations into a corresponding system of algebraic equations. The solution of this system approximates the solution to the original partial differential equations at certain locations in space and time. As for compressible flows, the mass conservation is a transport equation for density. With an additional energy equation $p$ can be constructed from a thermodynamic relation (ideal gas law).
\\\\
And for incompressible flows, density variation is not correlated to the pressure field. Mass conservation is a constraint on the velocity field. Combined with the momentum equation,  an equation for the pressure can be  derived analytically. Consider the continuity and Navier-Stokes equations
\begin{equation}
\nabla \cdot \boldsymbol{\vec{V}} =0
\label{eq5_1}
\end{equation}
\begin{equation}
\frac{\partial }{\partial t}\boldsymbol{\vec{V}}+\nabla \cdot \boldsymbol{\vec{V}} \boldsymbol{\vec{V}}=-\frac{1 }{\rho}\nabla p +\nu \nabla^2 \boldsymbol{\vec{V}}
\label{eq5_0}
\end{equation}
where $\rho$ is the density, $p$ is the pressure, $\nu$ is the kinematic viscosity. The principle of conservation of energy will yield the equation of energy for negligible viscous dissipation:
\begin{equation}
\frac{\partial T}{\partial t}+\nabla \cdot T \boldsymbol{\vec{V}}  =\alpha \nabla^2 T
\label{eq5_2}
\end{equation}
where $\alpha= k/\rho c_p$ is thermal diffusivity.
\subsection{INTEGRAL FORM OF THE EQUATION}
If $V$ is a closed region in space enclosed by a surface $S$, then
$$
\int_V \nabla \cdot \Gamma dV=\int_S \boldsymbol{\vec{n}} \cdot \Gamma dS
$$
where $ \boldsymbol{\vec{n}}$ is the outward normal surface vector and $\Gamma$ can represent any tensor field.  The finite volume method performs well on the physical conservation and is adopted in the present study. Taking the volume integral on equations (\ref{eq5_1}) and (\ref{eq5_0}) and transforming equations into the surface integral forms using Gauss divergence theorem, we can get the integral form of the equations as follows
\begin{equation}
\int_S \boldsymbol{\vec{n}} \cdot \boldsymbol{\vec{V}} dS=0
\label{eq5_3}
\end{equation}

\begin{equation}
\frac{\partial }{\partial t}\int_V \boldsymbol{\vec{V}}dV+\int_S \boldsymbol{n} \cdot  \boldsymbol{\vec{V}}\boldsymbol{\vec{V}} dS=-\frac{1 }{\rho}\int_S \boldsymbol{n} \cdot  p dS +\nu \int_S \boldsymbol{n} \cdot \nabla \boldsymbol{\vec{V}}dS
\label{eq5_4}
\end{equation}

\begin{equation}
\frac{\partial }{\partial t}\int_V T dV+\int_S \boldsymbol{n} \cdot  T\boldsymbol{\vec{V}} dS= \alpha \int_S \boldsymbol{n} \cdot \nabla TdS
\label{eq5_5}
\end{equation}

\subsection{FINITE VOLUME DISCRETIZATION}
When solving the Navier-Stokes equations, the region is often discretized using a staggered grid, in which the different unknown variables are not located at the same grid points. In the grid we shall use, the pressure $p$ is located in the cell centers, the horizontal velocity $u$ in the midpoints of the vertical cell edges, and the vertical velocity $v$ in the midpoints of the horizontal cell edges.\\
\begin{figure}[!htb]
\centering
\includegraphics[width= 9cm]{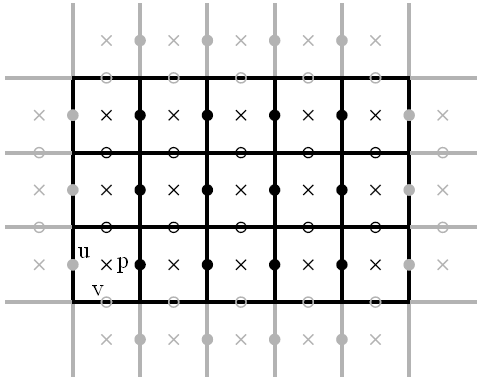}
\caption{Staggered grid with boundary cells}
\end{figure}

The computational domain is divided into a set of discrete volumes which do not overlap and fill the computational domain completely. The above equations are then volume-integrated over each individual finite volume. To convert the divergence terms into surface-integrated flux terms,  Gauss's theorem is used to reduce the problem. The divergence term is discretized to one of finding difference approximations for the fluxes at the surface of the control volume based on the known cell-center values. The temporal derivatives can be discretized using finite-difference approximations. The integrals can be replaced in the sum terms:
\begin{subequations}
\begin{gather}
\int_f \boldsymbol{\vec{n}} dS=\boldsymbol{S}\\
\int_S \boldsymbol{\vec{n}} dS=\sum_f \boldsymbol{S}\\
\int_S \boldsymbol{\vec{n}}\cdot \boldsymbol{\vec{V}} dS=\sum_f \boldsymbol{S}\cdot \boldsymbol{\vec{V}}=\sum_f F\\
\int_VdV = V_P
\end{gather}
\end{subequations}

where $f$ is one face in the polyhedral cells and $P$ represent the cell. Equations (\ref{eq5_3}) and (\ref{eq5_4}) are descretized as follows,
\begin{equation}
\sum_f \boldsymbol{S}\cdot \boldsymbol{\vec{V}}_f=0
\label{eq5_5}
\end{equation}

\begin{equation}
\frac{\partial }{\partial t}\boldsymbol{\vec{V}}_P V_P+\sum_f F\boldsymbol{\vec{V}}_f=-\frac{1 }{\rho}\sum_f \boldsymbol{S}  p  +\nu \sum_f (\boldsymbol{S} \cdot \nabla) \boldsymbol{\vec{V}}dS
\label{eq5_6}
\end{equation}

\begin{equation}
\frac{\partial }{\partial t}T_P V_P+\sum_f FT_f=\alpha \sum_f (\boldsymbol{S} \cdot \nabla) TdS
\label{eq5_7}
\end{equation}

Equations (\ref{eq5_5}) to (\ref{eq5_7}) are linearized by fixing the flux $F$ because of the linear of the variable. As a consequence the vector equation can be decomposed into three component equations. From the equation (\ref{eq5_6}), the discretized $\boldsymbol{\vec{V}}$ can be expressed:

\begin{equation}
\boldsymbol{\vec{V}}_P=\frac{\boldsymbol{H}(\boldsymbol{\vec{V}})}{a_P}-\frac{1}{a_P}\nabla p
\label{1_7}
\end{equation}
and also
\begin{equation}
\boldsymbol{\vec{V}}_f=\left(\frac{\boldsymbol{H}(\boldsymbol{\vec{V}})}{a_P}\right)_f-\left(\frac{1}{a_P}\nabla p\right)_f
\label{1_8}
\end{equation}
\\
It is noticed that equation (\ref{eq5_6}) divided by the finite volume is then converted to the equation (\ref{1_7}) but the pressure field is not discretized. The N-S equation is dependent on the pressure through the pressure gradient term in the momentum equation but we do not have a dependent pressure equation. If the flow is compressible the continuity equation can be used to obtain the density field which can be applied to solve the pressure from an equation of state. \\\\
On the contrary, for incompressible flows, the continuity equation becomes an additional constraint on the velocity field. One way to overcome this difficulty is to build up a pressure field such that velocity satisfies the continuity equation. From the continuity equation (\ref{eq5_1}), the divergence of equation (\ref{1_7}) results in the pressure equation:
\begin{equation}
\nabla \cdot \left(\frac{1}{a_P}\nabla p\right)=\nabla \cdot  \left(\frac{\boldsymbol{H}(\boldsymbol{\vec{V}})}{a_P}\right)
\label{1_9}
\end{equation}

The equation (\ref{1_9}) discretization must use the face interpolation of ${\boldsymbol{H}(\boldsymbol{\vec{V}})}/{a_P}$ and
${1}/{a_P}$, which results in the following expression:

\begin{equation}
\sum_f \boldsymbol{S}\cdot  \left(\frac{1}{a_P}\nabla p\right)_f=\sum_f \boldsymbol{S}\cdot   \left(\frac{\boldsymbol{H}(\boldsymbol{\vec{V}})}{a_P}\right)_f
\label{1_10}
\end{equation}

\section{OPENFOAM}
OpenFOAM (Open Source Field Operation and Manipulation) is a flow solver of choice because it has a pre-existing, robust mesh motion capability that satisfies the GCL, and its source code is freely available through the GNU General Public License. OpenFOAM is a free, open source CFD software package produced by a commercial company, OpenCFD Ltd. It is an object-oriented library written in the C++ language, developing for the customized numerical solvers, and pre-/post-processing utilities for the solution of continuum mechanics problems, including computational fluid dynamics (CFD) \cite{openfoam}.
\\\\
In the commercial and academic organizations, it has a large number of user groups across the engineering and scientific fields. OpenFOAM can be applied to a wide range of capabilities to solve complex fluid flow, involving chemical reactions, turbulence and heat transfer. It has a set of third-party packages ParaView, which is used to post-process the CFD geometry and display analysis using the GUI \cite{koranne2010handbook}. OpenFOAM versions 1.6-dev has been used for this research.
\\
\begin{figure}[htb]
\begin{center}
\includegraphics[width= 14cm]{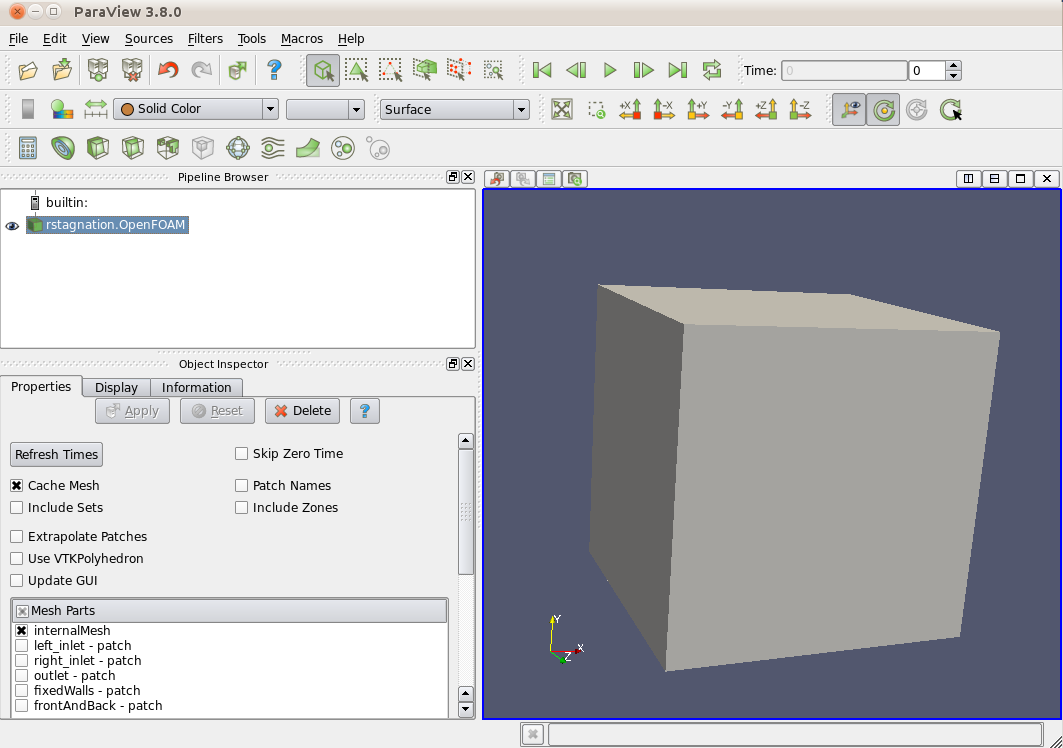}
\caption{ParaView as frontend to OpenFOAM}
\label{para}
\end{center}
\end{figure}

\begin{figure}[!htp]
\begin{center}
\mbox{
\subfigure[OpenFOAM mesh]{\includegraphics[height= 6cm]{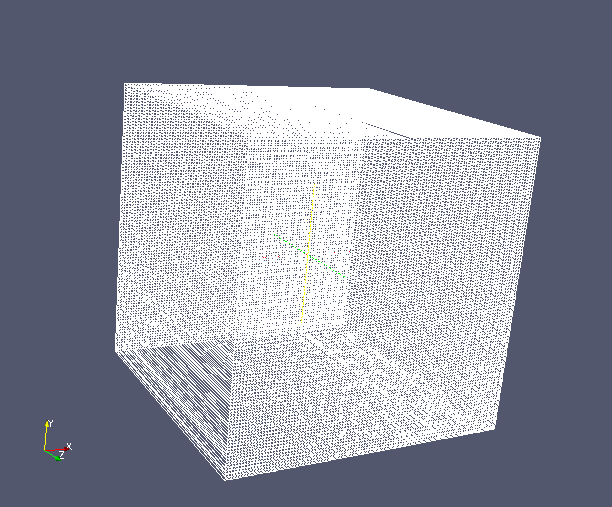}}
\subfigure[OpenFOAM surface with edges]{\includegraphics[height= 6cm]{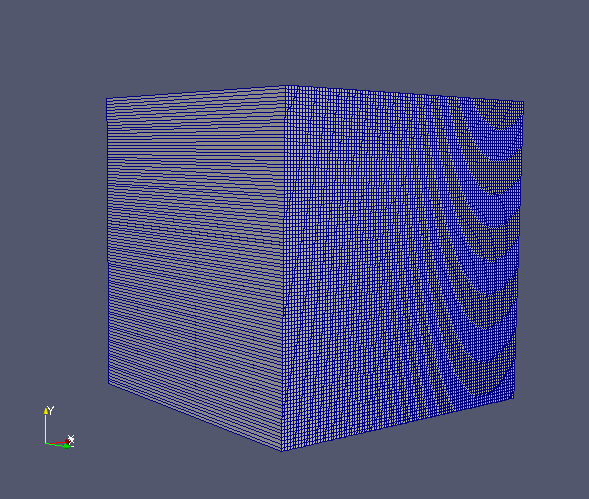}} }
\caption{OpenFOAM block mesh}
\label{subo}
 \end{center}
\end{figure}

\begin{enumerate}
\item
Pre-processing:\\
OpenFOAM provides a mesh generator $\mathbf{blockMESH}$ which the user can divide the geometry into many meshes. Figure (\ref{para}) shows the geometry of a square plate and we saw the use of $\mathbf{blockMESH}$ for mesh generation, shown in Figure (\ref{subo}). OpenFOAM also support converting the format of other CFD packages to the OpenFOAM format.
\item
Solvers:\\
OpenFOAM contains solvers for incompressible flow, channel flow, combustion and stress analysis. In addition, users can create custom solvers without having to modify and recompile the source code with the existing solver.
\item
Post-processing:\\
A plug-in ParaView is used to process the results of simulation cases, provide graphical post-processing and display analysis using the GUI.
\end{enumerate}

\section{ICOFOAM}

The flow problems to be modeled in this research are treated as incompressible and transient, and therefore OpenFOAM’s $\mathbf{icoFoam}$ solver provides a good starting point for the solver development. Since OpenFOAM provides a segregated algorithm to solve the coupled continuity (\ref{eq5_1}) and momentum equations (\ref{eq5_1}), which requires developing equations for each dependent variable and solved sequentially, an iterative method is required to solve the systems of algebraic equations. The $\mathbf{icoFoam}$ solver uses the PISO (Pressure Implicit with Splitting of Operators) algorithm to handle the pressure-velocity coupling. It relates to a momentum predictor and a correction loop, in which a pressure equation based on the volumetric continuity equation is solved and the momentum is corrected based on the pressure change. The PISO algorithm can be described as follows:

\begin{enumerate}
\item
The momentum equation (\ref{1_7}) is solved first by applying the estimated value of pressure field. Accurate source of the pressure gradient at this stage is unknown and the pressure field at the previous time-step is replaced. This stage is called the momentum predictor. The solution of the momentum equation gives an approximation of the new velocity field.
\item
Using the predicted velocities, the $\boldsymbol{H}(\boldsymbol{\vec{V}})$ operator can be substituted  and the pressure equation (\ref{1_10}) can be evaluated. The pressure equation solution provides the first estimate of the new pressure field. This step is known as pressure solution.
\item
It provides a new set of pressure field, which has always been a conservative flux. As a consequence of a new pressure distribution the velocity field should be corrected explicitly by a velocity correction. This is the explicit velocity correction stage.
\end{enumerate}

The velocity corrector consists of two parts: a correction due to the change in the pressure gradient and the transported influence of corrections of neighboring velocities. The fact that the velocity correction is explicit means that the latter part is neglected. The whole velocity error is assumed to come from the error in the pressure term. It is, however, not true and therefore is necessary to correct the $\boldsymbol{H}(\boldsymbol{\vec{V}})$ term, formulating another pressure equation and repeating the procedure. In other words, the PISO loop consists of an implicit momentum predictor followed by a series of pressure solutions and explicit velocity correctors. This loop is repeated until the total variation in the velocity field from one time level to the next is less than a pre-determined tolerance.
\\

\section{MYICOFOAM}

However, the flow is treated as nonisothermal which requires solving the Navier-Stokes equations coupled to the energy equation. A transient solver for incompressible, laminar flow of Newtonian fluids $\mathbf{myicoFoam}$ is configured to model the nonisothermal reversed stagnation-point flow in OpenFOAM. The $\mathbf{myicoFoam}$ solver is an extension of $\mathbf{icoFoam}$ such that it enables solving the Navier-Stokes equations coupled to the energy equation (\ref{eq5_2}).
\\

The main framework is same as the incompressible flow, thereby the steps is as follows:

\begin{enumerate}
\item
The momentum equation (\ref{1_7})  is solved first by applying the estimated value of pressure field. Accurate source of the pressure gradient at this stage is unknown and the pressure field at the previous time-step is replaced. This stage is called the momentum predictor. The solution of the momentum equation gives an approximation of the new velocity field.
\item
The energy equation is solved in which the flux is from solving the previous the momentum equation (\ref{1_7}). The corresponding solution is within the PISO loop, which implies that the energy equation is solved again using the new flux when the new flux is evaluated. This stage is called the energy solution. It is speculated that the thermal coupling is as important as the coupling between the same pressure and velocity.
\item
Using the predicted velocities, the $\boldsymbol{H}(\boldsymbol{\vec{V}})$ operator can be substituted  and the pressure equation (\ref{1_10}) can be evaluated. The pressure equation solution provides the first estimate of the new pressure field. This step is known as pressure solution.
\item
It provides a new set of pressure field, which has always been a conservative flux. As a consequence of a new pressure distribution the velocity field should be corrected explicitly by a velocity correction. This is the explicit velocity correction stage.
\end{enumerate}

\begin{figure}[htb]
\begin{center}
\includegraphics[scale=0.75]{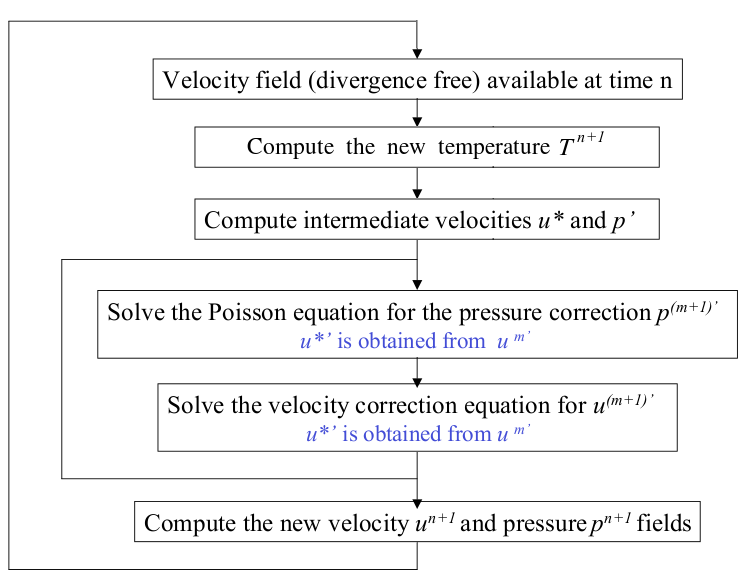}
\caption{Flow diagram of $\mathbf{myicoFoam}$ solver}
\label{piso}
\end{center}
\end{figure}

A flow diagram of $\mathbf{myicoFoam}$ solver is shown in Figure (\ref{piso}). The coupling of pressure and velocity is more important than the coupling with the density and temperature. Before the momentum predictor the density predictor is performed. Pressure solution needs the density change so the energy equation is solved first to update the density which is the main driving force for flow. 

\section{MESH DISTRIBUTION}
While simulating the flow near the wall, due to the viscosity effect near the wall, it is necessary to divide much more mesh in the region near the wall. In OpenFOAM, the mesh distribution can be selected either uniform or non-uniform. The mesh definitions are contained in a list named blocks, consisting of a list of vertex labels, the number of cells in each direction and the cell expansion ratio in each direction.

The meshes are defined as follows:\\
\lstset{language=C++,
         breaklines=true,
         extendedchars=false,
         showstringspaces=false,
         numbers=left,
         numberstyle=\ttfamily\scriptsize,
         frame=trbl,framesep=5pt,framexleftmargin=8mm,
         frameround=tttt,
         keywordstyle=\ttfamily\bf\color{violet},
         ndkeywordstyle=\ttfamily\bf\color{brown},
         commentstyle=\color{blue},
         identifierstyle=\ttfamily\color{black}\bfseries,
         stringstyle=\color{red}\ttfamily
}
\begin{lstlisting}[label=C++,caption=]
convertToMeters 1;
vertices
(
    (0 0 0)
    (1 0 0)
    (1 1 0)
    (0 1 0)
    (0 0 1)
    (1 0 1)
    (1 1 1)
    (0 1 1)
);
blocks
(
    hex (0 1 2 3 4 5 6 7)
    (200 400 1)
    simpleGrading (1 5 1)
);
\end{lstlisting}

The $blockMesh$ dictionary defines a block and the mesh from the vertices. $hex$ means that it is a structured hexahedral block. $(0~1~2~3~4~5~6~7)$ is the vertices used to define a $1~m \times 1~m\times 1~m$ block. These sequences are very important - they should follow the right-hand system. $(200~400~1)$ is the number of mesh cells in each direction.

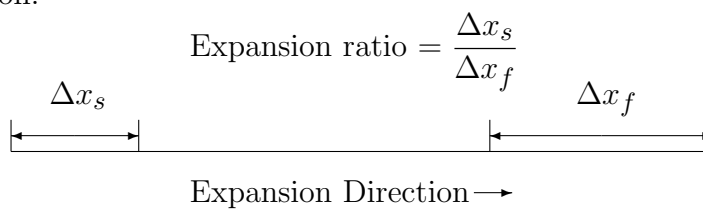
\begin{figure}[!htb]
\begin{picture}(300, 45)

\put(85, 15){\vector(-1, 0){20}}
\put(85, 15){\vector(1, 0){20}}
\put(65, 10){\line(0, 1){10}}
\put(105, 10){\line(0, 1){10}}
\put(250, 15){\vector(-1, 0){35}}
\put(250, 15){\vector(1, 0){35}}
\put(215, 10){\line(0, 1){10}}
\put(285, 10){\line(0, 1){10}}
\put(65, 10){\line(1,0){220}}
\put(77, 25){$\Delta x_s$}
\put(242, 25){$\Delta x_f$}
\put(121, 40){Expansion ratio $= \displaystyle\frac{\Delta x_s}{\Delta x_f}$}
\put(121, -6){Expansion Direction}
\put(210, -3){\vector(1, 0){12}}
\end{picture}
\caption{Expansion ratios in a given direction}
\label{expansion_ratio}
\end{figure}

In order to simulate the flow near the wall, the mesh applied in the model is chosen to be non-uniform.  In OpenFOAM, $simpleGrading~(1~5~1)$ is the expansion ratio.  The ratio is that of the width of the final mesh $\Delta x_f$ along one edge of a block to the width of the start mesh $\Delta x_s$ along that edge, as shown in Figure \ref{expansion_ratio}. The expansion ratio allows a mesh refinement in particular direction. In our model the ratio of mesh widths along $x-$ and $z-$ axis is $1$, along $y-$ is $5$.

\begin{figure}[!htb]
\begin{center}
\includegraphics[scale=1]{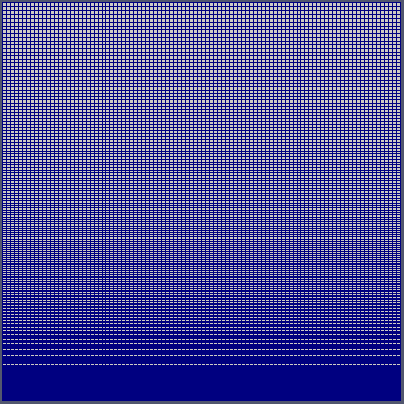}
\caption{Mesh distribution in the model}
\end{center}
\end{figure}

\section{CONVERGENCY}
In general, to obtain a more accurate solution, more meshes should be used in a numerical simulation. One should be kept in mind is that, more time is required to compute a solution if the domain of problem is divided into more meshes. In practical numerical simulations, although an accurate solution is desired, the number of meshes cannot be indefinitely increased because of the limitation of computing facilities and time constraints. To ensure the accuracy and efficiency of a practical numerical simulation, it is necessary to increase the number of meshes until no significant difference of the solutions is obtained by two consecutive simulations. Unfortunately, this is rather difficult to estimate the optimum number of meshes. It takes time and patience to anticipate the mesh number. 
\\\\
On the other side, in numerical simulation the Courant-Friedrichs-Lewy condition (CFL condition) is a necessary condition for convergence in solving hyperbolic PDEs numerically \cite{courant2011partial}. It is applied when explicit method is required in the numerical solution. When a CFD program is running, in order to achieve time accuracy and numerical stability, it requires that the Courant number $Co$ in the flow field is always smaller than 1. The Courant number is defined for one cell as:
$$
Co=\frac{u\Delta t}{\Delta x}
$$
where $u=0.5~m/s$ is the flow velocity in the model and $\Delta x$ is the length interval.  We therefore select based on the worst case which is the maximum $\Delta t$ corresponding to the combined effect of a large flow velocity and small length interval $\Delta x$. In our model the maximum mesh size occurs near the outlet and is equal to the width of the final mesh $\Delta x_f$ along $y-$axis:
$$
\Delta x=\frac{block~length}{number~of~mesh}\times {expansion~ratio} =\frac{(1~m)(5)}{400}=0.0125~m
$$
As a consequence, to achieve a Courant number less than or equal to 1 throughout the domain, the time step $\Delta t$ must be less than a specified time in the time-marching computer simulations, otherwise the simulation will produce incorrect results. The time step $\Delta t$ must be set to less than or equal to:
$$
\Delta t=\frac{Co \Delta x }{u_0}\leq \frac{(1) (0.0125~m) }{u_0}=\frac{0.0125}{u_0}~s
$$
After the progress of the simulation, we hope to get accurate results in the early time interval, so we can later view with a post-processing package. 
The factor of $0.08$ is taken from the experience that can be advantageous for accuracy and thus in our analysis the times step is equal to
$$
\Delta t=\frac{Co \Delta x }{u_0}\leq \frac{(1) (0.0125~m) }{u_0}=\frac{0.0125}{u_0}~s.
$$

\section{BOUNDARIES}
\begin{figure}[!htb]
\begin{center}
\includegraphics[width= 9cm]{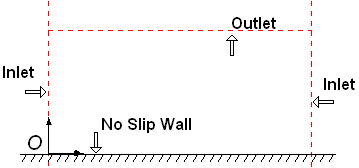}
\caption{Boundaries of reversed stagnation-point flow in a control volume}
\label{gib}
\end {center}
\end{figure}

In this section we discuss the way in which boundaries are treated in reversed stagnation-point flow. In order to solve the governing equations by the numerical method described in the previous section, boundary conditions must be prescribed. The boundaries involved in our model are not only simple geometric boundary conditions, but also the integral part of the solution and numerical simulation through boundary conditions or inter-boundary connections.
\\\\
We first need to consider setting up a numerical configuration of the simulation; the boundary has to be specified. The conditions consist of two inflow boundaries, an outflow boundary, and a symmetry plane on one of the two faces parallel to the plane of the paper and no-slip walls for the remaining boundaries. A schematic diagram of the problem is given in figure \ref{gib}. In OpenFOAM the boundary conditions of our problem is defined in the $blockMeshDict$ dictionary:

\begin{lstlisting}[label=C++,caption=]
patches
(
    patch left_inlet    ((2 6 5 1))
    patch right_inlet   ((0 4 7 3))
    patch outlet        ((3 7 6 2))
    wall fixedWalls     ((1 5 4 0))
    empty frontAndBack  ((0 3 2 1)
                         (4 5 6 7))
);
\end{lstlisting}

We select the uniform velocity profile for the inflow boundary. Discretization (\ref{1_10}) of the momentum equation (\ref{eq5_7}) involves the values of velocity on the boundary.  These velocity values are obtained from a discretization of the boundary conditions of the continuous problem.

\begin{enumerate}
\item
Inflow conditions:\\
On an inflow boundary the velocities are explicitly given; we impose this for the velocities normal to the boundary by directly fixing the values on the boundary line.
\item
Outflow conditions:\\
In the outflow boundary condition the normal derivatives of both velocity components are set to zero at the boundary, which means that the total velocity does not change in the direction normal to the boundary, i.e.,
$$ \frac{\partial u}{\partial x}=\frac{\partial v}{\partial y}=0$$
\item
No-slip condition: \\
The continuous velocities should vanish at the wall boundary to satisfy the no-slip condition. For the values laying directly on the wall boundary we thus set both velocity component to zero.
$$u(x,0)=v(x,0)=0$$
\item
Symmetry plane: \\
Our problem is a two-dimensional problem which is symmetric about $z-$axis. This means that boundary condition refers to a planar boundary surface. Values lying directly on the boundary are not required to calculate.
\end{enumerate}

The expression of the velocity values on the boundary is shown as following:
\begin{lstlisting}[label=C++,caption=Velocity Boundary conditions]
boundaryField
{
    left_inlet   {type    fixedValue;
                  value   uniform (-1 0 0);}
    right_inlet  {type    fixedValue;
                  value   uniform (1 0 0);}
    outlet       {type    zeroGradient;}
    fixedWalls   {type    fixedValue;
                  value   uniform (0 0 0);}
    frontAndBack {type    empty;}
}
\end{lstlisting}

\begin{figure}[!htb]
\begin{center}
\includegraphics[width = 15cm]{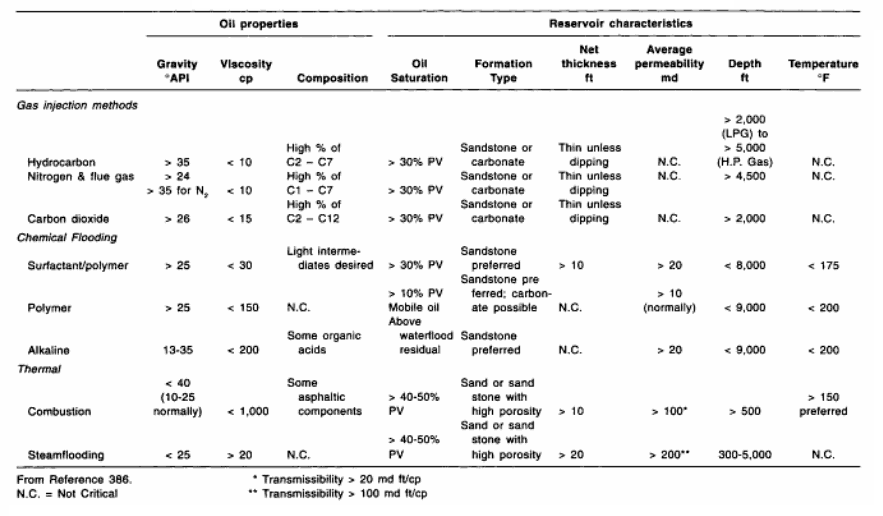}
\caption{Properties of crude oil in enhanced recovery methods \cite{lyons2005standard}}
\label{tg2}
\end {center}
\end {figure}

Each patch defines a type, a name, and a list of boundary faces. The patch is defined by three sides of the block based on the vertex numbers. The order of the vertex numbers is such that they are marched clockwise when looking inside from the control volume. For example, type of the boundary $fixedWalls$ is defined as $wall$. For two-dimensional flow the flow field on the boundary $frontAndBack$ is not required to evaluate, thereby is defined as $empty$.
\\\\
For the temperature profile in nonisothermal flow, there are essentially two different boundary conditions; to impose these, we divide the boundary into two parts:
\begin{enumerate}
\item
Dirichlet boundary conditions:\\
Using this boundary condition, the constant wall temperature $T_w$ is prescribed at the wall. The temperature of the fluid from a wall may be described in the form
$$T(x,0)=T_w$$
\item
Neumann boundary conditions:\\
This boundary condition describes how much heat is passed on to the wall by the fluid. This is determined by both the material properties of the wall and the temperature difference across the wall. For a constant fluid's thermal conductivity $k$ and heat flux $q_w$ across the wall, it may be described in the form
$$-k\frac{\partial T}{\partial y}=q_w$$
\end{enumerate}
In our model, the external flow temperature and wall temperature are constants and Dirichlet boundary conditions are required. The expression of the temperature values on the boundary is shown as following:
\begin{lstlisting}[label=C++,caption=Temperature Boundary conditions]
boundaryField
boundaryField
{
    left_inlet   {type    fixedValue;
                  value   uniform 373;}
    right_inlet  {type    fixedValue;
                  value   uniform 373;}
    outlet       {type    zeroGradient;}
    fixedWalls   {type    fixedValue;
                  value   uniform 273;}
    frontAndBack {type    empty;}
}
\end{lstlisting}
On the other side, the selection of fluid is difficult because crude oil is predominantly a mixture of hydrocarbons. Under surface pressure and temperature conditions, the lighter hydrocarbons methane, ethane, propane and butane occur as gases, while the heavier ones from pentane and up are in the form of liquids or solids.
\\\\
It is, however, in the underground oil reservoir the proportion which is gas or liquid varies depending on the subsurface conditions. This represents that the flow system is not single-phase, but is multiphase. As a result, in order to simplify the difficulties in simulation, it is possible to select another fluid to replace the crude oil. After simulation, the numerical result can be analyzed in the crude oil situation, by comparing Reynolds number. We create a fluid that has similar physical properties of crude oil. In Fig.~(\ref{tg2}), the viscosity of crude oil is approximately 10 $cp$ , which equals to 0.01 $Pa\cdot s$.

\section{DIMENSIONESS NUMBER}
Besides, it is required to know fluid physical properties. For our transient solver $myicoFoam$ the physical properties $nu$ and $DT$ are stored in the $transportProperties$ file which is a dictionary for the dimensioned scalar. The first items loaded is the kinematic viscosity from the $transportProperties$ dictionary file and is equal to $\nu$. Another transport property related to the thermal diffusion denoted as $DT$ equals to $DT ={\nu}/{Pr}$
\\

In this thesis, effect of reversed stagnation-point on the nonisothermal flow field behavior is studied. The two parameters $u_0$ and $DT$, which are the inflow velocity at both the left and right boundary and molecular thermal diffusivity respectively, are investigated in the numerical simulation. \\

One of the contributions in this thesis is to acquire the relationship of Reynolds number and flow velocity, where the reversed stagnation-point flow exists under the condition of ensuring the flow is laminar.  Reynolds number is a dimensionless flow property. It gives a measure of the ratio of inertial forces to viscous forces and, consequently, it quantifies the relative importance of these two types of forces for given flow conditions. Reynolds number is defined as:
$$Re =  \displaystyle \frac{V L}{\nu} $$
where $V$ is the mean velocity and $L$ is the characteristic length, equals to half of the length of wall. Reynolds number can also describe the property of the flow, whether it is laminar, transition or turbulent flow. For a smooth flat plate with a uniform free stream, the transition process begins at a critical Reynolds number, $\textup{Re}_{\textup{critical}}\approx 1\times 10^5$, and continues until to the turbulent at
the transition Reynolds number, $\textup{Re}_{\textup{transition}}\approx 3\times 10^6$. The flow is said to be laminar flow when $\textup{Re}<\textup{Re}_{\textup{critical}}\approx 1\times 10^5$. Several cases of simulations with varied Reynolds number were performed, in which the value of $u_0$ is chosen from $0.5$ to $20$ with fixed value of $\nu$ in part of the simulations. \\

Reynolds number can be obtained as applying the nondimensional form of the incompressible Navier-Stokes equations:
\begin{equation}
\rho\left[\frac{\partial\boldsymbol{\vec{V}}}{\partial t}+(\boldsymbol{\vec{V}} \cdot \nabla )\boldsymbol{\vec{V}}\right]=-\nabla p+\mu \nabla^2 \boldsymbol{\vec{V}}+\boldsymbol{\vec{f}}
\end{equation}
When the equations undergo the dimensionless analysis, that is when it is multiplied by a factor with inverse units of the origin equation, we acquire a form which does not depend directly on the physical sizes. One possible way to get a nondimensional equation is to multiply the whole equation by the factor ${L}/{\rho V^2}$
and to set
$$ \boldsymbol{\vec{V'}} = \frac{\boldsymbol{\vec{V}}}{V},\ p' = p\frac{1}{\rho V^2}, \ \boldsymbol{\vec{f'}} = \boldsymbol{\vec{f}}\frac{L}{\rho V^2}, \ \frac{\partial}{\partial t'} = \frac{L}{V} \frac{\partial}{\partial t}, \ \nabla' = L \nabla, $$
The Navier-Stokes equation can be rewritten without dimensions:
\begin{equation}
\frac{\partial \boldsymbol{\vec{V'}}}{\partial t'} + (\boldsymbol{\vec{V'}} \cdot \nabla )\boldsymbol{\vec{V'}} = -\nabla' p' +\frac{\nu}{V L} \nabla'^2 \boldsymbol{\vec{V'}} +\boldsymbol{\vec{f'}}
\end{equation}
Finally, dropping the primes, we have
\begin{equation}
\frac{\partial\boldsymbol{\vec{V}}}{\partial t}+(\boldsymbol{\vec{V}} \cdot \nabla )\boldsymbol{\vec{V}}=-\nabla p+\frac{1}{\mathrm{Re}} \nabla^2 \boldsymbol{\vec{V}}+\boldsymbol{\vec{f}}
\end{equation}
Therefore, all flows with the same Reynolds number are comparable mathematically. It is noted that, in the above equation, as $\mathrm{Re} \rightarrow \infty$ the viscous terms vanish.  High Reynolds number flows are approximately inviscid in the external flow. Meanwhile, the velocity components and nondimensional variable of our problem can be rewritten in the form:
\begin{subequations}
  \begin{gather}
u= -xf_{\eta}({\eta})\\
v=  {\mathrm{Re}}^{-\frac{1}{2}}f({\eta})\\
\eta={\mathrm{Re}}^{\frac{1}{2}}y\\
\tau=t
  \end{gather}
\label{sed}
\end{subequations}
and the governing similarity equation remains unchanged:
\begin{equation}
f_{\eta\tau}-(f_{\eta})^2+ff_{\eta\eta}-f_{\eta\eta\eta}=-\frac{3}{4\tau^2}
\end{equation}
That is why the results of direct numerical simulations and that of similarity analysis are comparable with the same Reynolds number. On the other side, studying the effect between different Prandtl number and temperature distribution in nonisothermal flow is the second goal in this thesis. In Chapter 2 the definition of Prandtl number is introduced as:
$$Pr =   \displaystyle \frac{\rho c_p \nu}{k} = \frac{\nu}{DT} $$
Prandtl number is the ratio of momentum diffusivity to thermal diffusivity. Several cases of simulations with different Prandtl number were performed, ranging from $0.003$ to $0.1$ with fixed value of $\nu$ in part of the simulations. The following table illustrates the parameters used in this study.

\begin{table}[!hbp]
\large
\caption{Reversed stagnation-point flow with parameters}
\begin{center}
\begin{tabular}{lcccccc}
\hline
 Case &$\nu$&$DT$& $u_0~$(m/s) & $\Delta t~$(s)& $Re$& $Pr$\\
\hline
1  &$0.01$ &$0.01$  &1    &$0.001$    & 50    &1\\
2  &$0.01$ &$0.01$  &2    &$0.0005$   & 100   &1\\
3  &$0.01$ &$0.01$  &5    &$0.0001$   & 250   &1\\
4  &$0.01$ &$0.01$  &10   &$0.00005$  & 500   &1\\
5  &$0.01$ &$0.01$  &20   &$0.000025$ & 1000  &1\\
6  &$0.01$ &$0.01$  &50   &$0.00001$  & 2500  &1\\
7  &$0.01$ &$0.01$  &100  &$0.000005$ & 5000  &1\\
8  &$0.01$ &$0.01$  &200  &$0.0000025$& 10000 &1\\
9  &$0.01$ &$0.033333333$ &20  &$0.000025$    & 1000    &0.3\\
10 &$0.01$ &$0.014285714$ &20    &$0.000025$    & 1000    &0.7\\
11 &$0.01$ &$0.003333333$  &20    &$0.000025$    & 1000   &3\\
12 &$0.01$ &$0.001428571$  &20    &$0.000025$    & 1000   &7\\
13 &$0.01$ &$0.001$   &20   &$0.000025$    & 1000   &10\\
\hline
\end{tabular}
\label{TPs}
\end{center}
\end{table}

\chapter{RESULT AND DISCUSSION}
In this chapter, the numerical results of reversed stagnation-point flow in OpenFOAM will be discussed. The results of direct numerical simulations are compared to the analytical solutions of the reversed stagnation-point flow to ensure validation of modeling in the simulations and to check the reliability of the numerical results.

\section{FLOW VISUALIZATION}
\begin{figure}[htbp]
\centering
\subfigure[$t=0.05$]{\includegraphics[width=7cm]{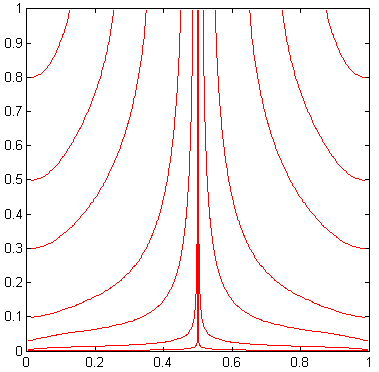}}
\subfigure[$t=0.1$]{\includegraphics[width=7cm]{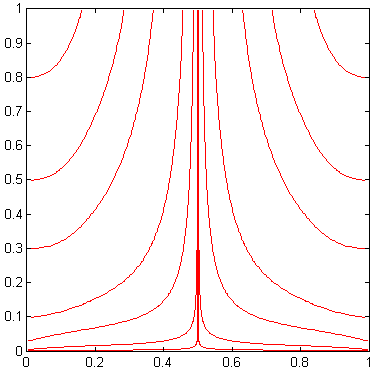}}
\subfigure[$t=0.2$]{\includegraphics[width=7cm]{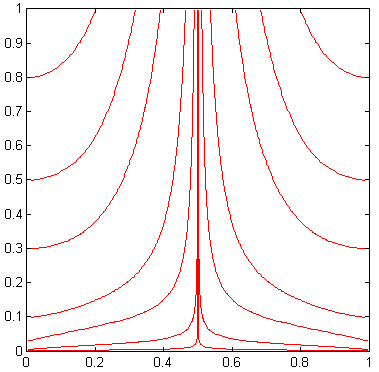}}
\subfigure[$t=0.3$]{\includegraphics[width=7cm]{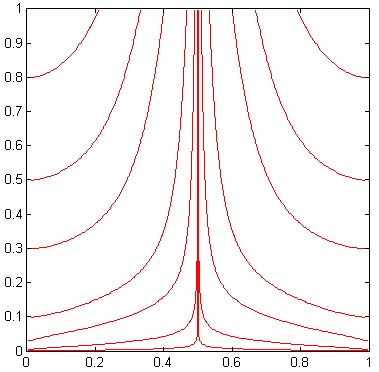}}
\subfigure[$t=0.4$]{\includegraphics[width=7cm]{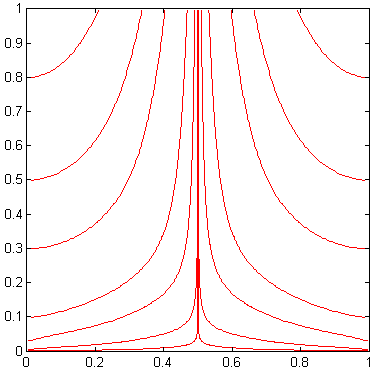}}
\subfigure[Steady]{\includegraphics[width=7cm]{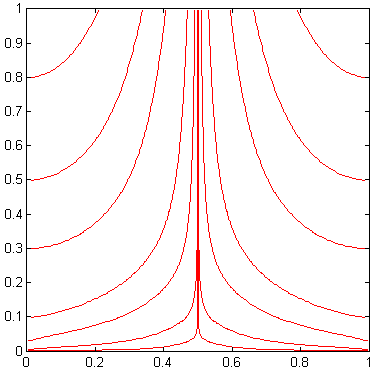}}
\caption{Stream Line, time evolution at $Re=50$}
\label{Splot9}
\end{figure}

\begin{figure}[htbp]
\centering
\subfigure[$t=0.01$]{\includegraphics[width=7cm]{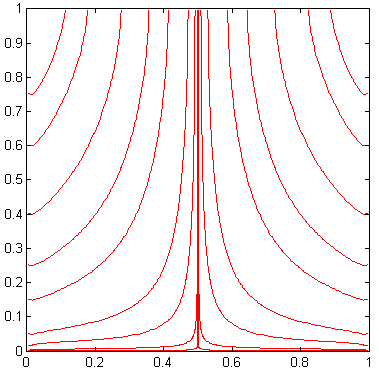}}
\subfigure[$t=0.1$]{\includegraphics[width=7cm]{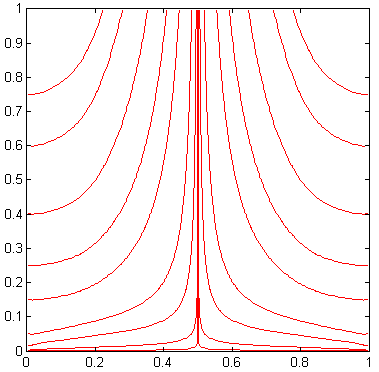}}
\subfigure[$t=0.2$]{\includegraphics[width=7cm]{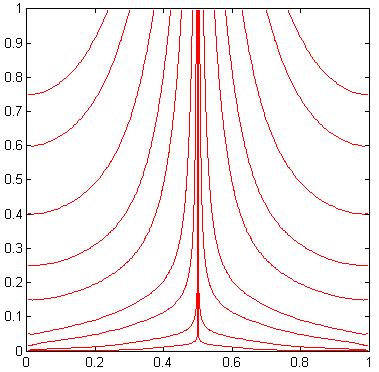}}
\subfigure[$t=0.23$]{\includegraphics[width=7cm]{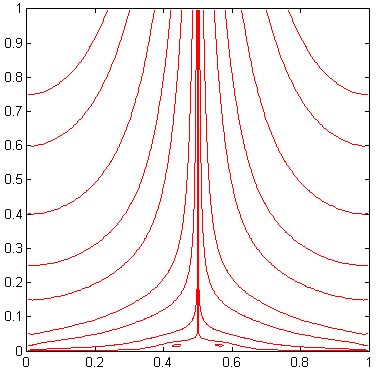}}
\subfigure[$t=0.25$]{\includegraphics[width=7cm]{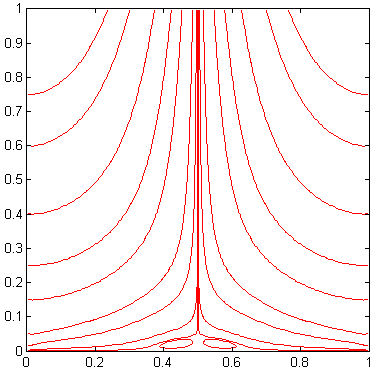}}
\subfigure[$t=0.28$]{\includegraphics[width=7cm]{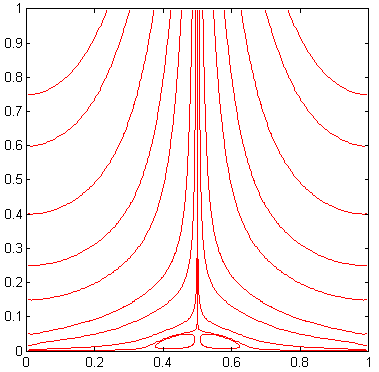}}

\caption{Stream Line, time evolution at $Re=100$}
\label{Splot7}
\end{figure}

We can plot the position of each particle in our simulation inside of the control volume to see the effects of the streamlines for various Reynolds number. The following pages (Figures \ref{Splot9} to \ref{Splot2}) show the stream lines, both evolving in time as well as at steady state, at various $u_0$. At $t=0$, the inflow velocity is instantaneously set from zero to $u_0$, thereby slowly setting in motion the fluid initially at rest.
\\

\addtocounter{figure}{-1}
\begin{figure}[htbp]
\addtocounter{subfigure}{6}
\centering
\subfigure[$t=0.3$]{\includegraphics[width=7cm]{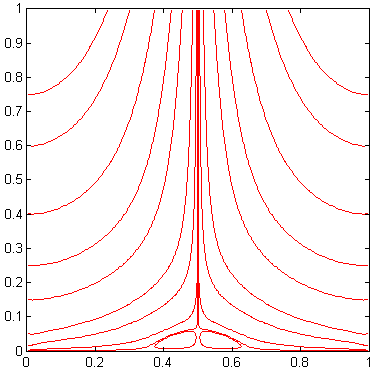}}
\subfigure[$t=0.35$]{\includegraphics[width=7cm]{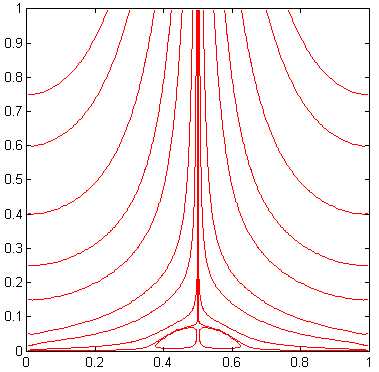}}
\subfigure[$t=0.4$]{\includegraphics[width=7cm]{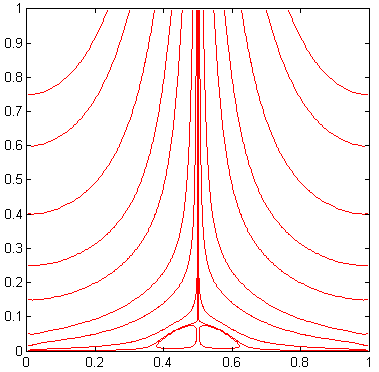}}
\subfigure[$t=0.5$]{\includegraphics[width=7cm]{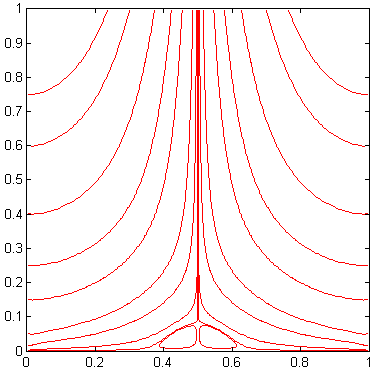}}
\subfigure[$t=0.6$]{\includegraphics[width=7cm]{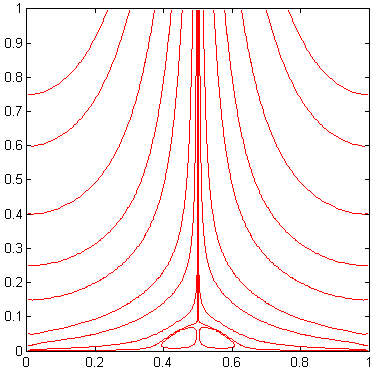}}
\subfigure[Steady]{\includegraphics[width=7cm]{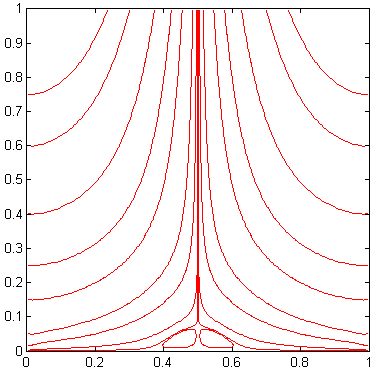}}
\caption{Stream Line, time evolution at $Re=100$}
\label{Splot8}
\end{figure}

\begin{figure}[htbp]
\centering
\subfigure[$t=0.01$]{\includegraphics[width=7cm]{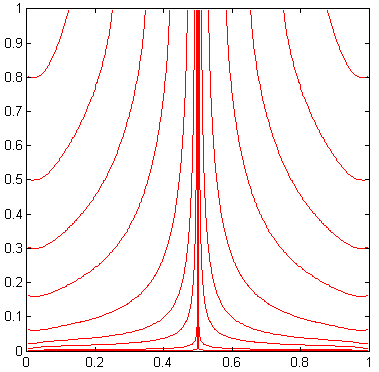}}
\subfigure[$t=0.05$]{\includegraphics[width=7cm]{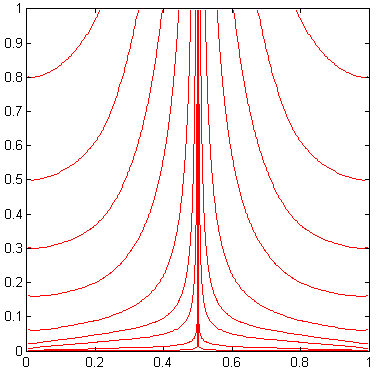}}
\subfigure[$t=0.07$]{\includegraphics[width=7cm]{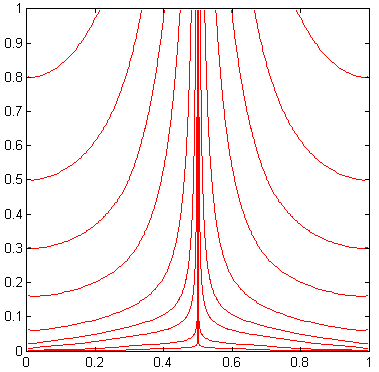}}
\subfigure[$t=0.075$]{\includegraphics[width=7cm]{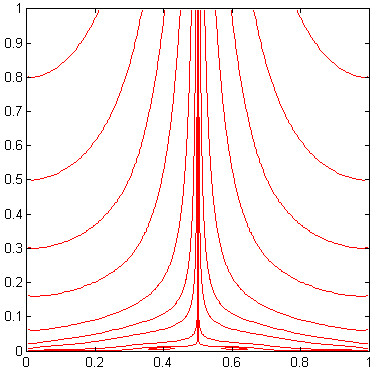}}
\subfigure[$t=0.08$]{\includegraphics[width=7cm]{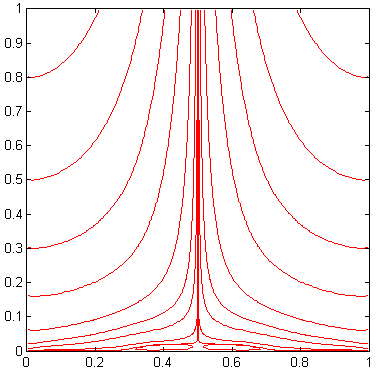}}
\subfigure[$t=0.09$]{\includegraphics[width=7cm]{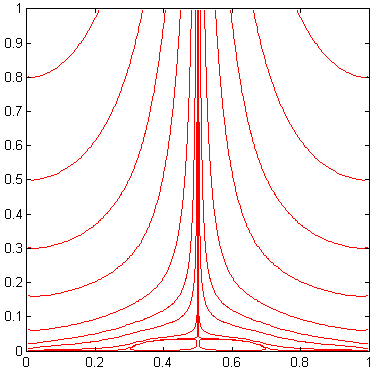}}
\caption{Stream Line, time evolution at $Re=250$}
\label{Splot5}
\end{figure}
\addtocounter{figure}{-1}
\begin{figure}[htbp]
\addtocounter{subfigure}{6}
\centering
\subfigure[$t=0.1$]{\includegraphics[width=7cm]{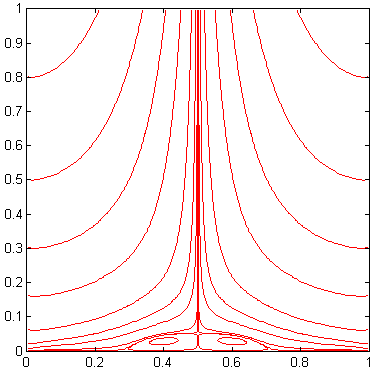}}
\subfigure[$t=0.11$]{\includegraphics[width=7cm]{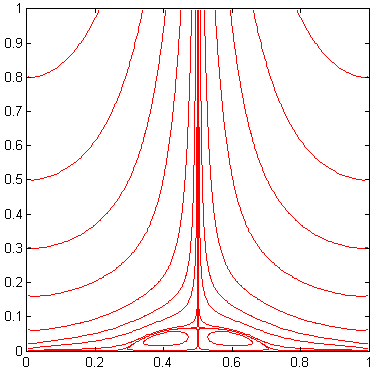}}
\subfigure[$t=0.13$]{\includegraphics[width=7cm]{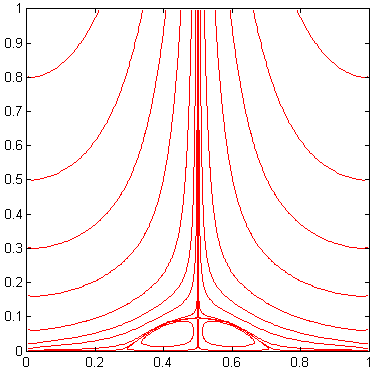}}
\subfigure[$t=0.15$]{\includegraphics[width=7cm]{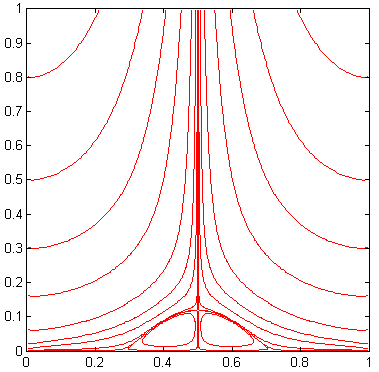}}
\subfigure[$t=0.2$]{\includegraphics[width=7cm]{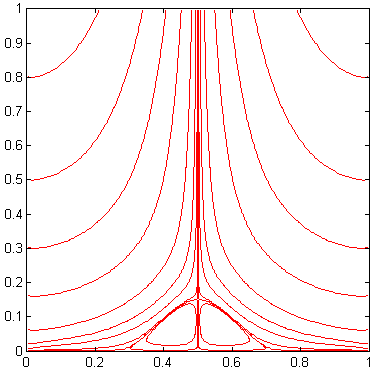}}
\subfigure[Steady]{\includegraphics[width=7cm]{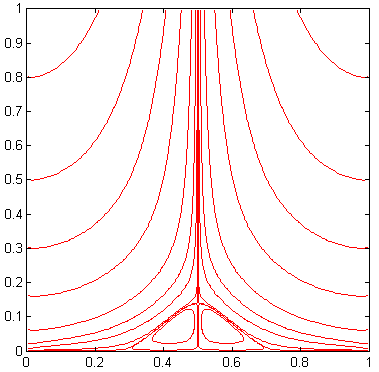}}
\caption{Stream Line, time evolution at $Re=250$}
\label{Splot6}
\end{figure}

\begin{figure}[htbp]
\centering
\subfigure[$t=0.02$]{\includegraphics[width=7cm]{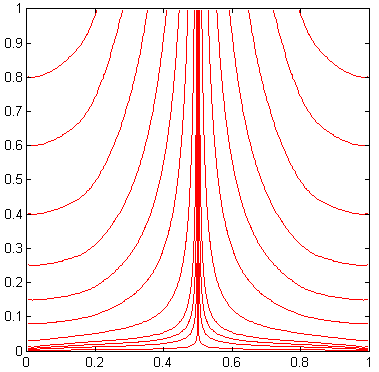}}
\subfigure[$t=0.04$]{\includegraphics[width=7cm]{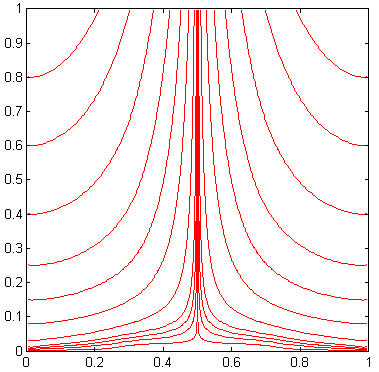}}
\subfigure[$t=0.0425$]{\includegraphics[width=7cm]{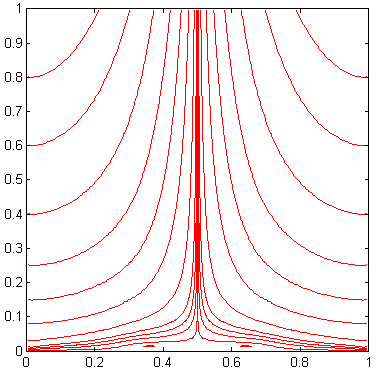}}
\subfigure[$t=0.045$]{\includegraphics[width=7cm]{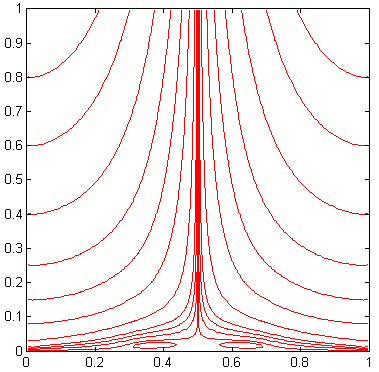}}
\subfigure[$t=0.05$]{\includegraphics[width=7cm]{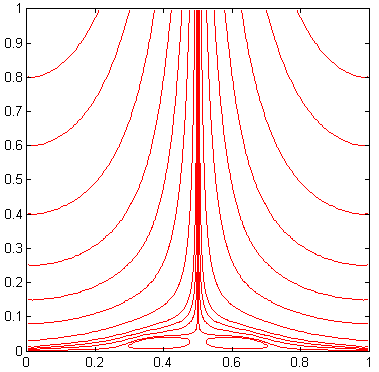}}
\subfigure[$t=0.055$]{\includegraphics[width=7cm]{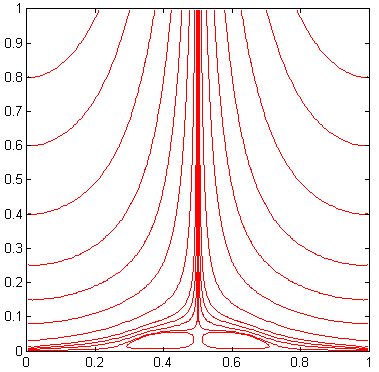}}
\caption{Stream Line, time evolution at $Re=500$}
\label{Splot3}
\end{figure}
\addtocounter{figure}{-1}
\begin{figure}[htbp]
\addtocounter{subfigure}{6}
\centering
\subfigure[$t=0.06$]{\includegraphics[width=7cm]{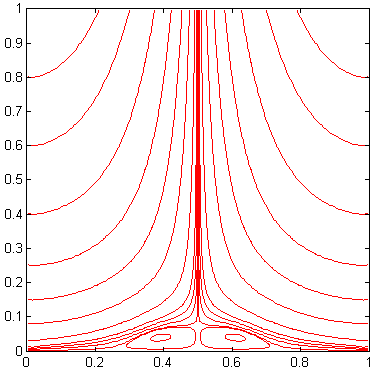}}
\subfigure[$t=0.07$]{\includegraphics[width=7cm]{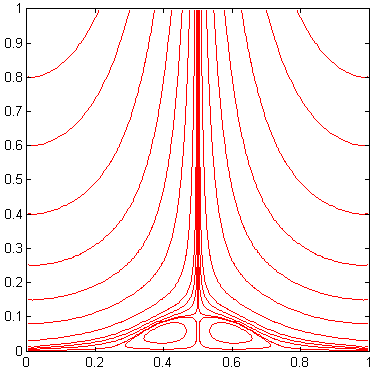}}
\subfigure[$t=0.08$]{\includegraphics[width=7cm]{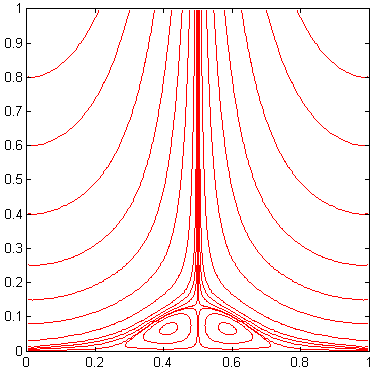}}
\subfigure[$t=0.09$]{\includegraphics[width=7cm]{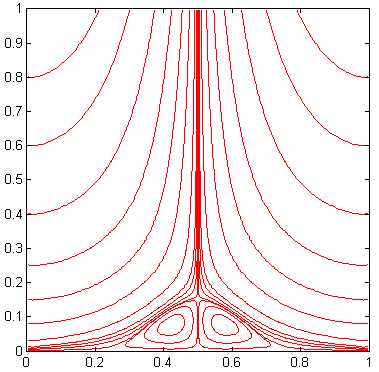}}
\subfigure[$t=0.1$]{\includegraphics[width=7cm]{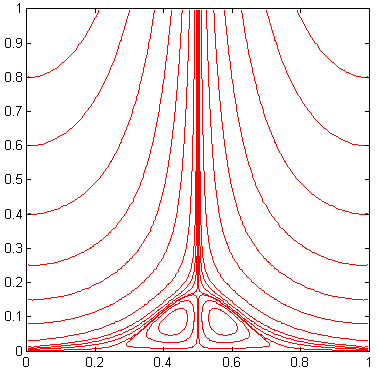}}
\subfigure[Steady]{\includegraphics[width=7cm]{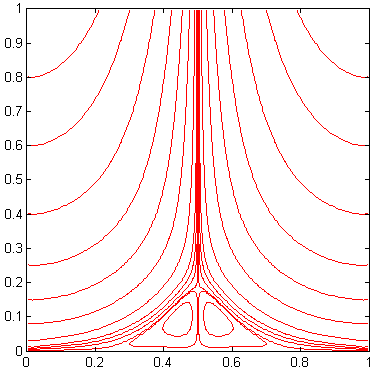}}
\caption{Stream Line, time evolution at $Re=500$}
\label{Splot4}
\end{figure}

\begin{figure}[htbp]
\centering
\subfigure[$t=0.005$]{\includegraphics[width=7cm]{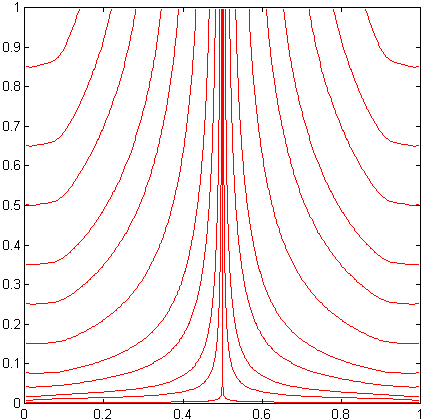}}
\subfigure[$t=0.015$]{\includegraphics[width=7cm]{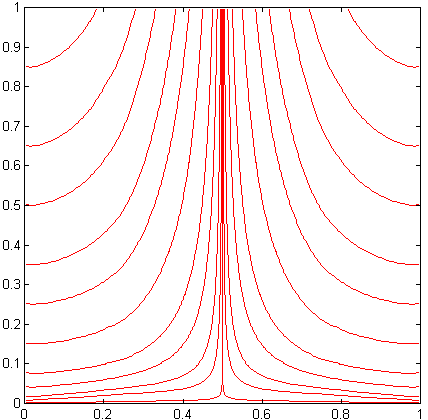}}
\subfigure[$t=0.02$]{\includegraphics[width=7cm]{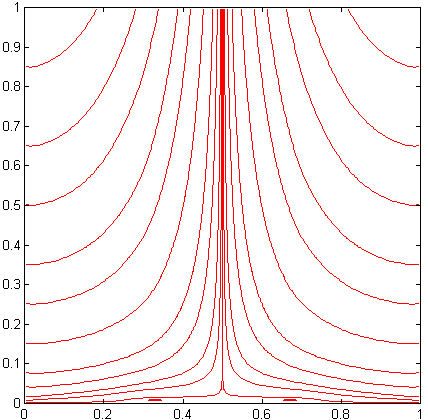}}
\subfigure[$t=0.0225$]{\includegraphics[width=7cm]{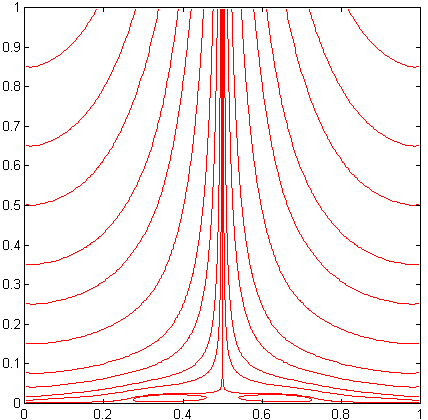}}
\subfigure[$t=0.025$]{\includegraphics[width=7cm]{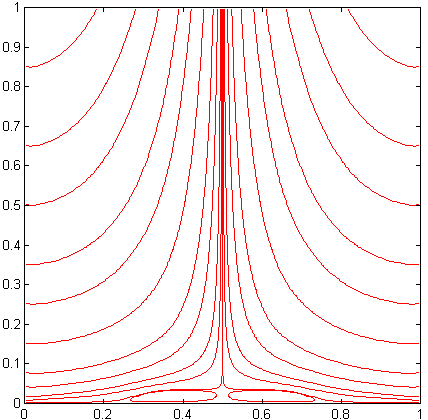}}
\subfigure[$t=0.0275$]{\includegraphics[width=7cm]{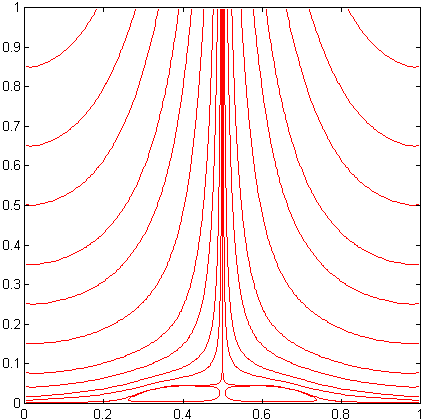}}
\caption{Stream Line, time evolution at $Re=1000$}
\label{Splot11}
\end{figure}
\addtocounter{figure}{-1}
\begin{figure}[htbp]
\addtocounter{subfigure}{6}
\centering
\subfigure[$t=0.03$]{\includegraphics[width=7cm]{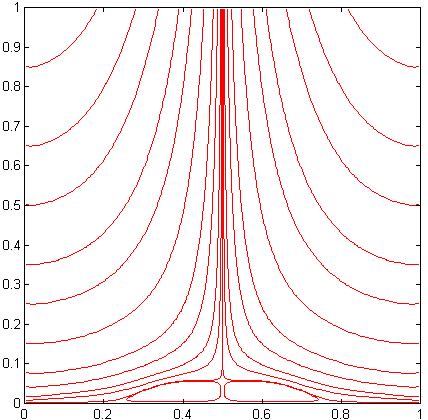}}
\subfigure[$t=0.0325$]{\includegraphics[width=7cm]{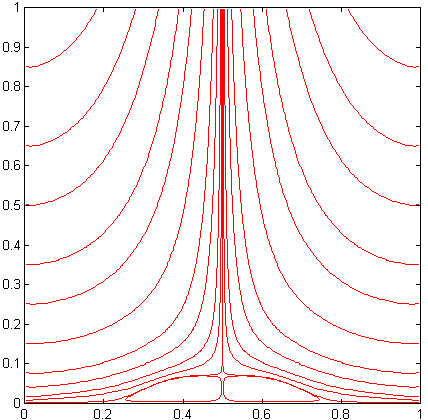}}
\subfigure[$t=0.035$]{\includegraphics[width=7cm]{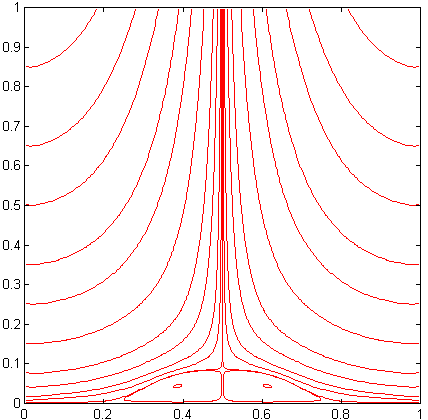}}
\subfigure[$t=0.04$]{\includegraphics[width=7cm]{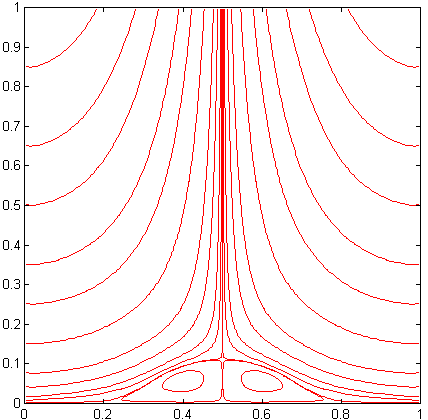}}
\subfigure[$t=0.045$]{\includegraphics[width=7cm]{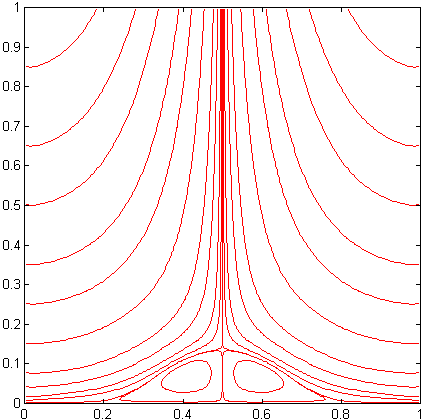}}
\subfigure[Steady]{\includegraphics[width=7cm]{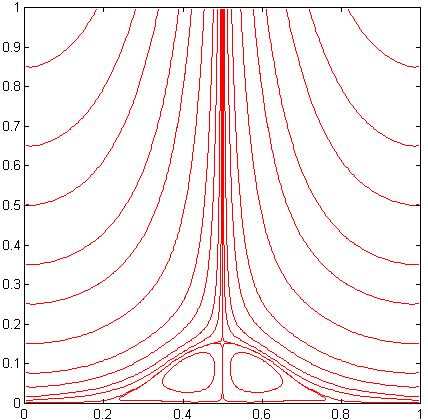}}
\caption{Stream Line, time evolution at $Re=1000$}
\label{Splot12}
\end{figure}

\begin{figure}[htbp]
\centering
\subfigure[$t=0.005$]{\includegraphics[width=7cm]{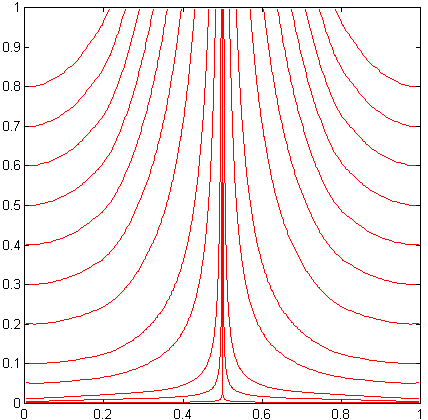}}
\subfigure[$t=0.0075$]{\includegraphics[width=7cm]{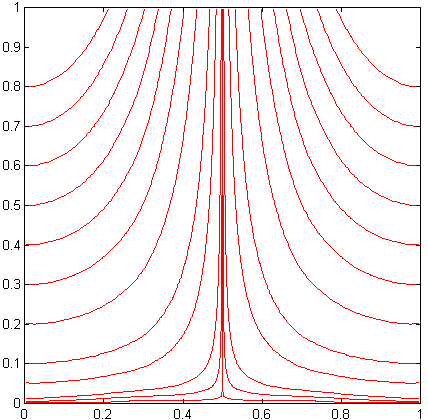}}
\subfigure[$t=0.01$]{\includegraphics[width=7cm]{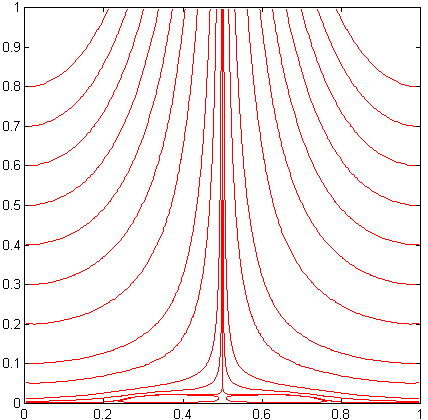}}
\subfigure[$t=0.015$]{\includegraphics[width=7cm]{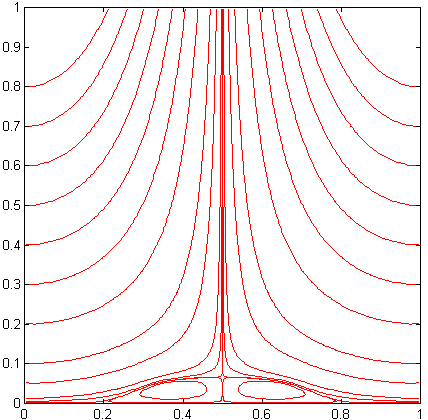}}
\subfigure[$t=0.0175$]{\includegraphics[width=7cm]{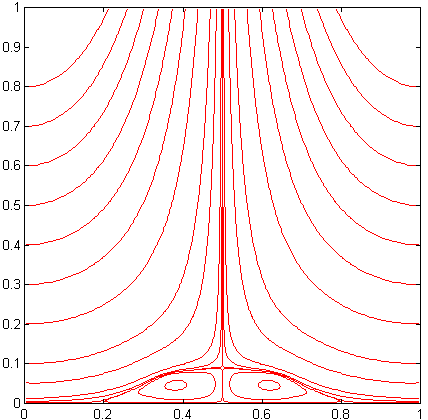}}
\subfigure[$t=0.02$]{\includegraphics[width=7cm]{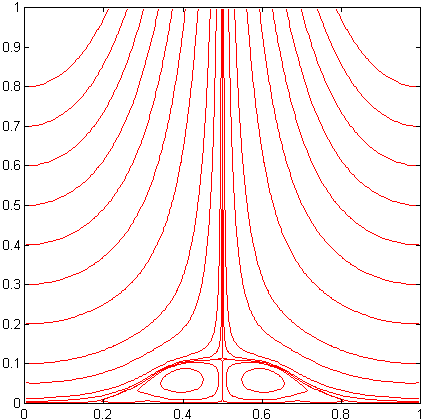}}
\caption{Stream Line, time evolution at $Re=2500$}
\label{Splot11}
\end{figure}
\addtocounter{figure}{-1}
\begin{figure}[htbp]
\addtocounter{subfigure}{6}
\centering
\subfigure[$t=0.0225$]{\includegraphics[width=7cm]{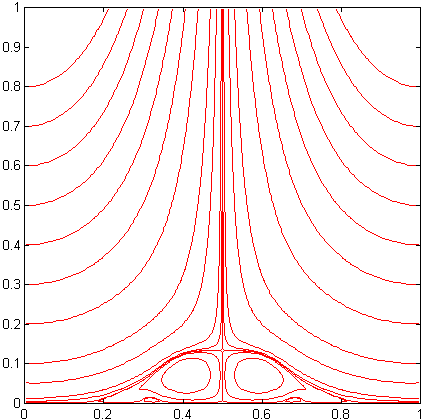}}
\subfigure[$t=0.025$]{\includegraphics[width=7cm]{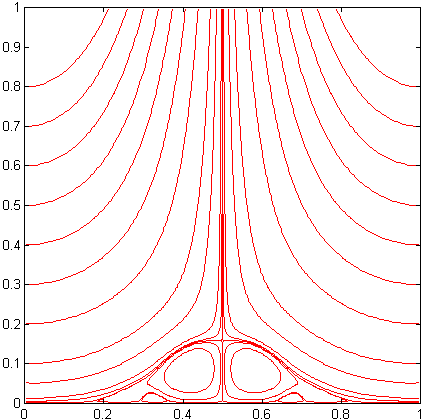}}
\subfigure[$t=0.03$]{\includegraphics[width=7cm]{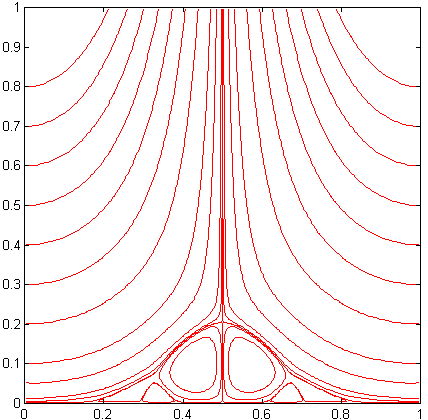}}
\subfigure[$t=0.035$]{\includegraphics[width=7cm]{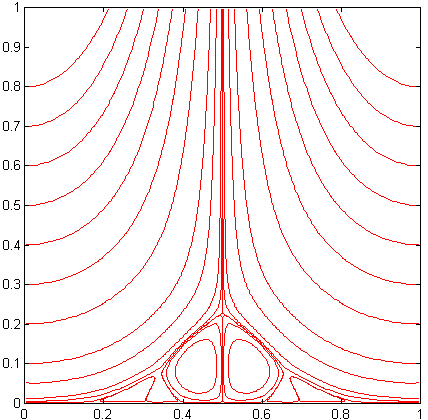}}
\subfigure[$t=0.4$]{\includegraphics[width=7cm]{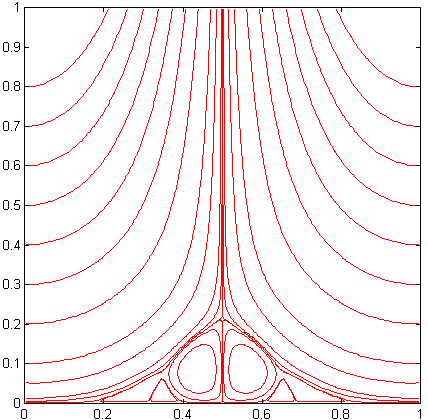}}
\subfigure[Steady]{\includegraphics[width=7cm]{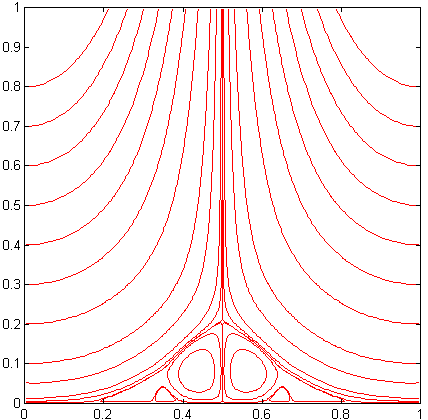}}
\caption{Stream Line, time evolution at $Re=2500$}
\label{Splot12}
\end{figure}

\begin{figure}[htbp]
\centering
\subfigure[$t=0.0025$]{\includegraphics[width=7cm]{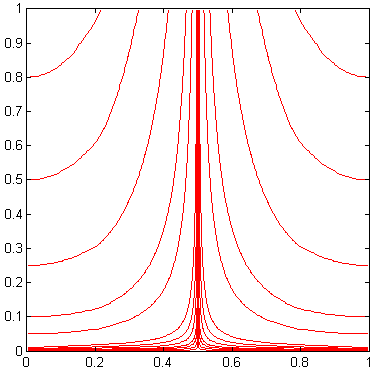}}
\subfigure[$t=0.00375$]{\includegraphics[width=7cm]{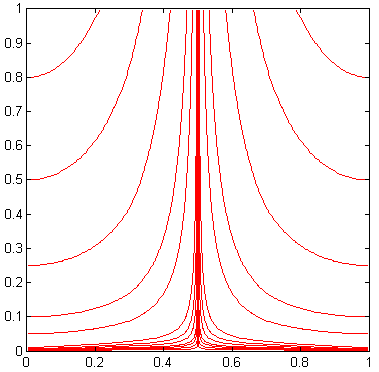}}
\subfigure[$t=0.005$]{\includegraphics[width=7cm]{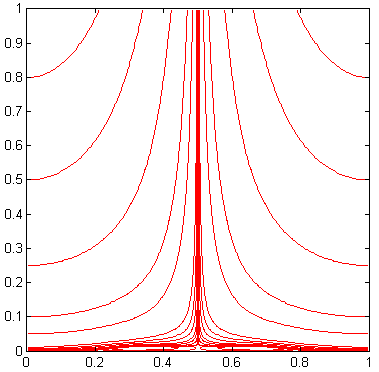}}
\subfigure[$t=0.00625$]{\includegraphics[width=7cm]{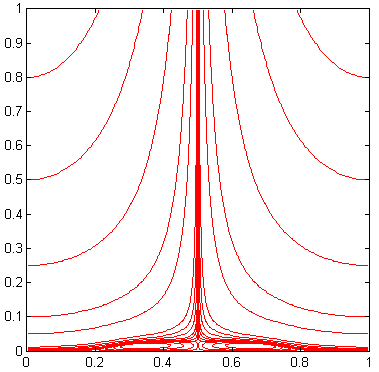}}
\subfigure[$t=0.0075$]{\includegraphics[width=7cm]{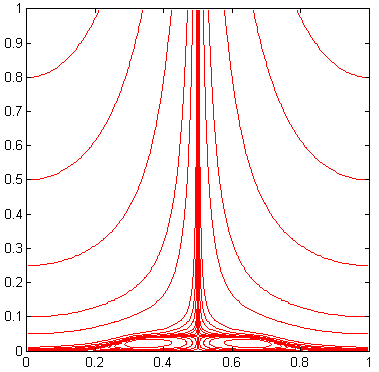}}
\subfigure[$t=0.0087$]{\includegraphics[width=7cm]{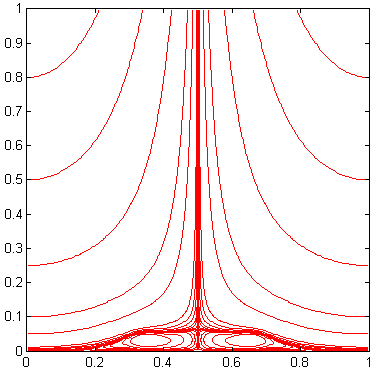}}
\caption{Stream Line, time evolution at $Re=5000$}
\label{Splot1}
\end{figure}
\addtocounter{figure}{-1}
\begin{figure}[htbp]
\addtocounter{subfigure}{6}
\centering
\subfigure[$t=0.01$]{\includegraphics[width=7cm]{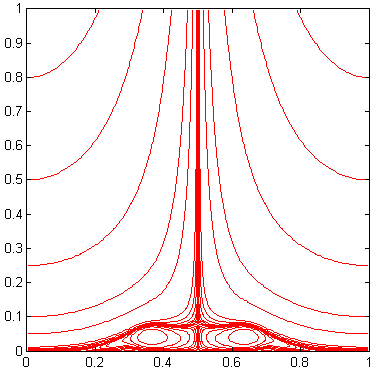}}
\subfigure[$t=0.01125$]{\includegraphics[width=7cm]{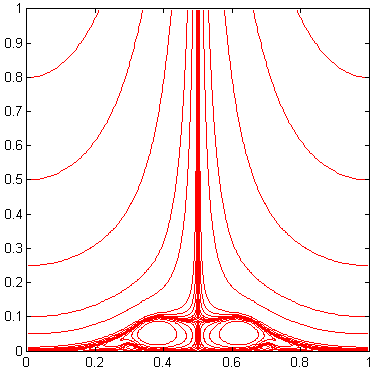}}
\subfigure[$t=0.0125$]{\includegraphics[width=7cm]{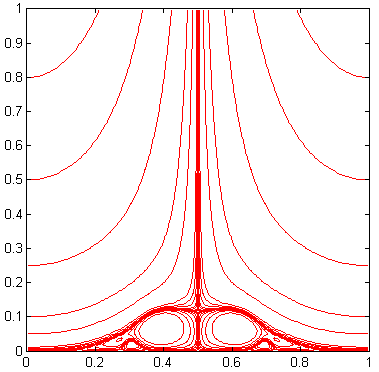}}
\subfigure[$t=0.01375$]{\includegraphics[width=7cm]{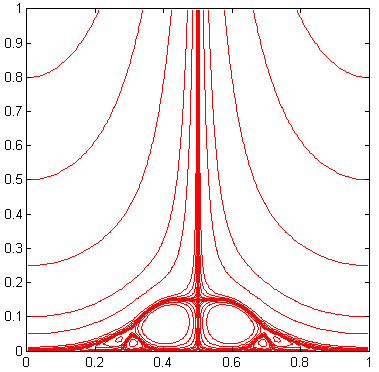}}
\subfigure[$t=0.015$]{\includegraphics[width=7cm]{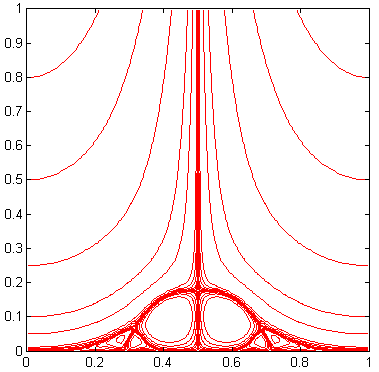}}
\subfigure[Steady]{\includegraphics[width=7cm]{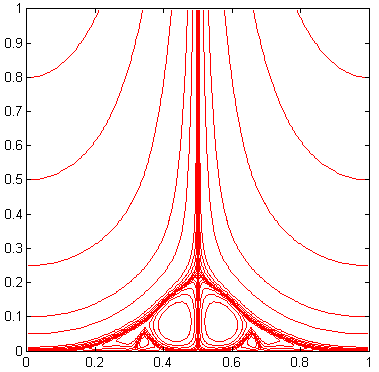}}
\caption{Stream Line, time evolution at $Re=5000$}
\label{Splot2}
\end{figure}

When Reynolds number is relatively small, say Re $<50$, convective forces can be neglected as compared to viscous forces and the laminar boundary layer separated from the wall at the reversed stagnation point. With an increase in Reynolds number, both convective forces and viscous forces are in the same order. The laminar boundary layer starts separating from the wall before the reversed stagnation point. At the same time, there emerges a symmetrical pair of stable vortices which create a back flow, and hence, a circulation region forms close to the reversed stagnation point. With a further increase in Reynolds number, the laminar boundary layer becomes thicker and the vortices extend. The corresponding steady state velocity profile indicates that near the wall most of the fluid has a reversal direction, allowing two steady symmetric eddies to form in the resulting gap. In all cases, however, the flow reaches steady state, hence streamlines coincide with streaklines.
\\\\
The origins of boundary-layer separation are associated with the frictional forces within the boundary layer and a positive or adverse pressure gradient occurs in the direction of flow. Near the wall region, some fluid energy is dissipated in overcoming friction in the boundary layer. When vortices are formed on the decelerated boundary layer, the flow tries to decelerate in a short manner. In the entire boundary layer, once the outer flow is accelerated by a pressures drop, the fluid elements will also move in the direction of motion, and hence, the flow will keep in its original direction along the surface. On the other hand, if the pressure of particles declines in the direction opposite to the flow, the outer flow is therefore decelerated. The remaining energy is not sufficient to overcome the increased pressure. Then slower fluid particles of the boundary layer are even more slowed down. Eventually, if the deceleration is large enough such that the flow particles stop in motion and start moving in the opposite direction, the flow separates from the wall and a backflow region emerges.
\begin{figure}[!htb]
\begin{center}
\includegraphics[width= 7.5cm]{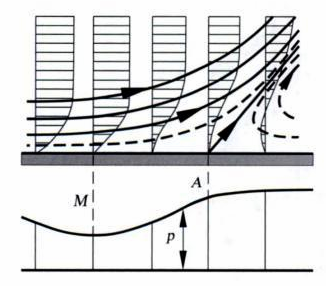}
\caption{Separation process (maximum velocity M, separation point A) \cite{oertel2004prandtl}}
\label{g4}
\end{center}
\end{figure}
\\

Fig.~(\ref{g4}) demonstrates the vortex formation in the pressure distribution $p$. When the streamline portrait of the boundary-layer flow is close to the separation position A, since the backflow is close to the wall, the separation rolls up into one or more vortices.  Soon after, a great thickening of the boundary layer exists near this region.
\\\\
At the separation point the wall streamline departs the wall at a certain angle. The position of the point of separation is that point on the wall where the velocity gradient perpendicular to the wall vanishes. In another words, the point where the walls shear stress becomes zero.

\begin{figure}[htbp]
\centering
\subfigure[$x=0.5$]{\includegraphics[scale=0.37]{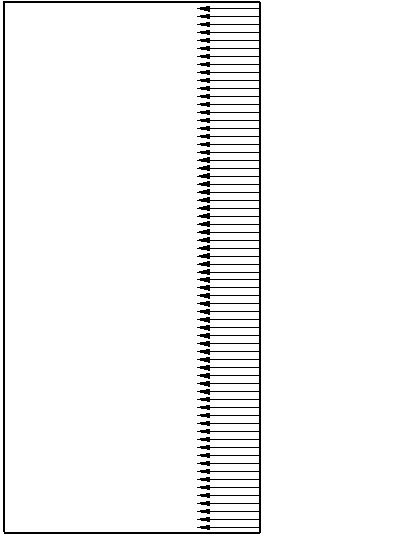}}
\subfigure[$x=0.4$]{\includegraphics[scale=0.37]{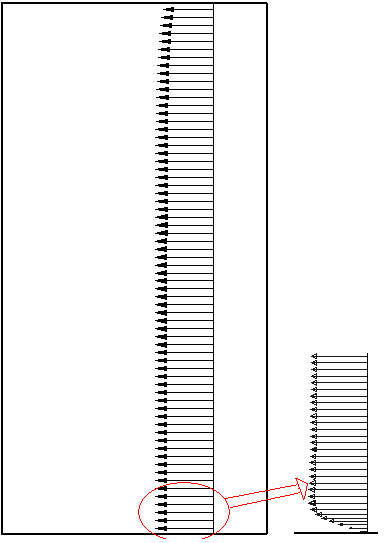}}
\subfigure[$x=0.3$]{\includegraphics[scale=0.37]{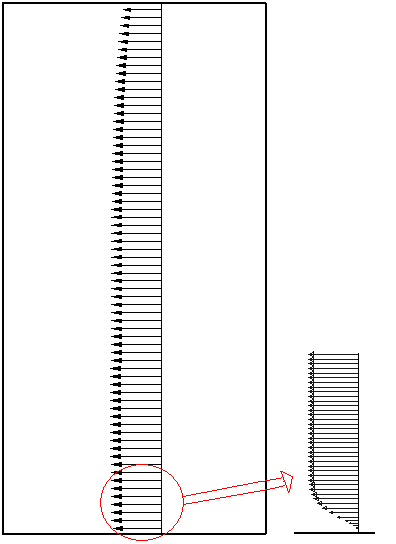}}
\subfigure[$x=0.2$]{\includegraphics[scale=0.37]{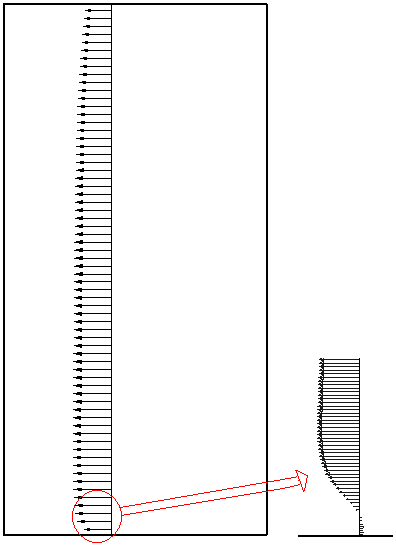}}
\subfigure[$x=0.1$]{\includegraphics[scale=0.37]{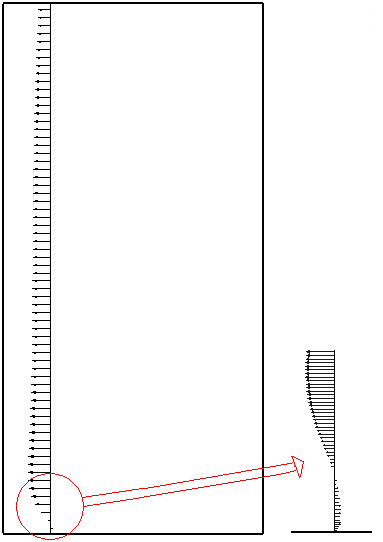}}
\caption{Similarity velocity field, at different values of $x$}
\label{Vprocedure}
\end{figure}

\section{VELOCITY PROFILE}
\begin{figure}[htbp]
\centering
\subfigure[$x=0$]{\includegraphics[width=15cm]{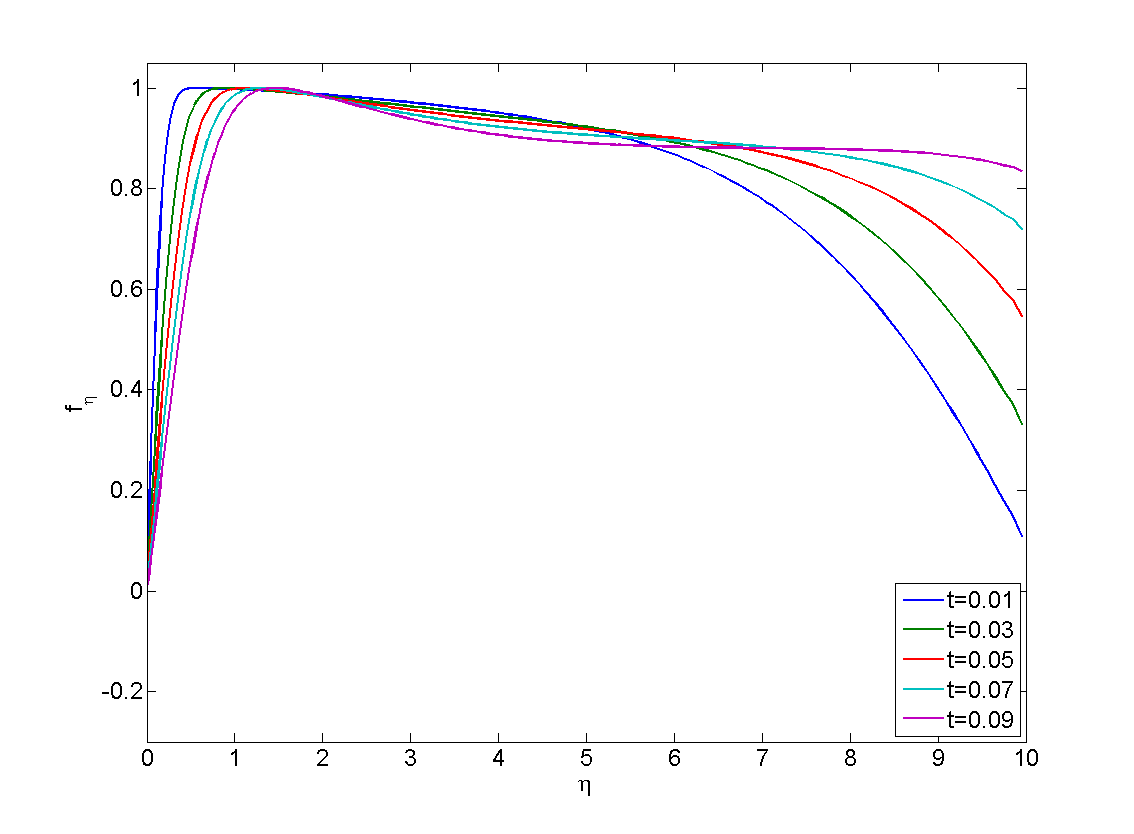}}
\subfigure[$x=0.1$]{\includegraphics[width=15cm]{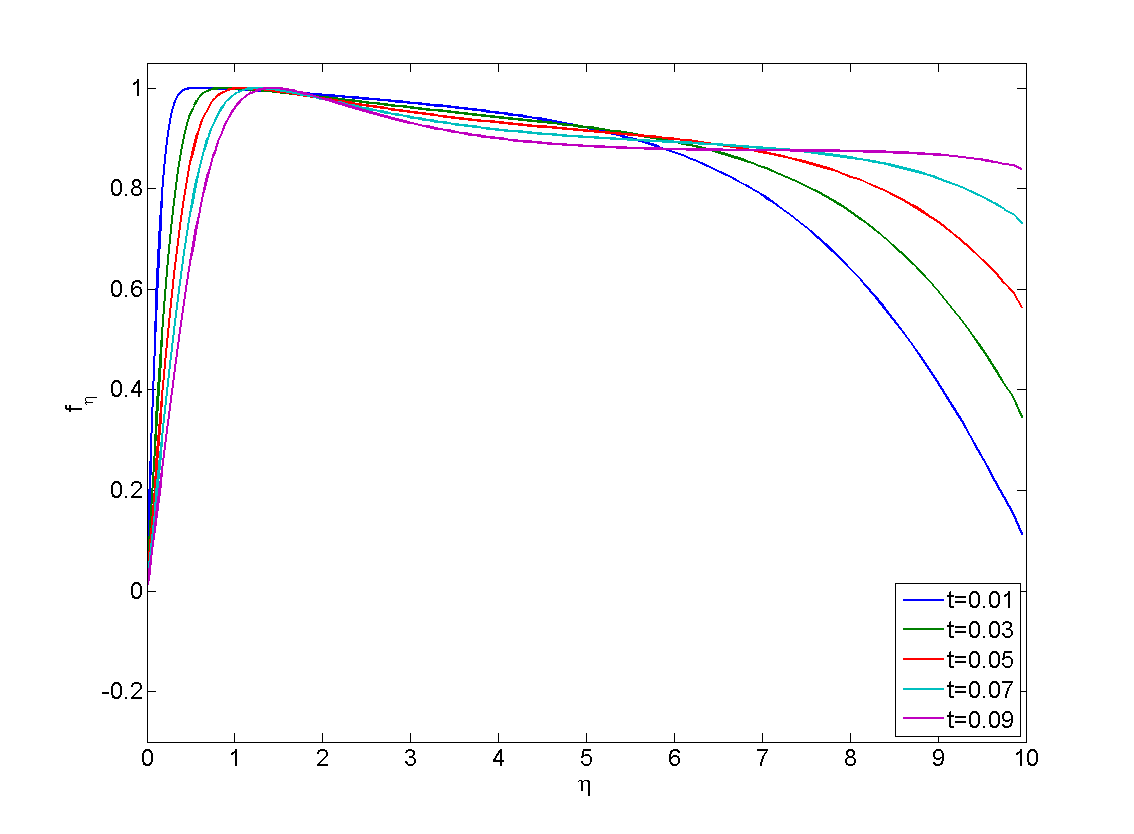}}
\caption{Similarity velocity field, time evolution at $Re=100$}
\label{Vplot}
\end{figure}

Next we discuss the velocity field $f_{\eta}$. Figures (\ref{Vplot}) to (\ref{Vplot8}) show the similarity velocity distribution along the $\eta$-direction at locations of $x = 0,~0.1,~0.2$ and $0.3$ respectively, evolving in time at various Reynolds number. Since the case of opposing flow can be mapped to a case of flow by $x\rightarrow -x$, we will not present here values of the negative value of $x$ for the case of opposing flow. 
\\

It can be observed from these figures that at the beginning of fluid motion, the minimum value of $f_{\eta}$ almost keeps to be zero when the value of Reynolds number ranging from 50 to 2500. No back flow is observed near the wall region. One of the reasons of this phenomenon is that, at the beginning of the fluid motion, in just a very short period of time, the viscous forces have propagated mostly into the fluid. As the fluid is at rest initially, the fluid flow has to overcome a large inertia, resulting in a fluid flow motion. It seems that the convective forces can be neglected as compared to viscous forces.
\\

We examine two phenomena here: the dependence of flow velocity on $x$¸ and the dependence of the external flow. One of the assumption in the analytical solution is that the velocity field $f_{\eta}$ is a function of $\eta$ only in the region near the reversed stagnation point, provided that the velocity field $f_{\eta}$ is independent of $x$. When the fluid flows near the origin or the location of $x$-coordinate is relatively small, say $x<0.1$, the distribution of $f_{\eta}$ is independent of $x$. On the other side, the numerical solutions show variation of velocity along the $x$-direction. Large discrepancy occurs as a larger value of $x$ is applied in the numerical simulation, which violate the assumption of no variation of velocity along the $x$-direction in the region far away from the reversed stagnation point. \\

\addtocounter{figure}{-1}
\begin{figure}[htbp]
\addtocounter{subfigure}{2}
\centering
\subfigure[$x=0.2$]{\includegraphics[width=15cm]{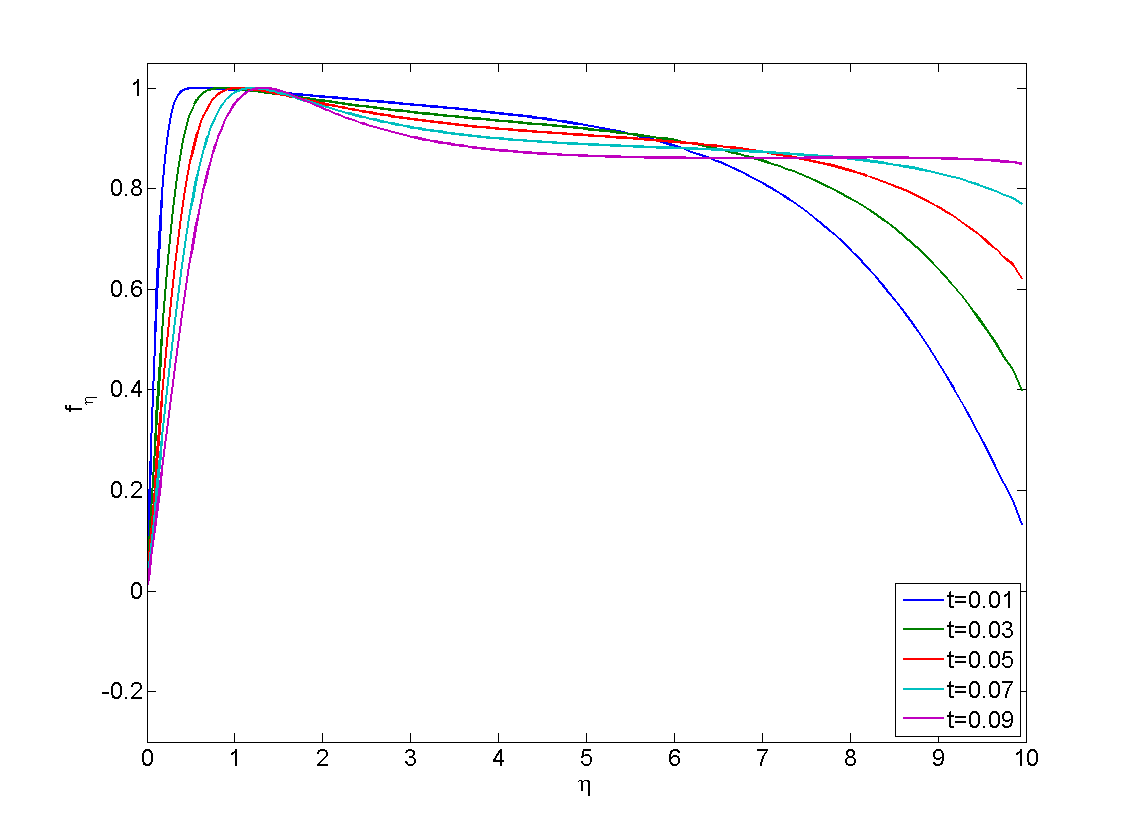}}
\subfigure[$x=0.3$]{\includegraphics[width=15cm]{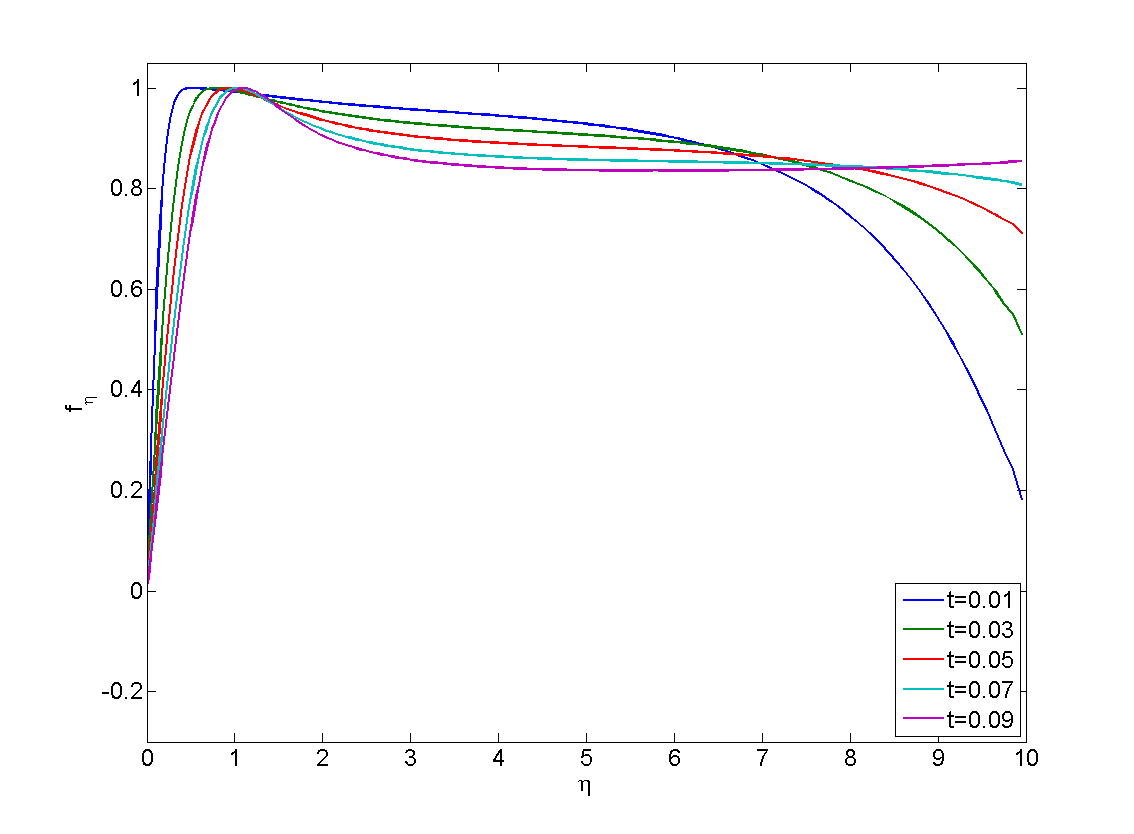}}
\caption{Similarity velocity field, time evolution at $Re=100$}
\label{Vplot2}
\end{figure}

\begin{figure}[htbp]
\centering
\subfigure[$x=0$]{\includegraphics[width=15cm]{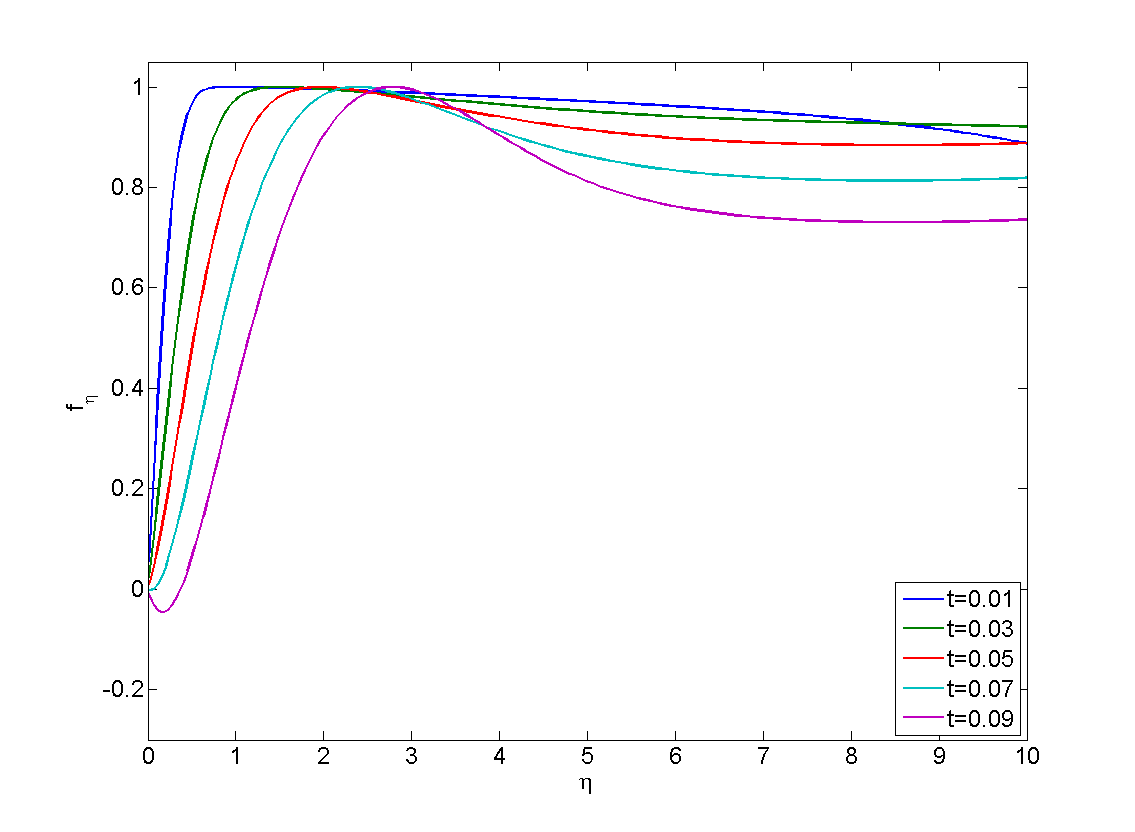}}
\subfigure[$x=0.1$]{\includegraphics[width=15cm]{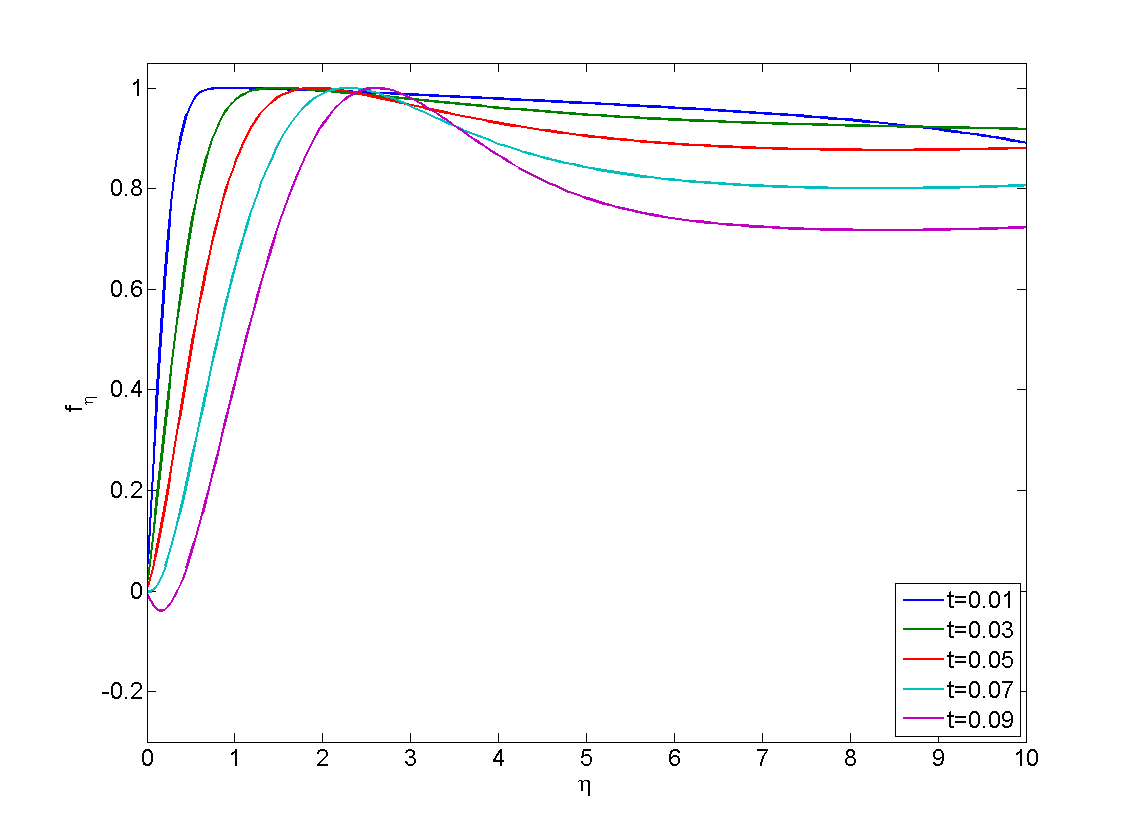}}
\caption{Similarity velocity field, time evolution at $Re=250$}
\label{Vplot3}
\end{figure}
\addtocounter{figure}{-1}
\begin{figure}[htbp]
\addtocounter{subfigure}{2}
\centering
\subfigure[$x=0.2$]{\includegraphics[width=15cm]{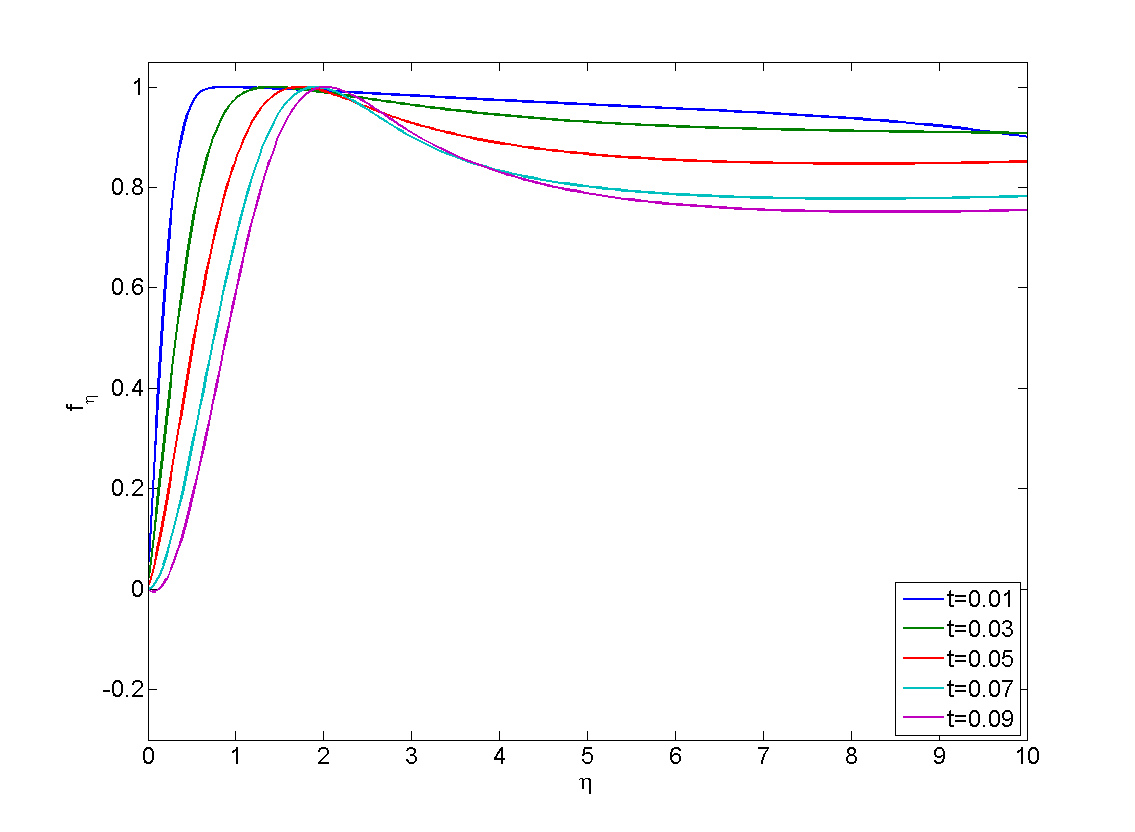}}
\subfigure[$x=0.3$]{\includegraphics[width=15cm]{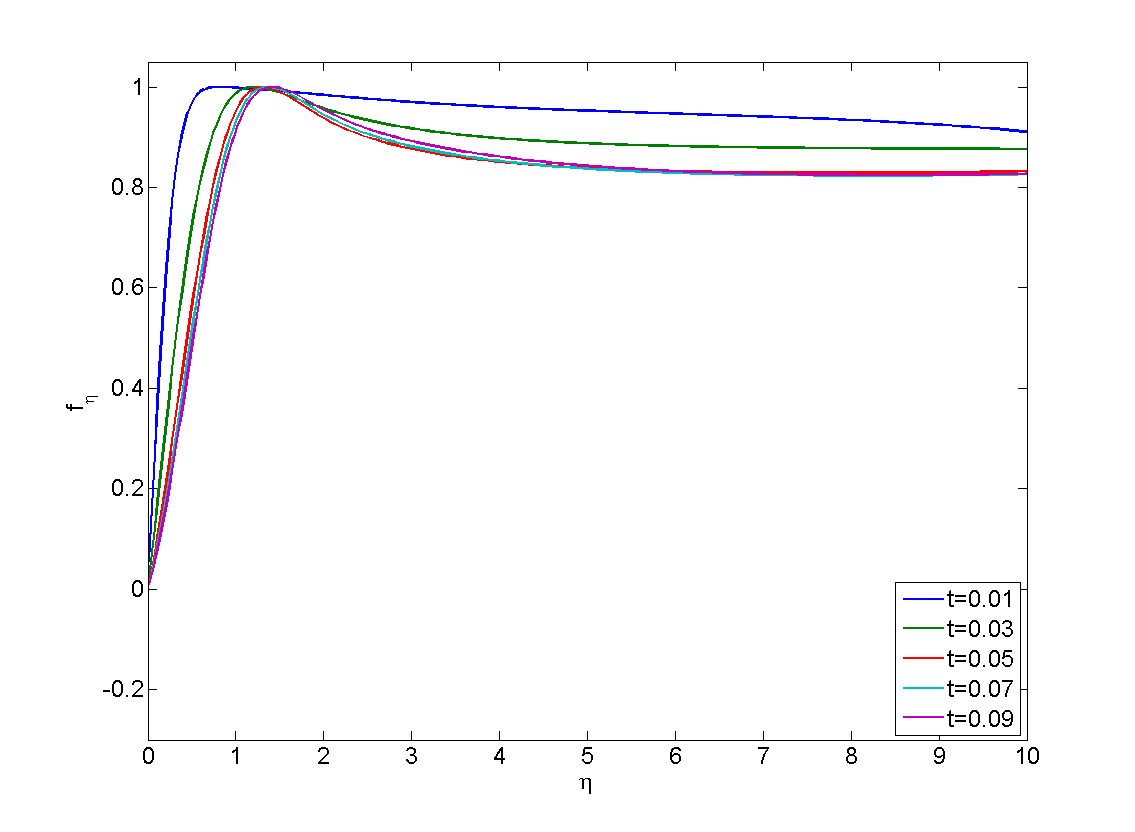}}
\caption{Similarity velocity field, time evolution at $Re=250$}
\label{Vplot4}
\end{figure}

\begin{figure}[htbp]
\centering
\subfigure[$x=0$]{\includegraphics[width=15cm]{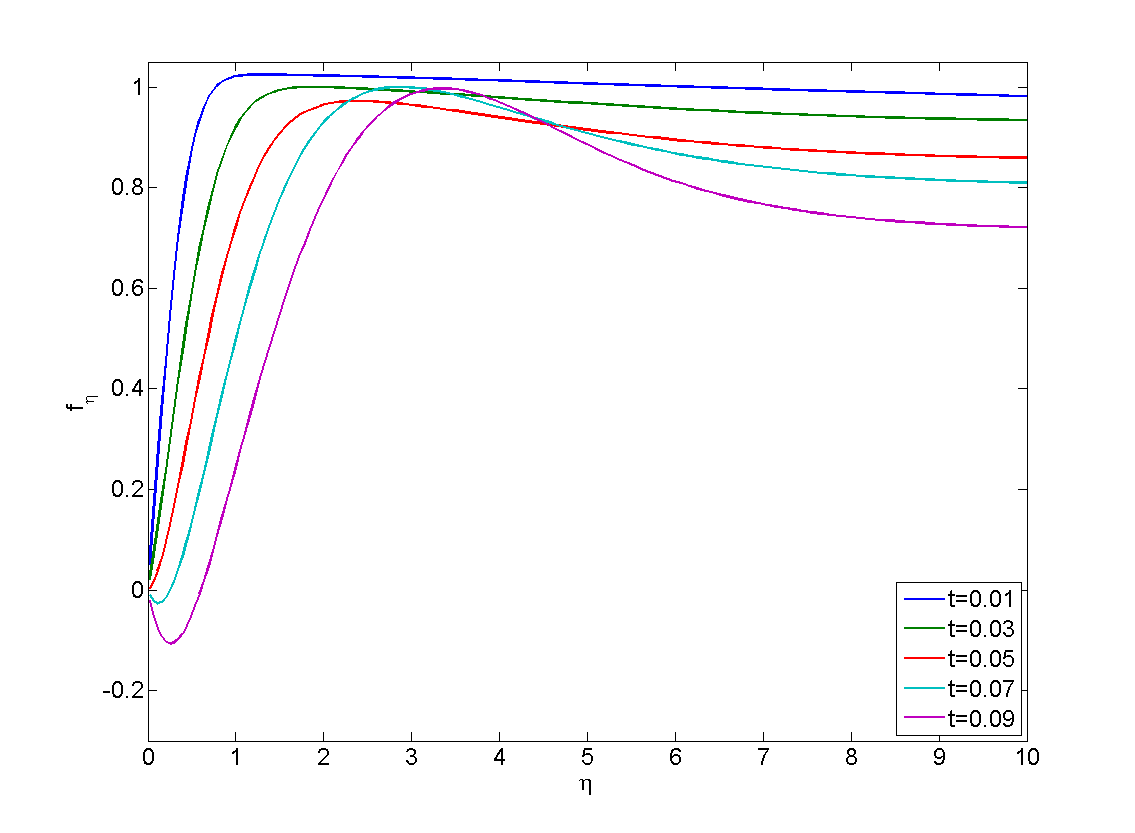}}
\subfigure[$x=0.1$]{\includegraphics[width=15cm]{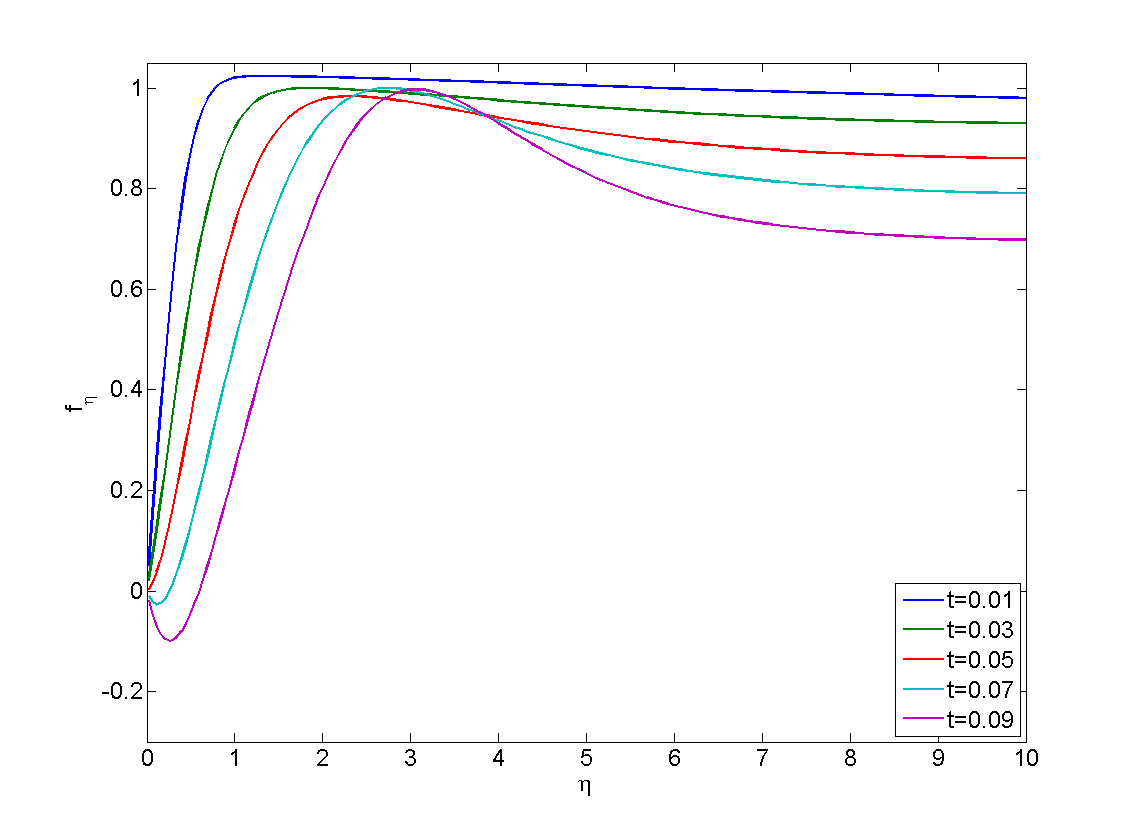}}
\caption{Similarity velocity field, time evolution att $Re=500$}
\label{Vplot5}
\end{figure}
\addtocounter{figure}{-1}
\begin{figure}[htbp]
\addtocounter{subfigure}{2}
\centering
\subfigure[$x=0.2$]{\includegraphics[width=15cm]{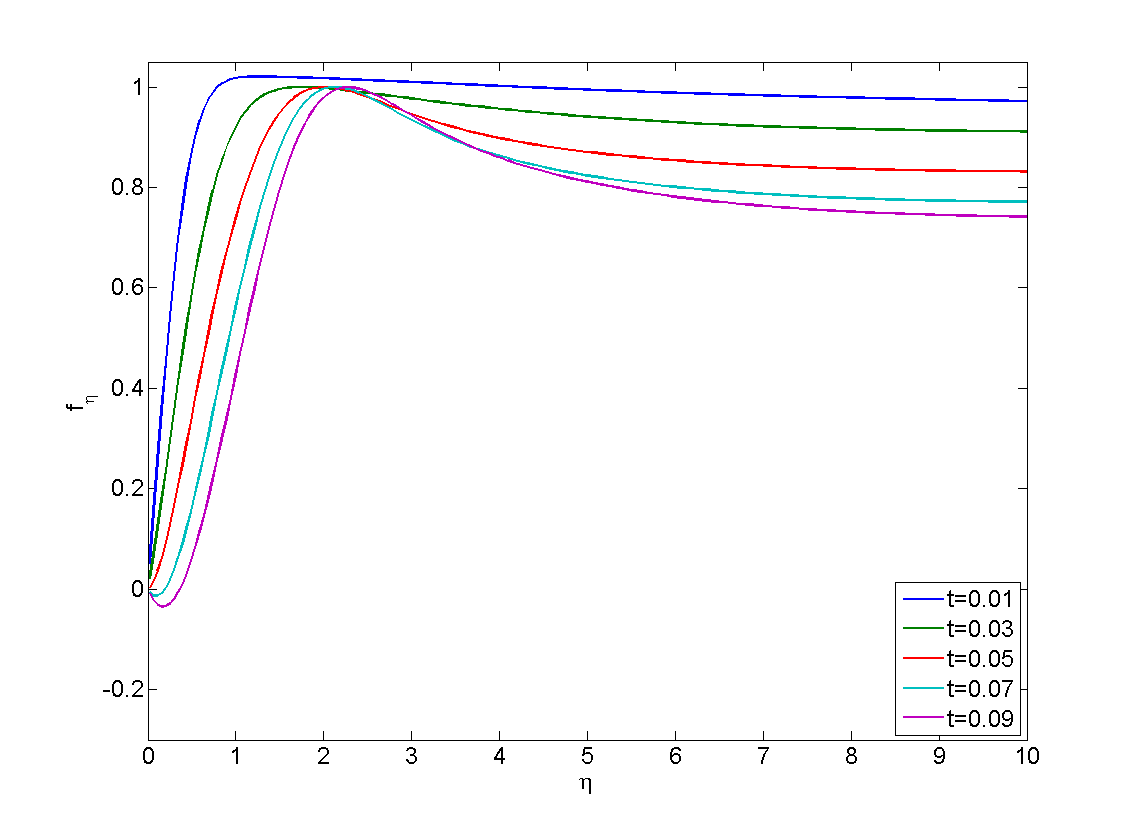}}
\subfigure[$x=0.3$]{\includegraphics[width=15cm]{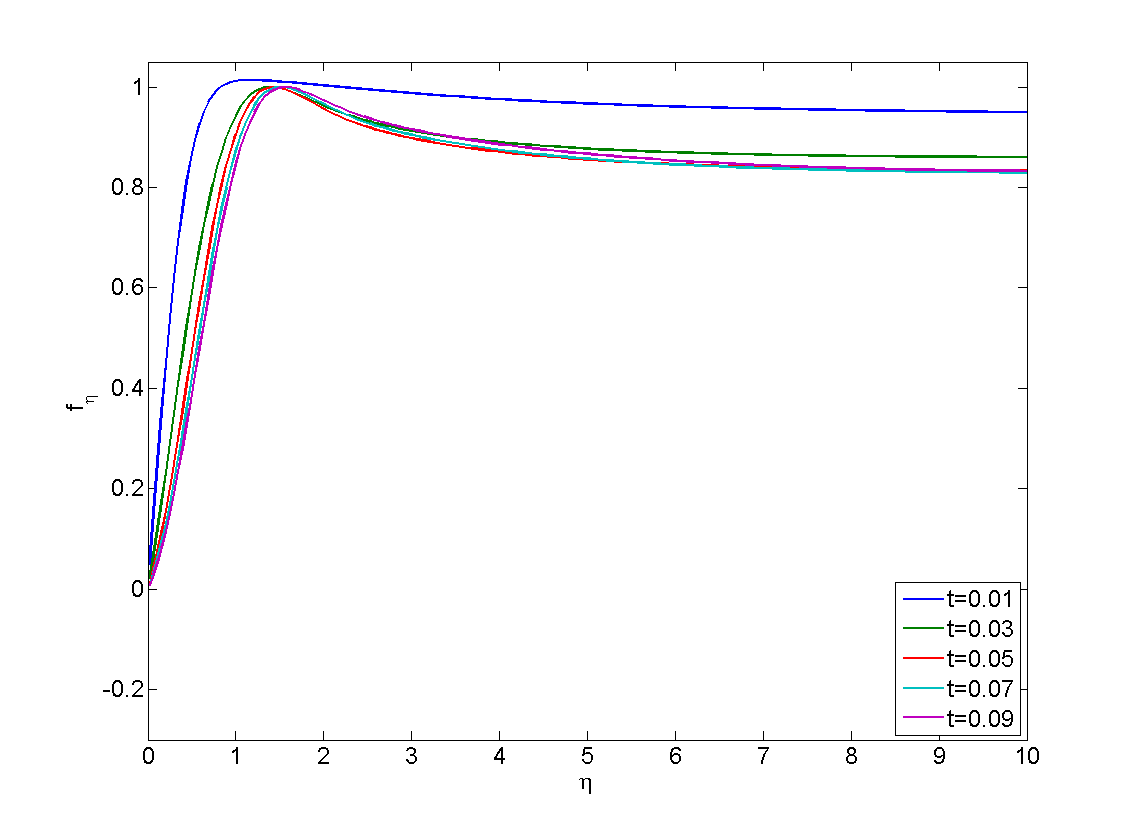}}
\caption{Similarity velocity field, time evolution at $Re=500$}
\label{Vplot6}
\end{figure}

\begin{figure}[htbp]
\centering
\subfigure[$x=0$]{\includegraphics[width=15cm]{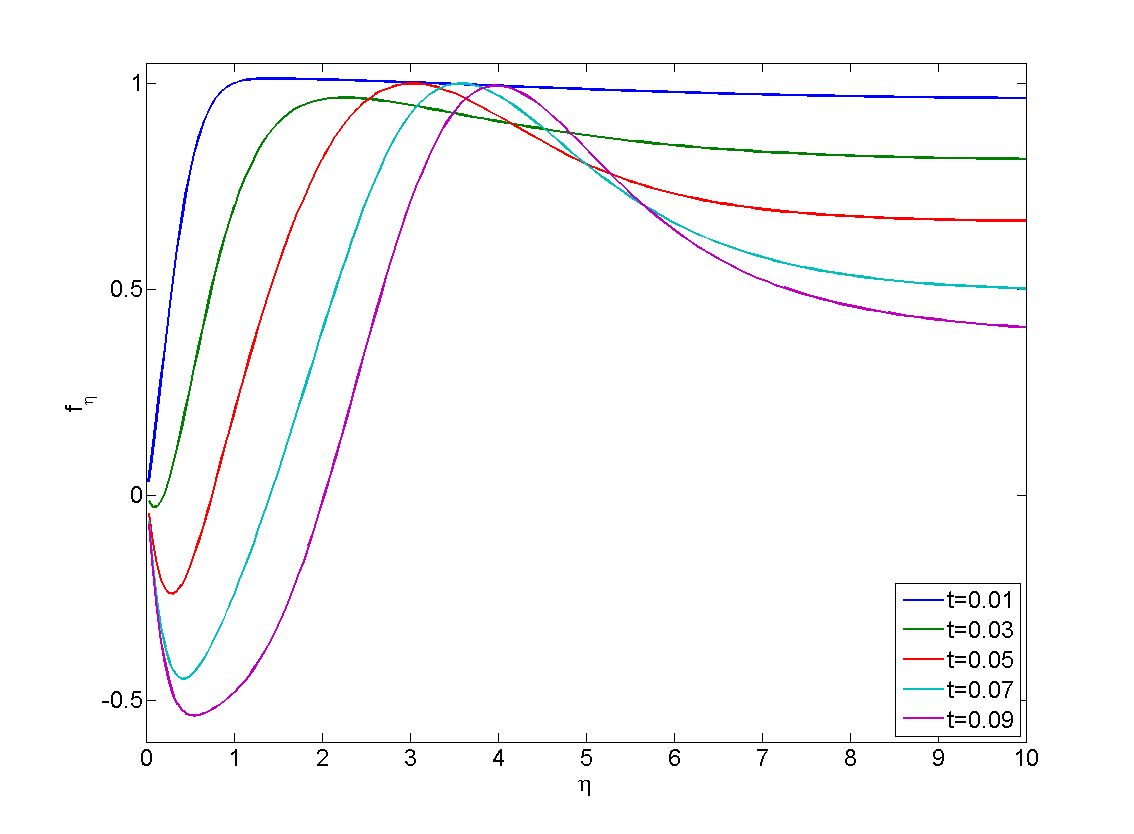}}
\subfigure[$x=0.1$]{\includegraphics[width=15cm]{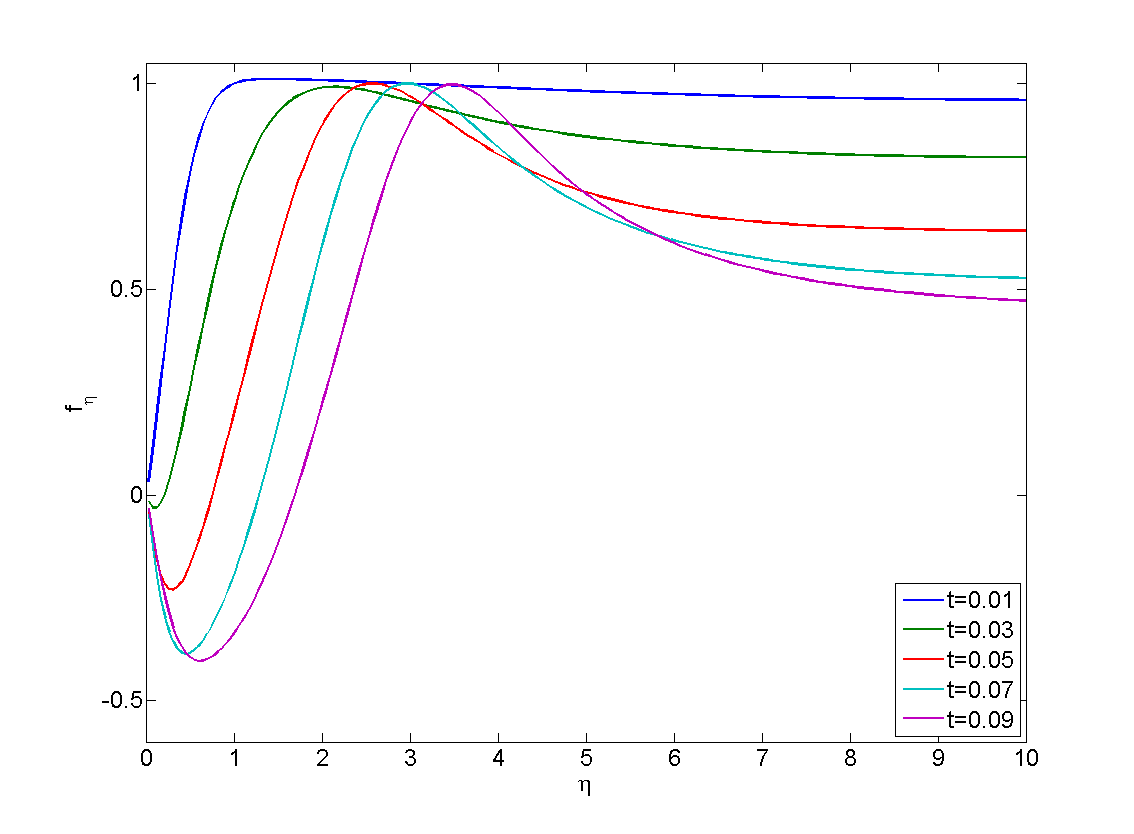}}
\caption{Similarity velocity field, time evolution at $Re=1000$}
\label{Vplot7}
\end{figure}
\addtocounter{figure}{-1}
\begin{figure}[htbp]
\addtocounter{subfigure}{2}
\centering
\subfigure[$x=0.2$]{\includegraphics[width=15cm]{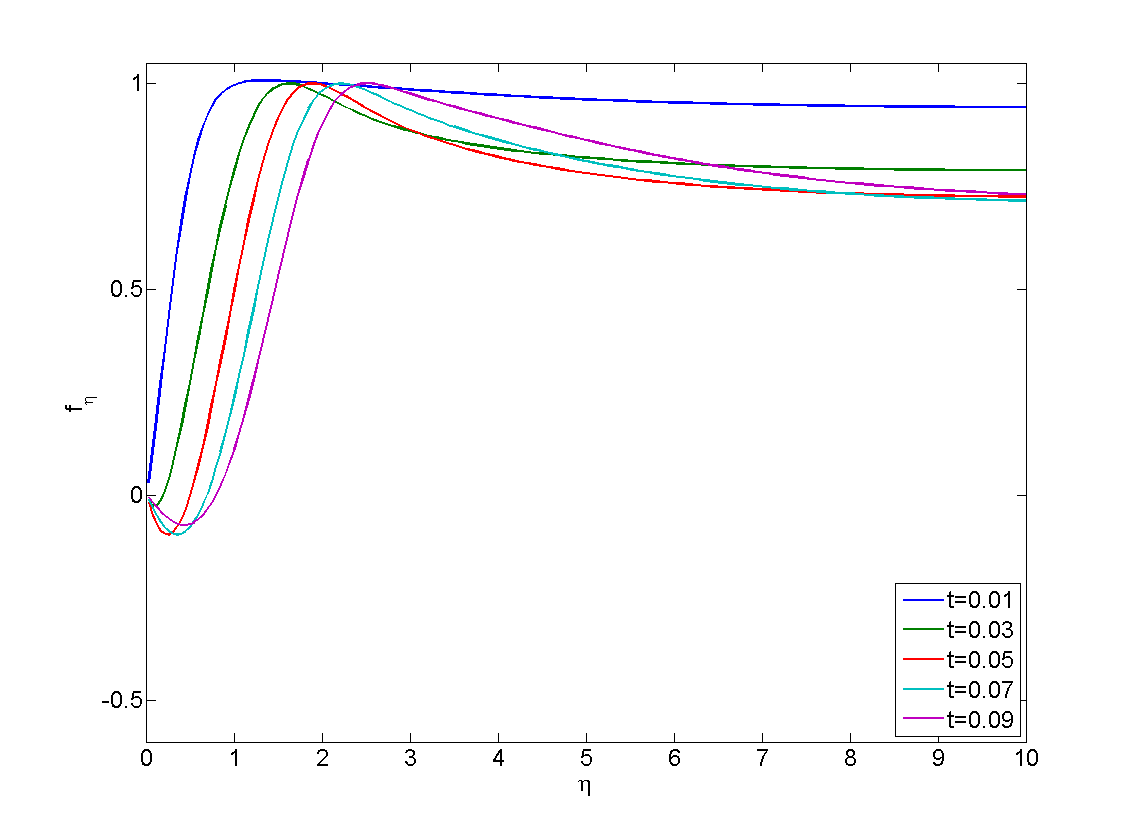}}
\subfigure[$x=0.3$]{\includegraphics[width=15cm]{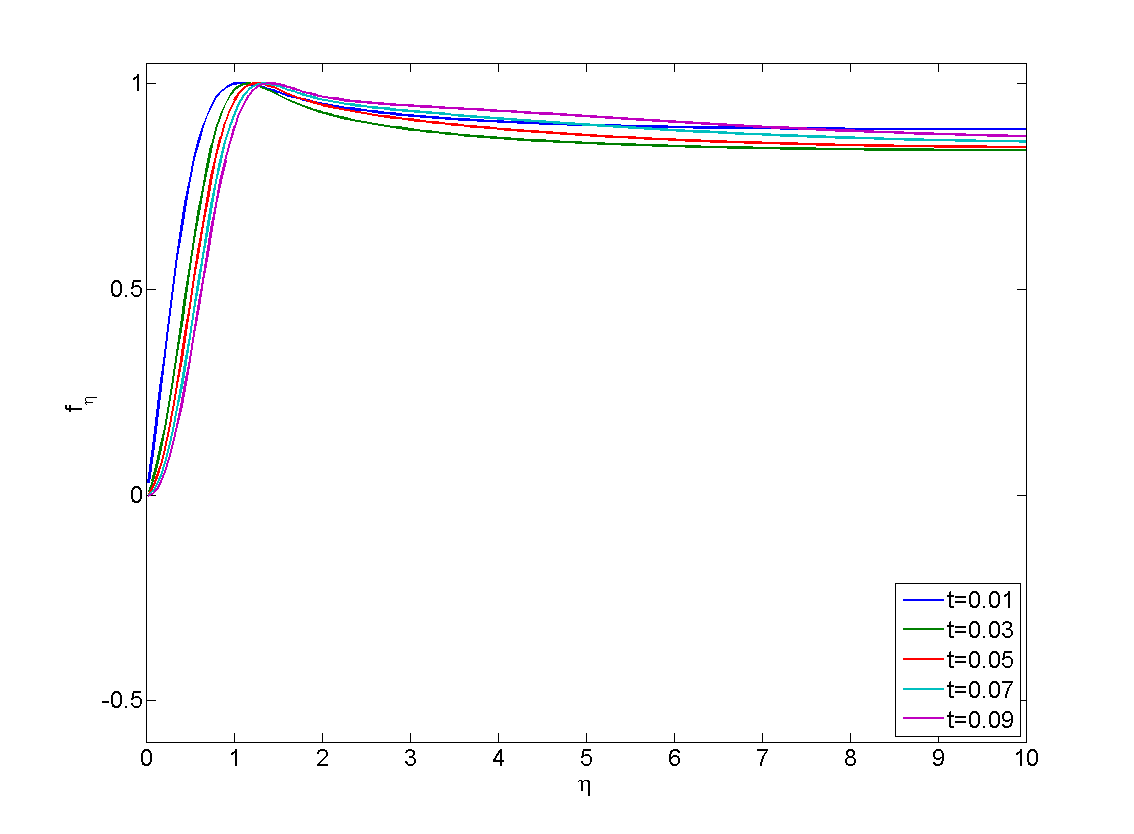}}
\caption{Similarity velocity field, time evolution at $Re=1000$}
\label{Vplot8}
\end{figure}

\newpage

\begin{figure}[!htbp]
\centering
\includegraphics[width=14cm]{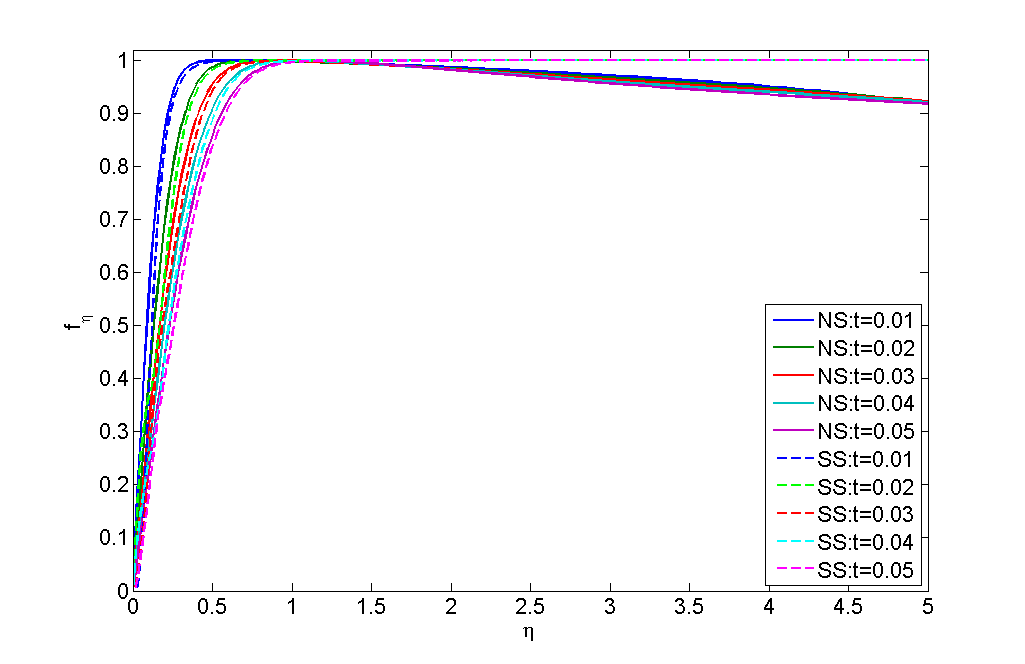}
\caption{Comparison between the numerical velocity profiles and similarity velocity at $Re=100$}
\label{Vplot10}
\end{figure}

\begin{figure}[!htbp]
\centering
\includegraphics[width=14cm]{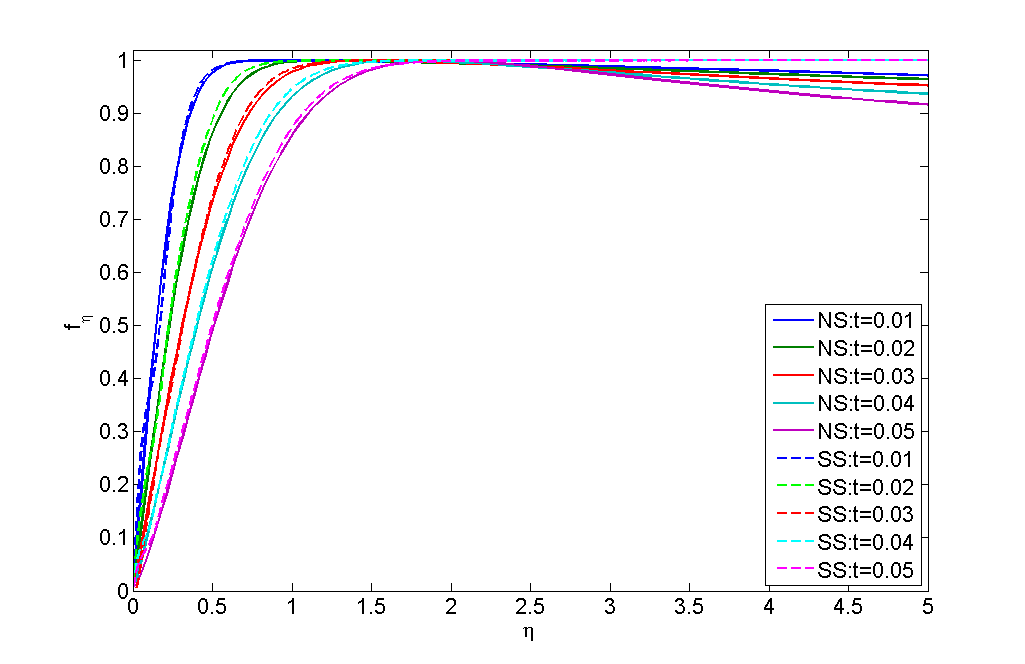}
\caption{Comparison between the numerical velocity profiles and similarity velocity at $Re=250$}
\label{Vplot11}
\end{figure}

\begin{figure}[!htbp]
\centering
\includegraphics[width=14cm]{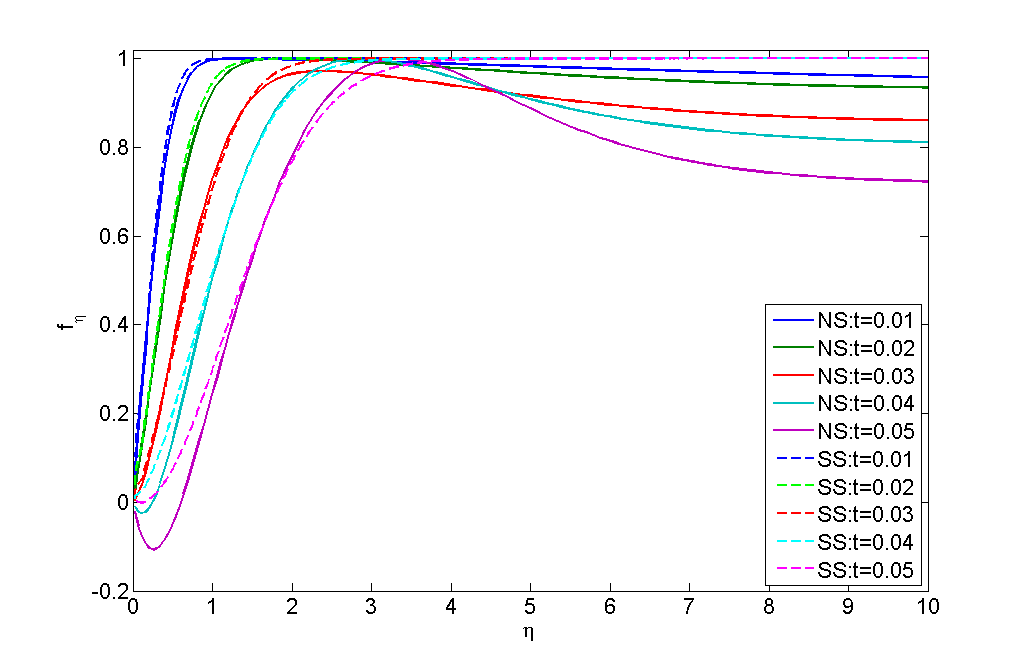}
\caption{Comparison between the numerical velocity profiles and similarity velocity at $Re=500$}
\label{Vplot12}
\end{figure}

\begin{figure}[!htbp]
\centering
\includegraphics[width=14cm]{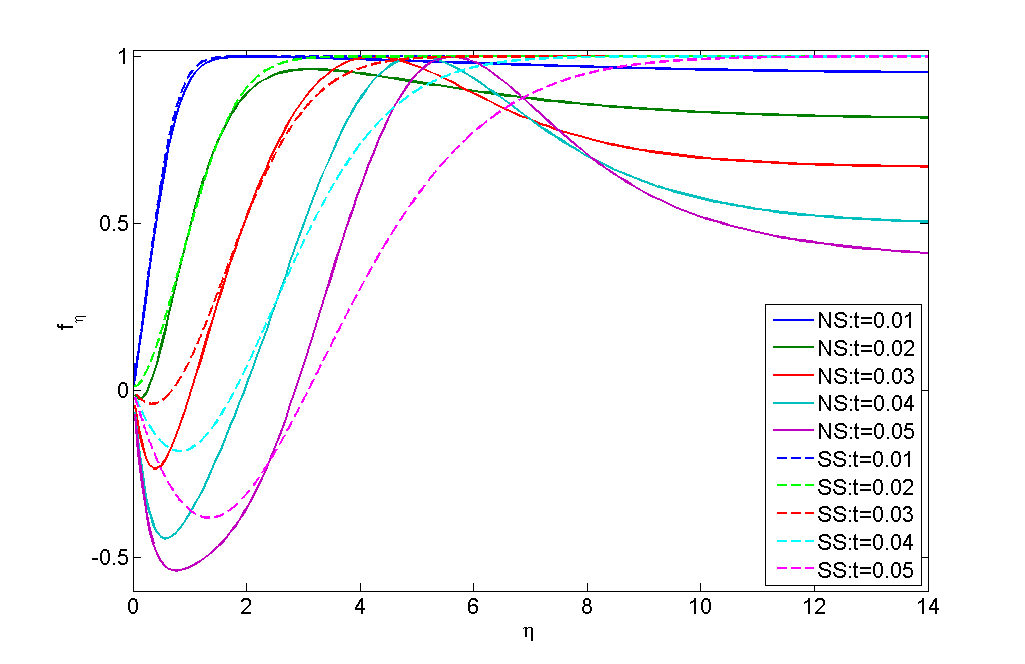}
\caption{Comparison between the numerical velocity profiles and similarity velocity at $Re=1000$}
\label{Vplot13}
\end{figure}

Figures (\ref{Vplot10}) to (\ref{Vplot13}) show comparisons between the numerical simulations and the similarity solutions of reversed stagnation-point flow for different values of $\tau$. Lines without markers denote results obtained from numerical simulation (NS), dotted lines are obtained from the finite-difference formulations (SS).  In the region near the reversed stagnation point the solution agrees remarkably well for smaller values of $\tau$ with the known similarity solution, thus confirming the predictions of the viscous Proudman-Johnson solution. On the other hand, discrepancy occurs as a larger value of $\tau$ is applied in the numerical simulation. However, far away from the wall region, a large difference is observed from the results obtained by these two solutions. The component of velocity normal to a wall is not outward the wall in the region near the reversed stagnation point.  The vorticity created at the wall will be convected outward the wall, which spreads the vorticity towards its source at the boundary.
\\

The other consideration is the behavior of external flow. Proudman and Johnson considered a constant potential flow outside the boundary layer. From Figures (\ref{Vplot}) to (\ref{Vplot13}), one may observe that as $\eta \rightarrow \infty$, the similarity velocity $f_{\eta}$ cannot ultimately approach to 1. It is clearly observed that $f_{\eta}$ gradually drops when the time step increases. As mentioned in previous section, near the wall region or in the boundary layer the phenomenon of reversed flow with boundary-layer separation occurred. There is no justification whatever outside the boundary layer for supposing that at large distances from the wall ($\eta \rightarrow \infty$) the velocity $v(x,\eta,\tau)$ should pass over smoothly into that for inviscid $V_0$. Also we proofed that $f_{\eta}$ cannot ultimately approach to 1 from above or below, nor in an oscillatory manner so that no solution to equation~(\ref{eq:e10}) exists in two-dimensional steady case.  As a consequence the assumption that the potential flow $V_0$ is restricted not to be a constant as well as a time dependent function is reasonable in reversed stagnation-point flow.\\


Moreover, two opposed vortices emerge in the regions are usually in the vicinity of the boundary of the fluid adjacent to wall where viscous forces are dominant. The most important implication of the solution contemplated is the growth of vortices near the wall in a main stream. According to the present similarity solution, separation of the main flow cannot start at any finite time $\tau$ in the limit as $\nu \rightarrow 0$. Moreover, the inviscid Proudman-Johnson solution implies a steady flow $f'\sim-1$ when $\tau \rightarrow \infty$ and the flow problem becomes the classic stagnation-point problem (Hiemenz \cite{hiemenz1911grenzschicht}) by changing the sign in $f$. The solution shown in Figure (\ref{fsf}) indicates that the region of reversed flow expands and has infinite dimensions as $\tau \rightarrow \infty$, which violates the results of numerical simulations that two finite-dimensional vortices appears near the wall in steady state.  Proudman-Johnson solution is only approximate but cannot guarantee that it is free from an infinite multiplicative error for large times. 


\section{TEMPERATURE PROFILE}
\begin{figure}[htbp]
\centering
\subfigure[$t=0.00025$]{\includegraphics[width=6.9cm]{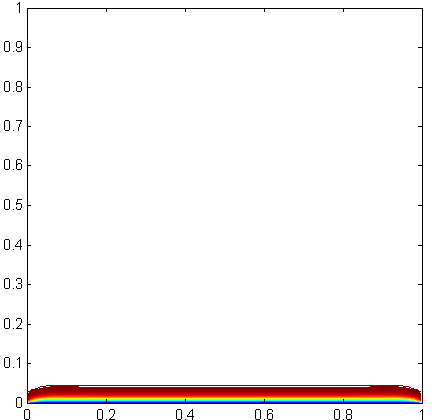}}
\subfigure[$t=0.0005$]{\includegraphics[width=6.9cm]{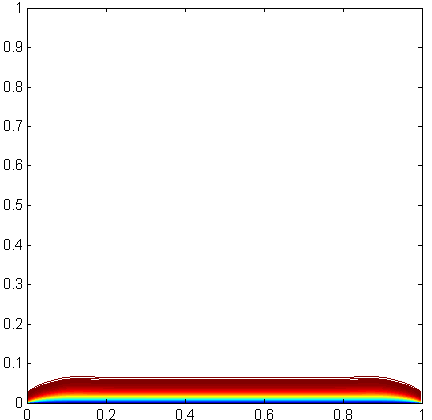}}
\subfigure[$t=0.001$]{\includegraphics[width=6.9cm]{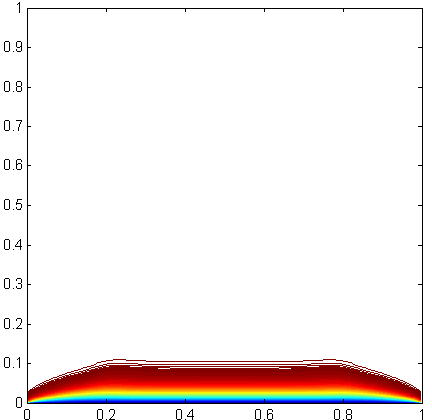}}
\subfigure[$t=0.0015$]{\includegraphics[width=6.9cm]{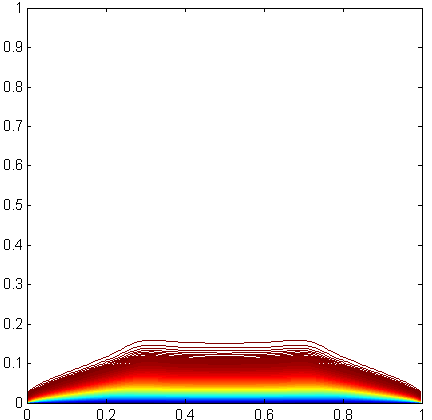}}
\subfigure[$t=0.002$]{\includegraphics[width=6.9cm]{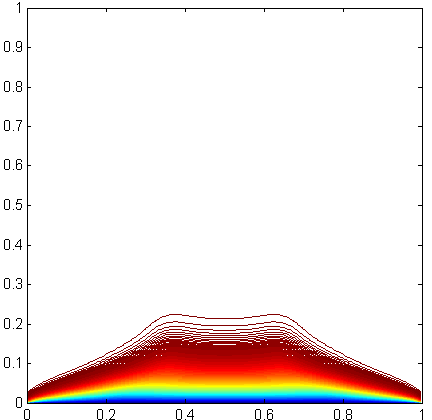}}
\subfigure[$t=0.0025$]{\includegraphics[width=6.9cm]{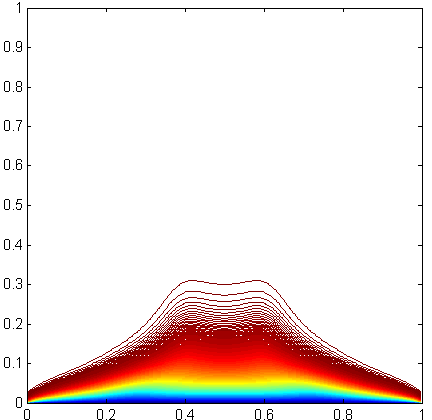}}
\subfigure{\includegraphics[width=8cm]{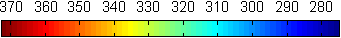}}
\caption{Heatline, time evolution at Pr$=0.3$}
\label{Tplot1}
\end{figure}

Now we return to the numerical results of the nonisothermal stagnation-point flow problem. The following pages (Figures \ref{Tplot1} to \ref{Tplot12}) show the heatlines, both evolving in time, at various Prandtl number. The thermal color was illustrated in the rainbow scale. Colors closer to red are hot areas and colors closer to blue are cold areas. At $t=0$, the inflow velocity is instantaneously set from zero to $u_0$, thereby slowly setting in motion the isothermal fluid initially at rest. The heatlines in these figures show heat flowing mainly from the cooled wall to the heated external flow by conduction in the beginning. The heated external flow passes though the wall and rises, and as it does, it cools down by conduction and convection of heat. After closing to the reversed stagnation point, under the motion of backflow, it sinks to the wall where it is prohibited from sinking further. This hot fluid has thermally contracted to become dense near the reversed stagnation point along the edges of the wall. It trapped in the region near the cooled wall starts to cool down. \\

\addtocounter{figure}{-1}
\begin{figure}[htbp]
\addtocounter{subfigure}{6}
\centering
\subfigure[$t=0.00275$]{\includegraphics[width=6.9cm]{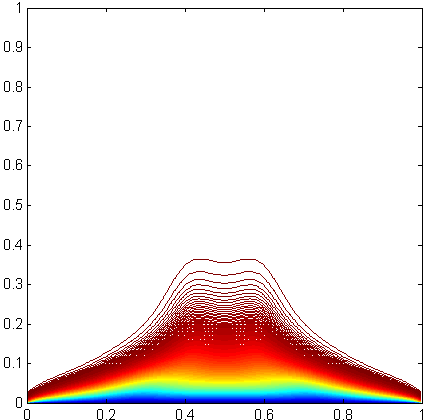}}
\subfigure[$t=0.003$]{\includegraphics[width=6.9cm]{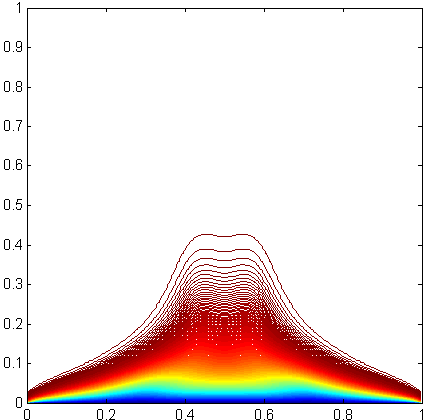}}
\subfigure[$t=0.0035$]{\includegraphics[width=6.9cm]{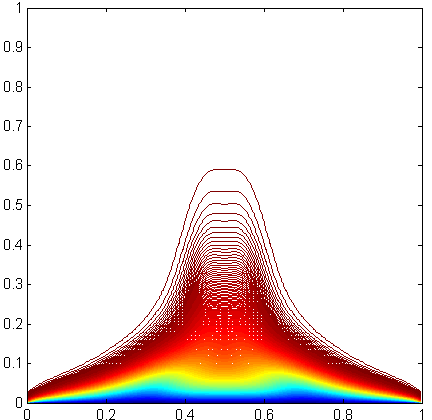}}
\subfigure[$t=0.004$]{\includegraphics[width=6.9cm]{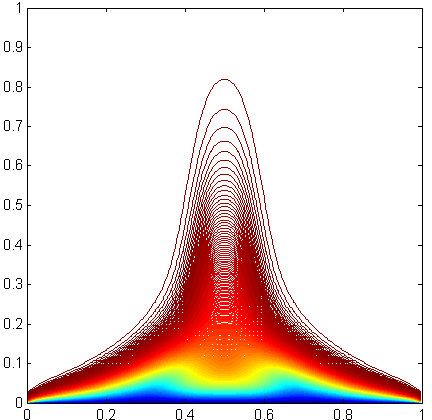}}
\subfigure[$t=0.0045$]{\includegraphics[width=6.9cm]{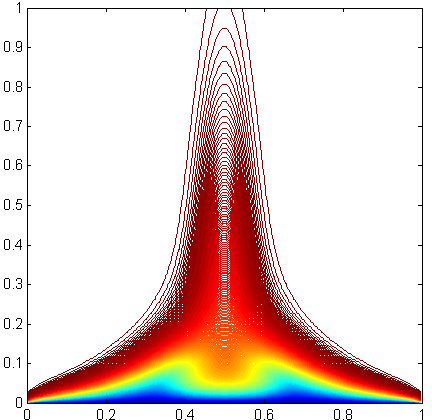}}
\subfigure[$t=0.005$]{\includegraphics[width=6.9cm]{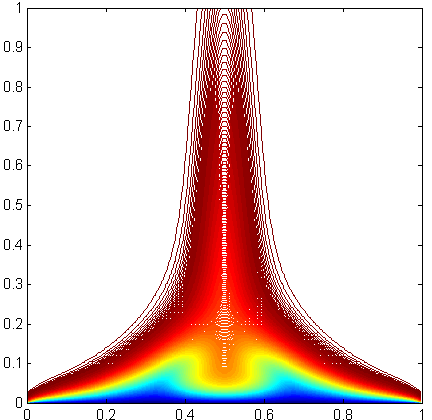}}
\subfigure{\includegraphics[width=8cm]{Bar_Pr1.PNG}}
\caption{Heatline, time evolution at Pr$=0.3$}
\label{Tplot2}
\end{figure}

\begin{figure}[htbp]
\centering
\subfigure[$t=0.00025$]{\includegraphics[width=6.9cm]{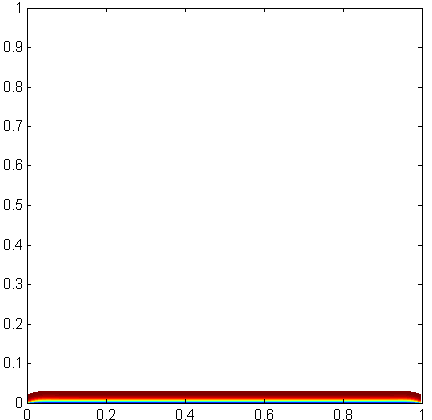}}
\subfigure[$t=0.0005$]{\includegraphics[width=6.9cm]{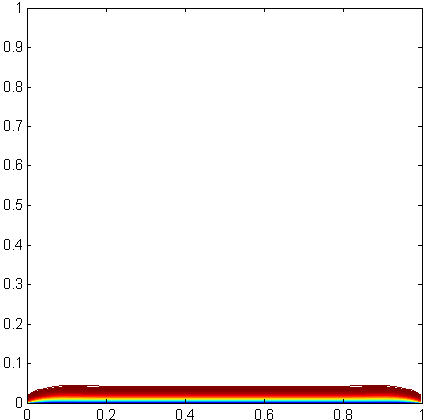}}
\subfigure[$t=0.001$]{\includegraphics[width=6.9cm]{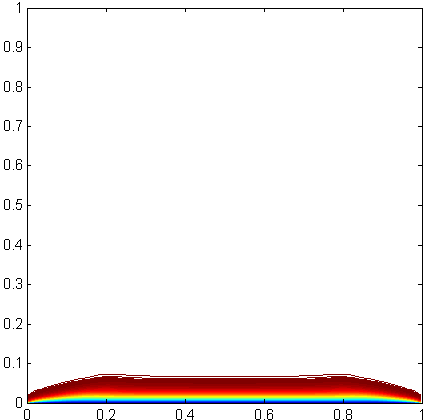}}
\subfigure[$t=0.0015$]{\includegraphics[width=6.9cm]{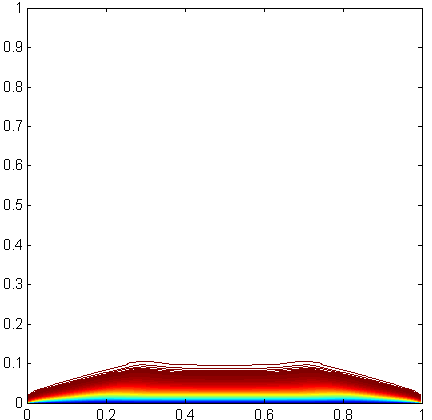}}
\subfigure[$t=0.002$]{\includegraphics[width=6.9cm]{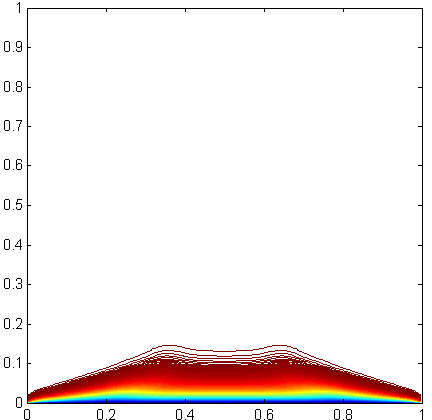}}
\subfigure[$t=0.0025$]{\includegraphics[width=6.9cm]{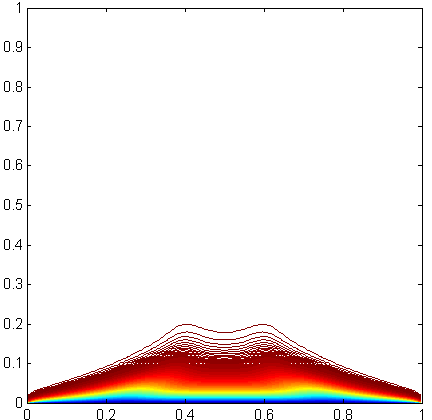}}
\subfigure{\includegraphics[width=8cm]{Bar_Pr1.PNG}}
\caption{Heatline, time evolution at Pr$=0.7$}
\label{Tplot3}
\end{figure}
\addtocounter{figure}{-1}
\begin{figure}[htbp]
\addtocounter{subfigure}{6}
\centering
\subfigure[$t=0.00275$]{\includegraphics[width=6.9cm]{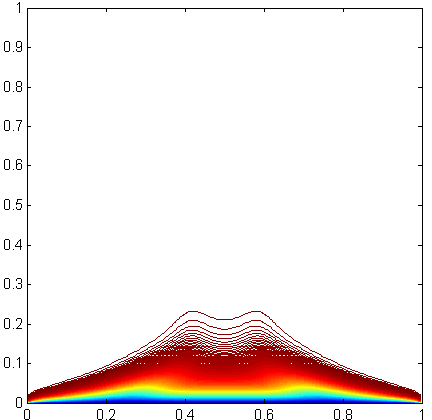}}
\subfigure[$t=0.003$]{\includegraphics[width=6.9cm]{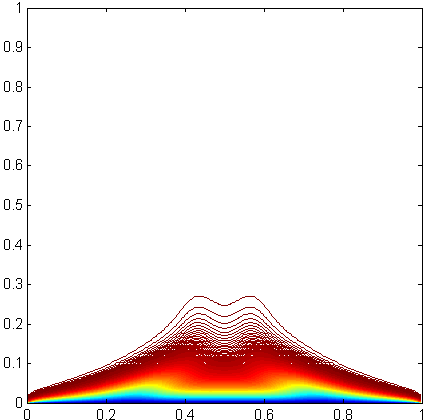}}
\subfigure[$t=0.0035$]{\includegraphics[width=6.9cm]{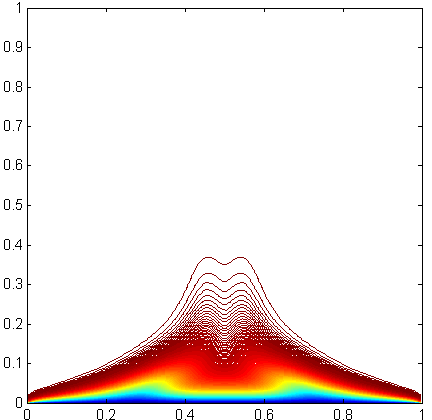}}
\subfigure[$t=0.004$]{\includegraphics[width=6.9cm]{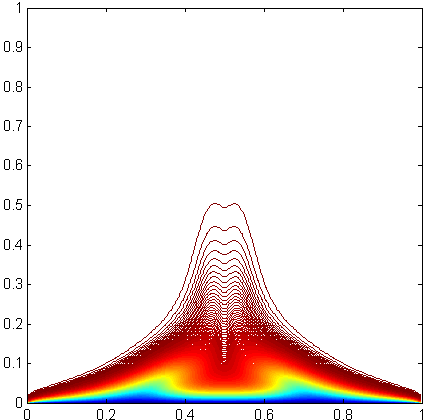}}
\subfigure[$t=0.0045$]{\includegraphics[width=6.9cm]{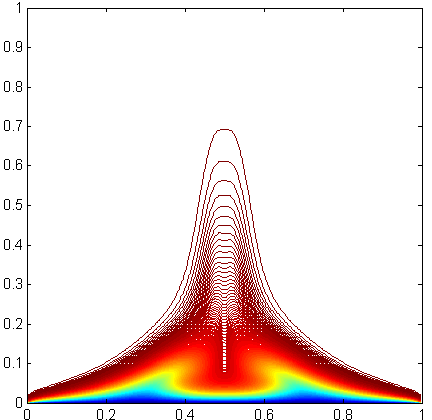}}
\subfigure[$t=0.005$]{\includegraphics[width=6.9cm]{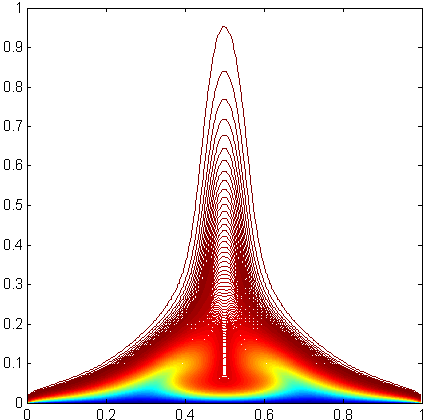}}
\subfigure{\includegraphics[width=8cm]{Bar_Pr1.PNG}}
\caption{Heatline, time evolution at Pr$=0.7$}
\label{Tplot4}
\end{figure}

\begin{figure}[htbp]
\centering
\subfigure[$t=0.00025$]{\includegraphics[width=6.9cm]{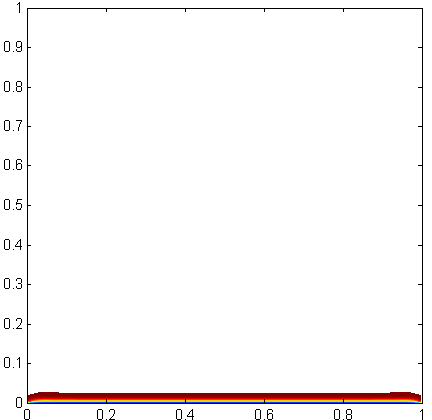}}
\subfigure[$t=0.0005$]{\includegraphics[width=6.9cm]{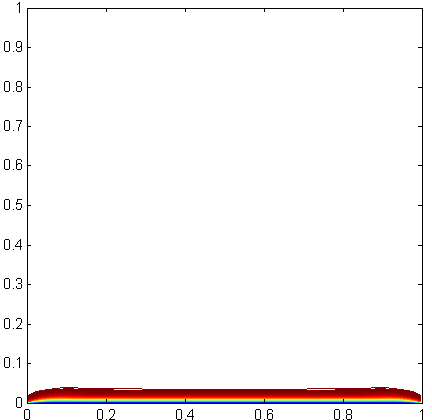}}
\subfigure[$t=0.001$]{\includegraphics[width=6.9cm]{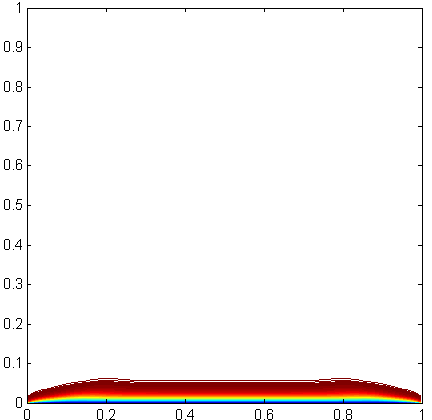}}
\subfigure[$t=0.0015$]{\includegraphics[width=6.9cm]{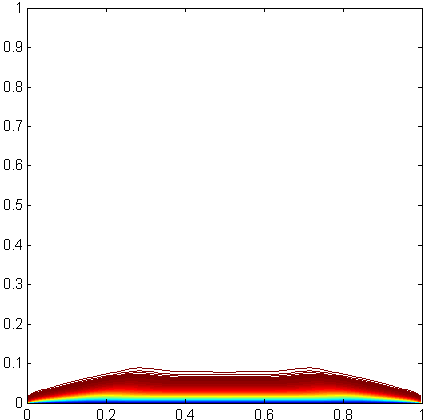}}
\subfigure[$t=0.002$]{\includegraphics[width=6.9cm]{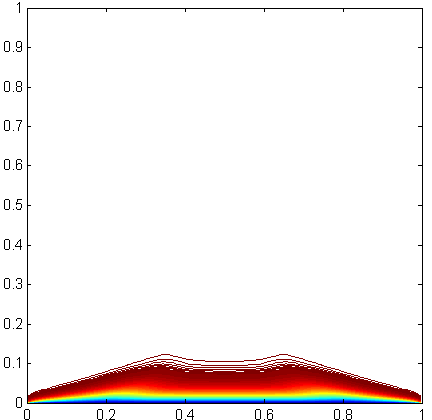}}
\subfigure[$t=0.0025$]{\includegraphics[width=6.9cm]{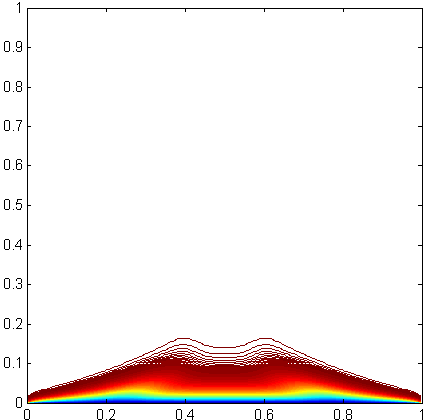}}
\subfigure{\includegraphics[width=8cm]{Bar_Pr1.PNG}}
\caption{Heatline, time evolution at Pr$=1$}
\label{Tplot5}
\end{figure}
\addtocounter{figure}{-1}
\begin{figure}[htbp]
\addtocounter{subfigure}{6}
\centering
\subfigure[$t=0.00275$]{\includegraphics[width=6.9cm]{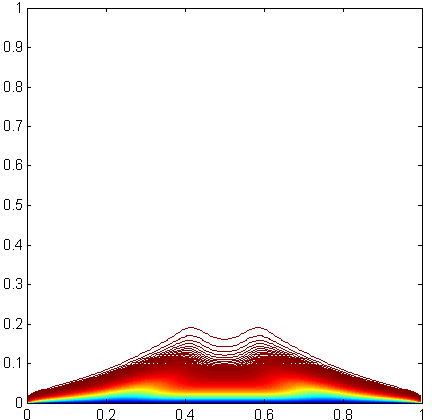}}
\subfigure[$t=0.003$]{\includegraphics[width=6.9cm]{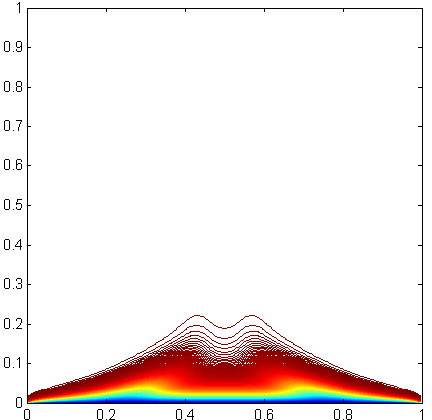}}
\subfigure[$t=0.0035$]{\includegraphics[width=6.9cm]{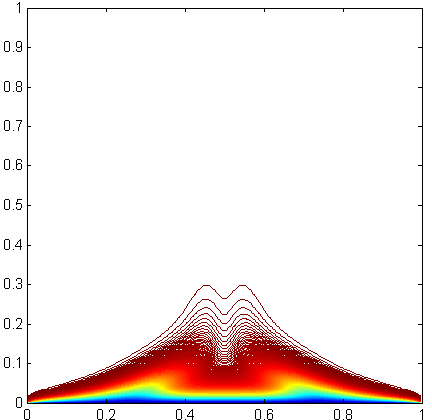}}
\subfigure[$t=0.004$]{\includegraphics[width=6.9cm]{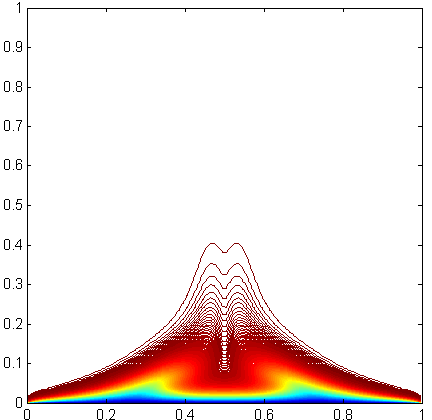}}
\subfigure[$t=0.0045$]{\includegraphics[width=6.9cm]{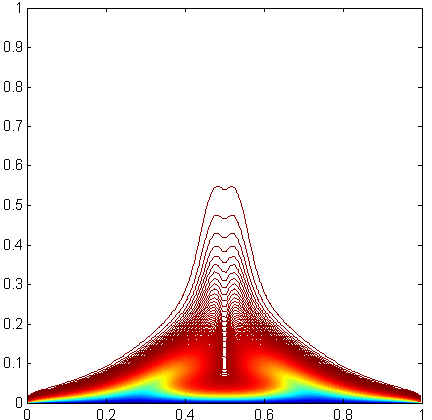}}
\subfigure[$t=0.005$]{\includegraphics[width=6.9cm]{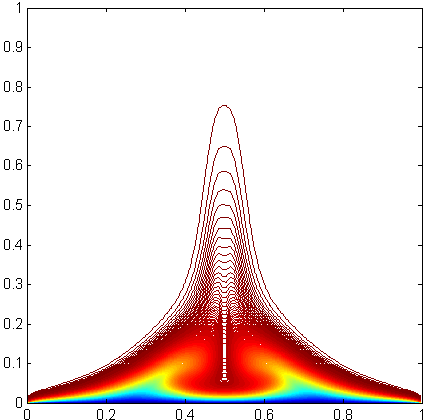}}
\subfigure{\includegraphics[width=8cm]{Bar_Pr1.PNG}}
\caption{Heatline, time evolution at Pr$=1$}
\label{Tplot6}
\end{figure}

\begin{figure}[htbp]
\centering
\subfigure[$t=0.00025$]{\includegraphics[width=6.9cm]{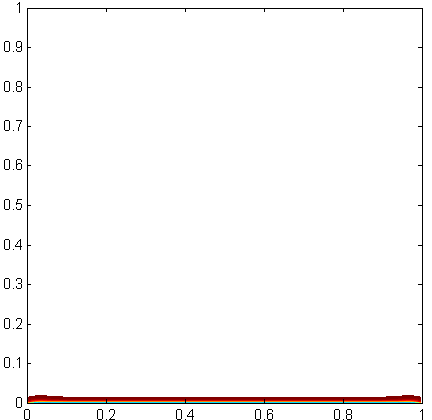}}
\subfigure[$t=0.0005$]{\includegraphics[width=6.9cm]{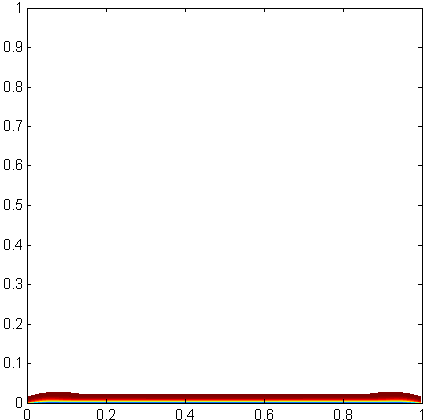}}
\subfigure[$t=0.001$]{\includegraphics[width=6.9cm]{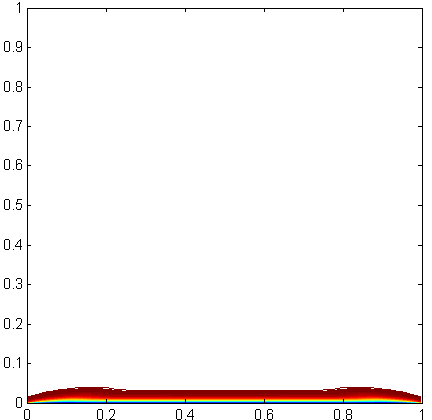}}
\subfigure[$t=0.0015$]{\includegraphics[width=6.9cm]{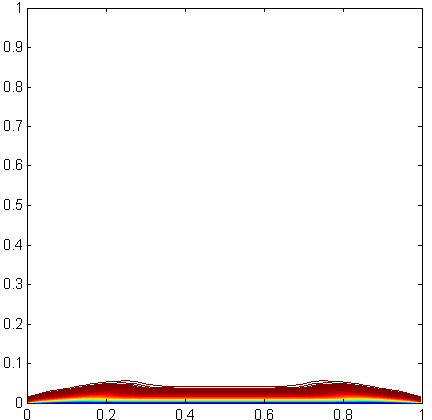}}
\subfigure[$t=0.002$]{\includegraphics[width=6.9cm]{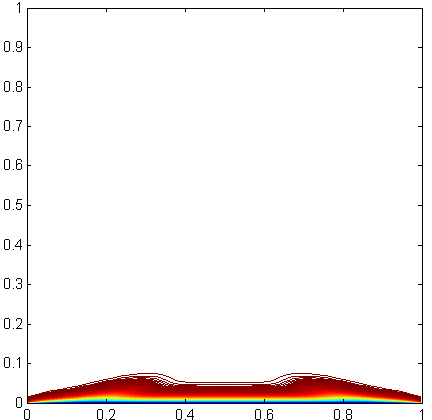}}
\subfigure[$t=0.0025$]{\includegraphics[width=6.9cm]{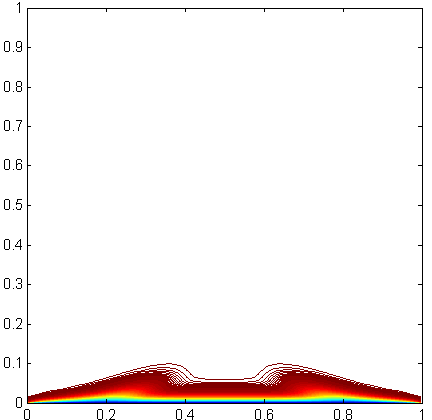}}
\subfigure{\includegraphics[width=8cm]{Bar_Pr1.PNG}}
\caption{Heatline, time evolution at Pr$=3$}
\label{Tplot7}
\end{figure}
\addtocounter{figure}{-1}
\begin{figure}[htbp]
\addtocounter{subfigure}{6}
\centering
\subfigure[$t=0.00275$]{\includegraphics[width=6.9cm]{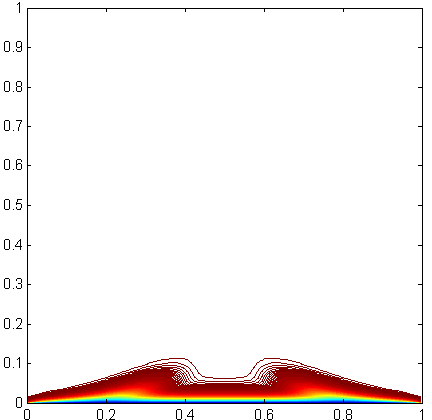}}
\subfigure[$t=0.003$]{\includegraphics[width=6.9cm]{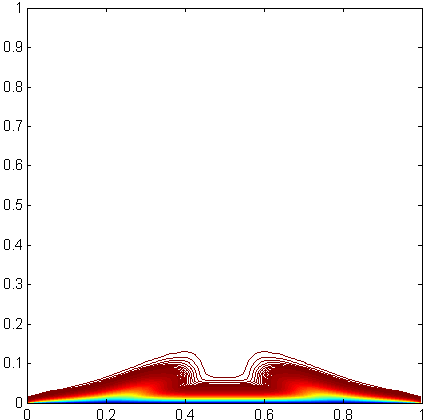}}
\subfigure[$t=0.0035$]{\includegraphics[width=6.9cm]{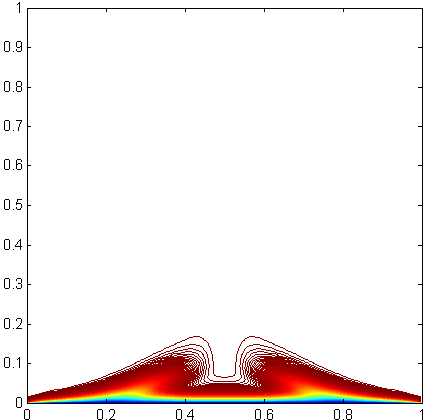}}
\subfigure[$t=0.004$]{\includegraphics[width=6.9cm]{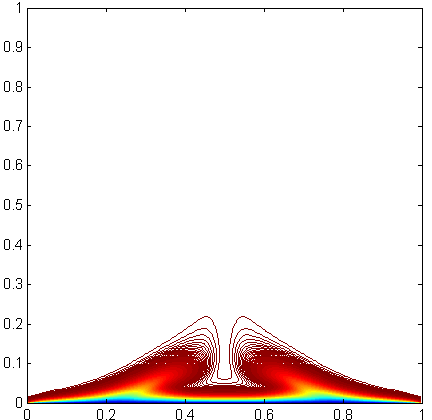}}
\subfigure[$t=0.0045$]{\includegraphics[width=6.9cm]{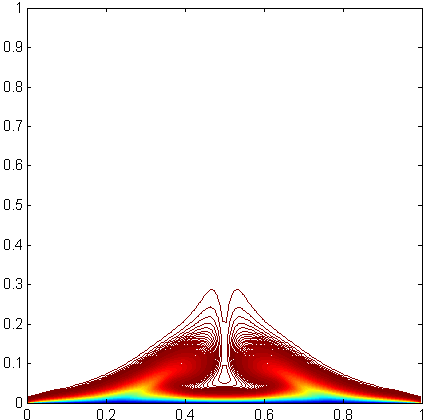}}
\subfigure[$t=0.005$]{\includegraphics[width=6.9cm]{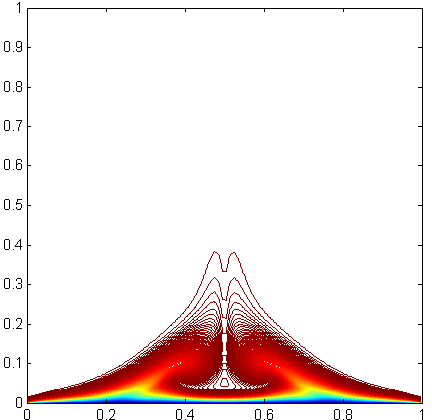}}
\subfigure{\includegraphics[width=8cm]{Bar_Pr1.PNG}}
\caption{Heatline, time evolution at Pr$=3$}
\label{Tplot8}
\end{figure}

\begin{figure}[htbp]
\centering
\subfigure[$t=0.00025$]{\includegraphics[width=6.9cm]{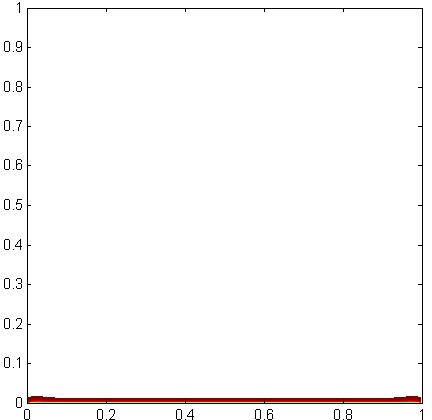}}
\subfigure[$t=0.0005$]{\includegraphics[width=6.9cm]{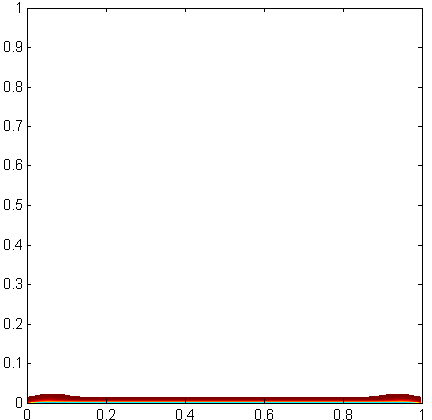}}
\subfigure[$t=0.001$]{\includegraphics[width=6.9cm]{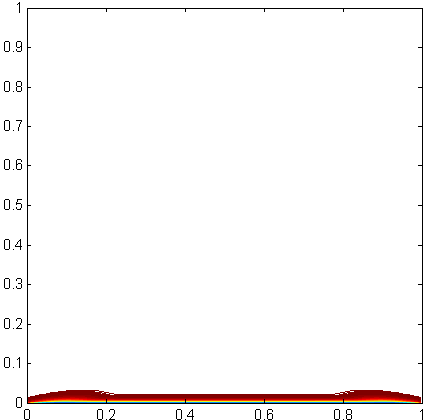}}
\subfigure[$t=0.0015$]{\includegraphics[width=6.9cm]{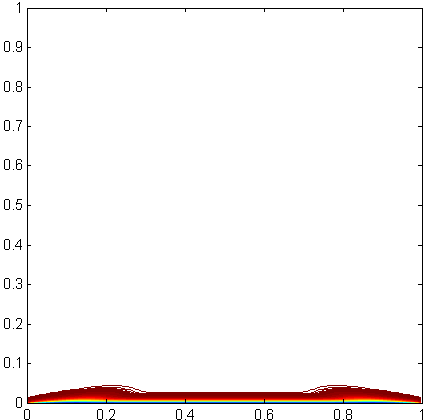}}
\subfigure[$t=0.002$]{\includegraphics[width=6.9cm]{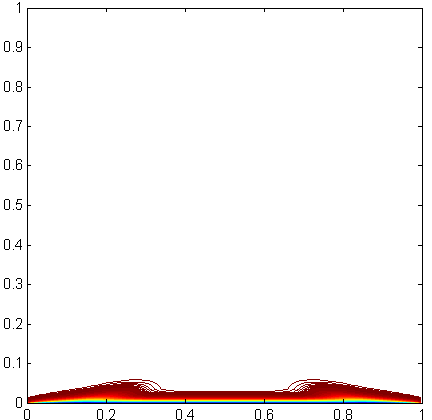}}
\subfigure[$t=0.0025$]{\includegraphics[width=6.9cm]{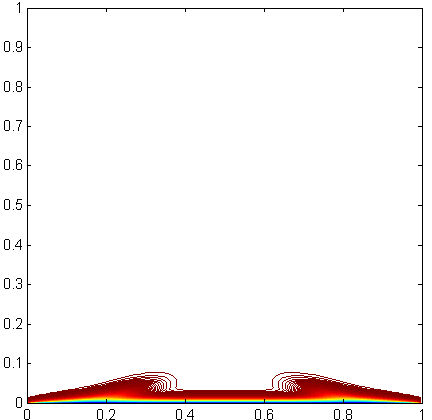}}
\subfigure{\includegraphics[width=8cm]{Bar_Pr1.PNG}}
\caption{Heatline, time evolution at Pr$=7$}
\label{Tplot9}
\end{figure}
\addtocounter{figure}{-1}
\begin{figure}[htbp]
\addtocounter{subfigure}{6}
\centering
\subfigure[$t=0.00275$]{\includegraphics[width=6.9cm]{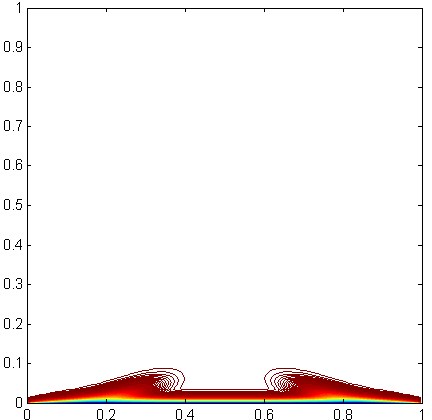}}
\subfigure[$t=0.003$]{\includegraphics[width=6.9cm]{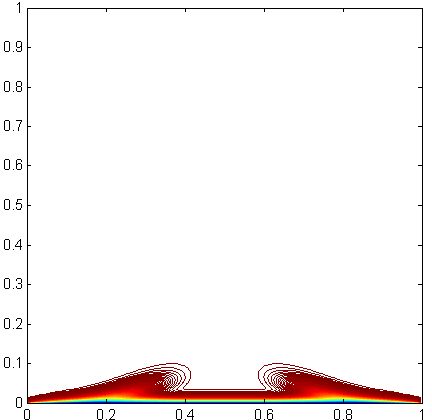}}
\subfigure[$t=0.0035$]{\includegraphics[width=6.9cm]{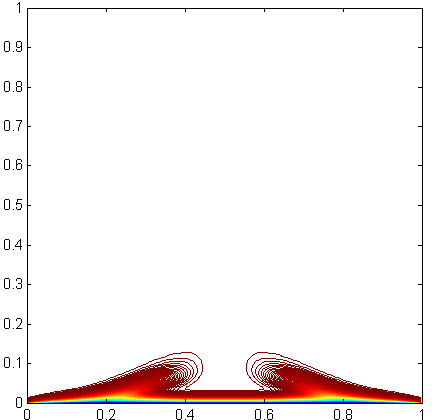}}
\subfigure[$t=0.004$]{\includegraphics[width=6.9cm]{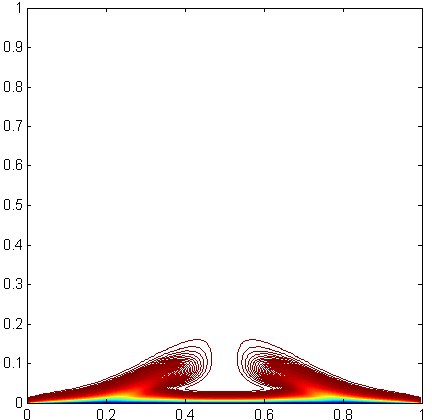}}
\subfigure[$t=0.0045$]{\includegraphics[width=6.9cm]{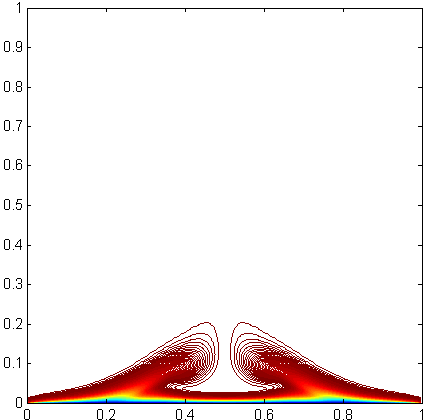}}
\subfigure[$t=0.005$]{\includegraphics[width=6.9cm]{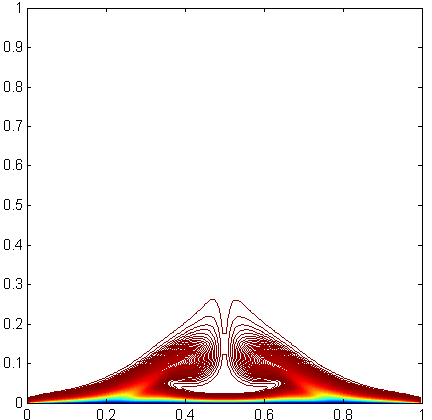}}
\subfigure{\includegraphics[width=8cm]{Bar_Pr1.PNG}}
\caption{Heatline, time evolution at Pr$=7$}
\label{Tplot10}
\end{figure}

\begin{figure}[htbp]
\centering
\subfigure[$t=0.00025$]{\includegraphics[width=6.9cm]{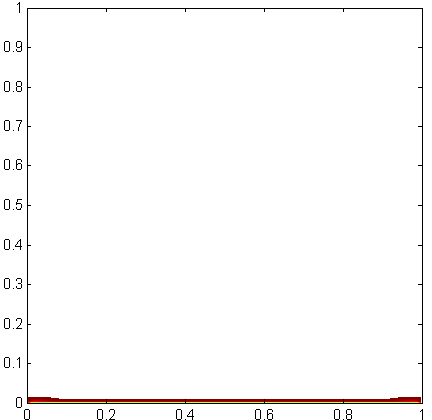}}
\subfigure[$t=0.0005$]{\includegraphics[width=6.9cm]{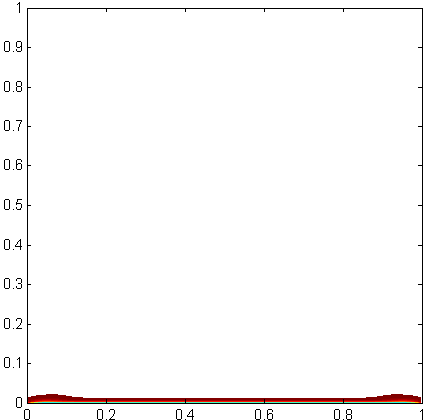}}
\subfigure[$t=0.001$]{\includegraphics[width=6.9cm]{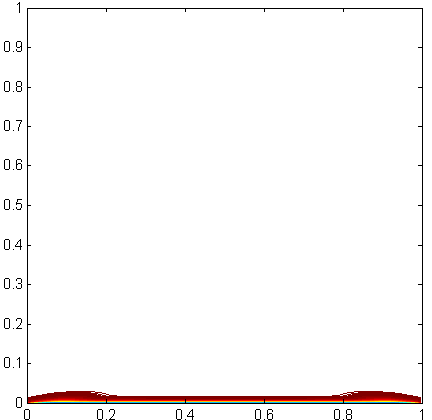}}
\subfigure[$t=0.0015$]{\includegraphics[width=6.9cm]{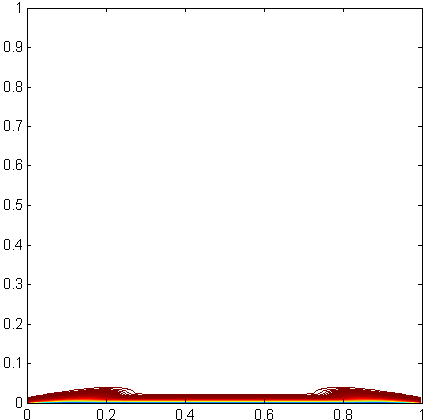}}
\subfigure[$t=0.002$]{\includegraphics[width=6.9cm]{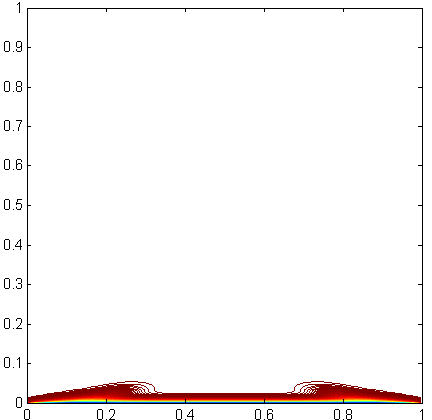}}
\subfigure[$t=0.0025$]{\includegraphics[width=6.9cm]{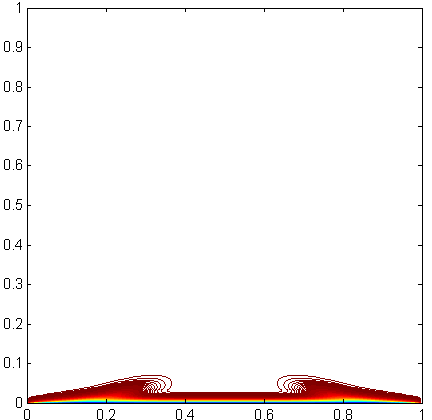}}
\subfigure{\includegraphics[width=8cm]{Bar_Pr1.PNG}}
\caption{Heatline, time evolution at Pr$=10$}
\label{Tplot11}
\end{figure}
\addtocounter{figure}{-1}
\begin{figure}[htbp]
\addtocounter{subfigure}{6}
\centering
\subfigure[$t=0.00275$]{\includegraphics[width=6.9cm]{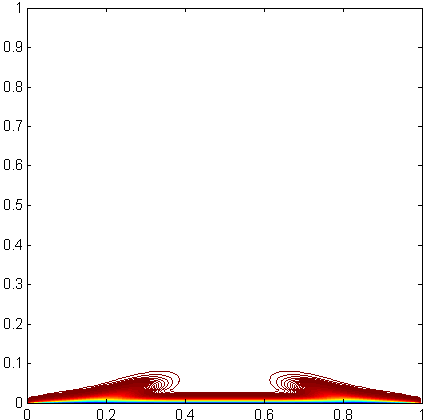}}
\subfigure[$t=0.003$]{\includegraphics[width=6.9cm]{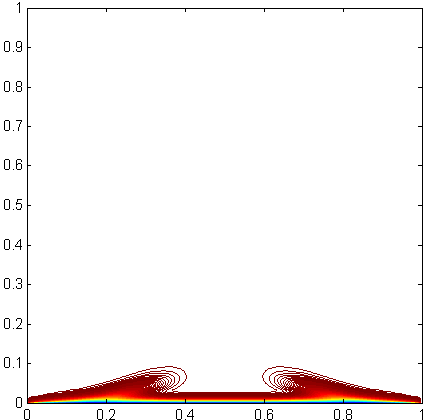}}
\subfigure[$t=0.0035$]{\includegraphics[width=6.9cm]{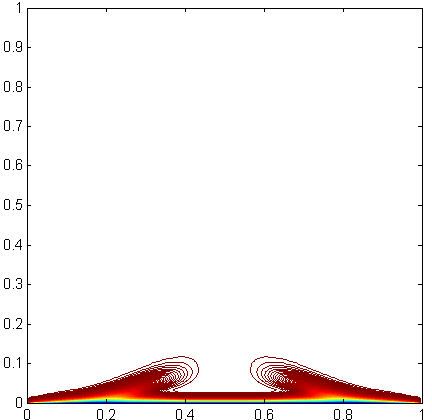}}
\subfigure[$t=0.004$]{\includegraphics[width=6.9cm]{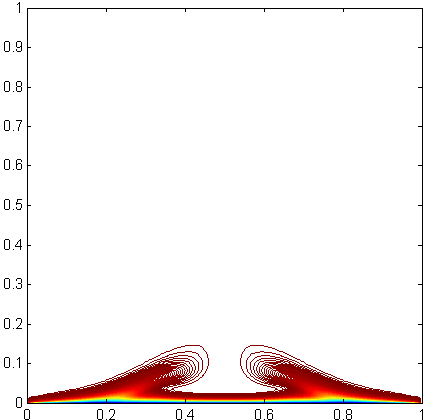}}
\subfigure[$t=0.0045$]{\includegraphics[width=6.9cm]{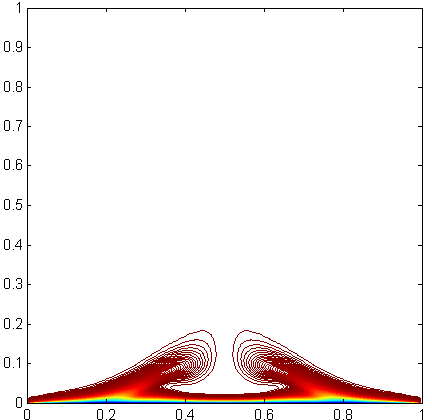}}
\subfigure[$t=0.005$]{\includegraphics[width=6.9cm]{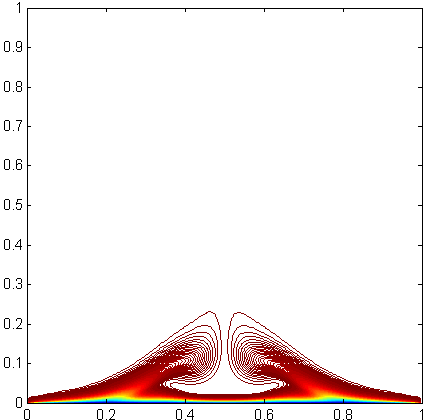}}
\subfigure{\includegraphics[width=8cm]{Bar_Pr1.PNG}}
\caption{Heatline, time evolution at Pr$=10$}
\label{Tplot12}
\end{figure}

It is worth talking into consideration that for liquid metals the Prandtl number is very small (Pr $<< 1$, generally in the range from 0.01 to 0.001. They have a high thermal conductivity and low viscosity. The value of Pr $= 1$ corresponds to diatomic gases, including air. For many fluids, including water, Prandtl number lies in the range from 1 to10. Large values of Pr $(>>1)$ correspond to high-viscosity oils and Pr $= 7$ corresponds to liquid water at room temperature. 
\newpage

\begin{figure}[htbp]
\centering
\subfigure[Pr$=0.3$]{\includegraphics[width=13cm]{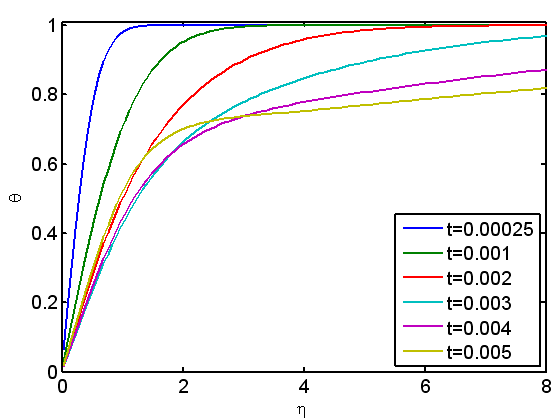}}
\subfigure[Pr$=0.7$]{\includegraphics[width=13cm]{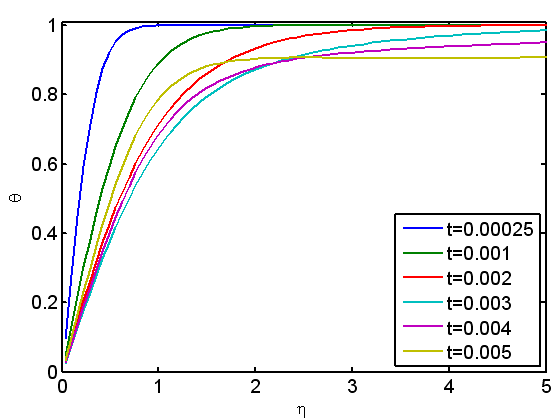}}
\caption{Similarity temperature field, time evolution}
\label{Tana1}
\end{figure}

Next we discuss the temperature distribution $\Theta$. Figures (\ref{Tana1}) show the similarity temperature distribution along the $\eta$-direction, evolving in time at various Prandtl number. As anticipated, since the temperature distribution is independent to $x$, we do not discuss the region where $x<0.1$. From Figures (\ref{Tana1}), the temperature distribution is monotonically increasing. $\Theta$ drops from a remote value to its value inside the thermal boundary layer adjacent to the wall. Near the backflow region, surprisingly, no discernible temperature signature appears between the dividing streamlines. The dimensionless temperature $\Theta$ is linear proportional to the dimensionless distance $\eta$ in the region close to the wall, which is consistent to our similarity temperature solution. \\

It is noticed that the dimensionless wall temperature gradient $\Theta'(0)$ raises with increase of Prandtl number, but the thermal boundary layer thickness decrease with increase of Prandtl number. Larger Prandtl numbers results in the thinner boundary layers and larger temperature gradients near the wall. When $\mathrm{Pr}$ is small, the heat diffuses very quickly compared to the velocity field and hence for liquid metals the thickness of the thermal boundary layer is much thicker than that of the velocity boundary layer. \\

\addtocounter{figure}{-1}
\begin{figure}[htbp]
\addtocounter{subfigure}{2}
\centering
\subfigure[Pr$=1$]{\includegraphics[width=13cm]{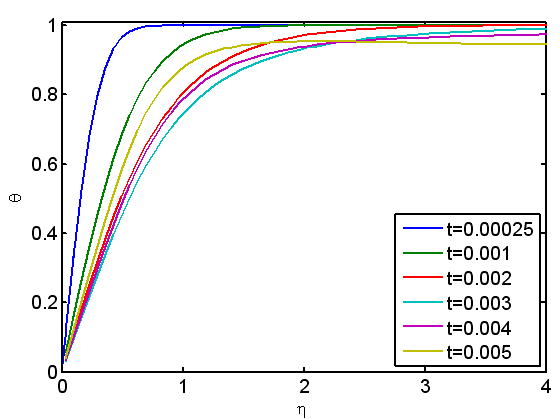}}
\subfigure[Pr$=3$]{\includegraphics[width=13cm]{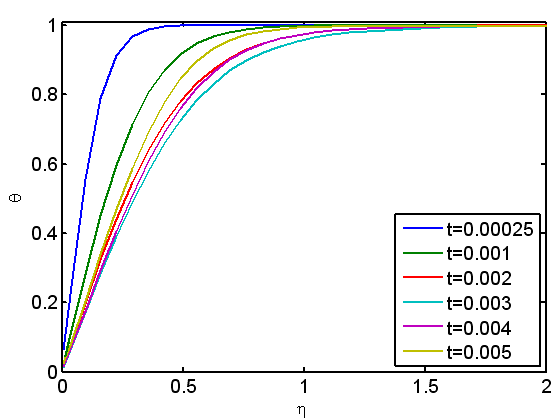}}
\caption{Similarity temperature field, time evolution}
\label{Tana2}
\end{figure}
\addtocounter{figure}{-1}
\begin{figure}[htbp]
\addtocounter{subfigure}{4}
\centering
\subfigure[Pr$=7$]{\includegraphics[width=13cm]{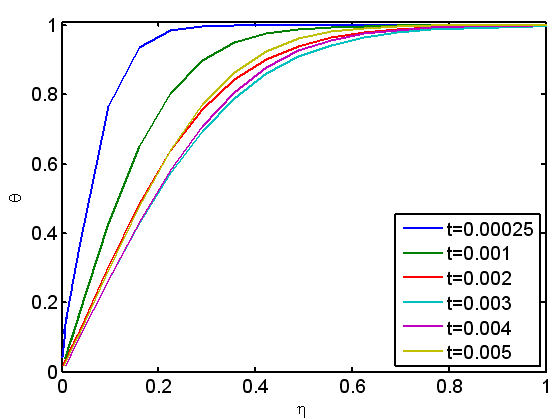}}
\subfigure[Pr$=10$]{\includegraphics[width=13cm]{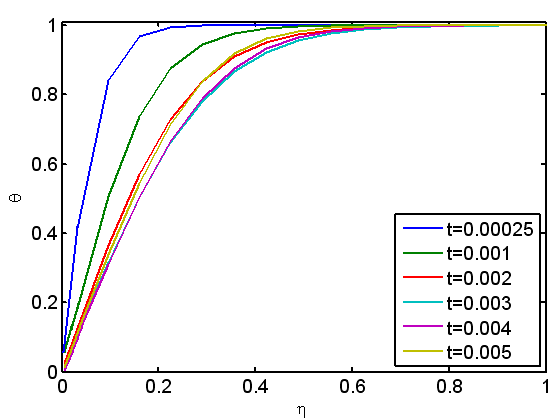}}
\caption{Similarity temperature field, time evolution}
\label{Tana3}
\end{figure}

\begin{figure}[htbp]
\centering
\subfigure[Pr$=0.3$]{\includegraphics[width=13cm]{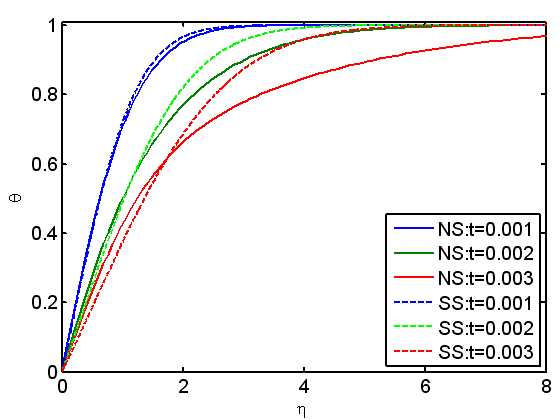}}
\subfigure[Pr$=0.7$]{\includegraphics[width=13cm]{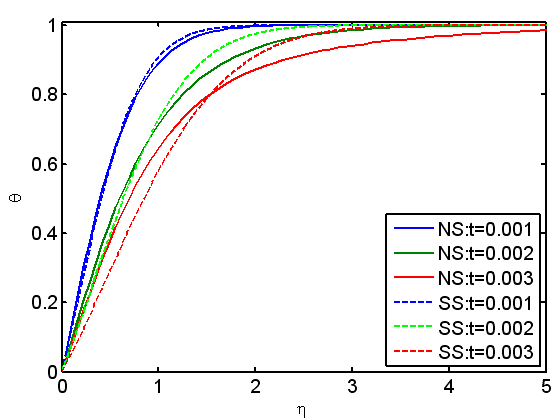}}
\caption{Comparison between the numerical temperature profiles and similarity temperature}
\label{Tana4}
\end{figure}

We are interested in comparison between the numerical simulation and the similarity result.  Figures (\ref{Tana4}) to (\ref{Tana6}) illustrate comparisons between the numerical simulations and the similarity solutions of nonisothermal reversed stagnation-point flow, when the dimensionless Reynolds number $Re=1000$. Lines without markers denote results obtained from numerical simulation (NS), dotted lines are obtained from the finite-difference formulations (SS).  It is shown that when Prandtl number is less than 1, our simulated results fall within the values obtained from the finite-difference formulations. 
\addtocounter{figure}{-1}
\begin{figure}[htbp]
\addtocounter{subfigure}{2}
\centering
\subfigure[Pr$=1$]{\includegraphics[width=13cm]{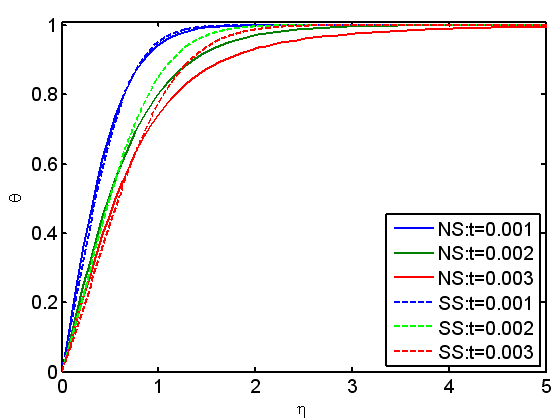}}
\subfigure[Pr$=3$]{\includegraphics[width=13cm]{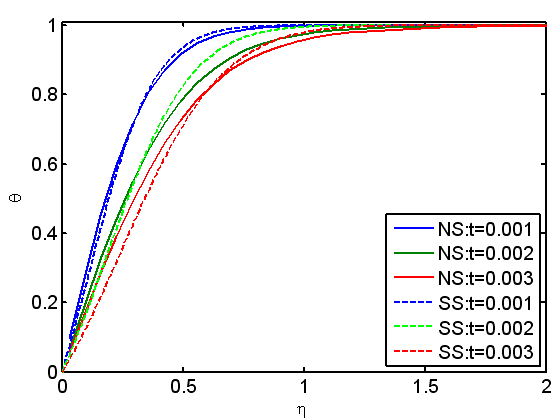}}
\caption{Comparison between the numerical temperature profiles and similarity temperature}
\label{Tana5}
\end{figure}
\addtocounter{figure}{-1}
\begin{figure}[htbp]
\addtocounter{subfigure}{4}
\centering
\subfigure[Pr$=7$]{\includegraphics[width=13cm]{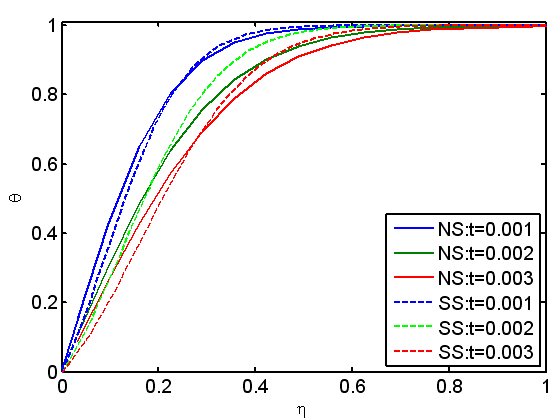}}
\subfigure[Pr$=10$]{\includegraphics[width=13cm]{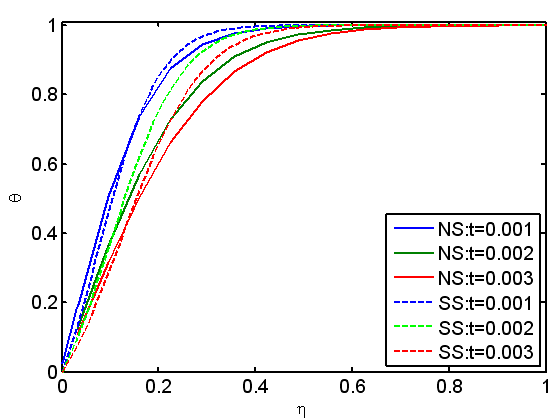}}
\caption{Comparison between the numerical temperature profiles and similarity temperature}
\label{Tana6}
\end{figure}

\chapter{PARTICULAR SOLUTION}
We complete the discussion of the Proudman-Johnson equation. Our objective is to obtain a similarity solution of the governing equation. Comparing to the results of numerical simulation, it is found that the potential flow $V_0$ may be expressed as a time dependent function.  In this chapter, rather than considering inviscid flow as external flow, it could instead be thought of a monotonic potential flow in balance of both viscous and convection terms in the total flow field. Let us discuss the similarity solution in a different manner.

\section{ANALYTICAL ANALYSIS}
In our two-dimensional model, the fluid remains at rest when time $t<0$ and is set in motion at $t>0$ such that at large distances far above the planar boundary the potential flow is a constant $V_0$ for all value of $t$. Both Proudman and Johnson \cite{proudman1962boundary}, and Robins and Howarth \cite{robins1972boundary} have set $V_0= 1$ and the corresponding boundary condition $f_{\eta}(\infty,\tau) = 1$. When the flow is in steady state such that $f_{\eta\tau}\equiv 0$, it was proven that the similarity velocity $f_\eta(\eta)$ cannot ultimately approach to 1. The differential equation has no solution. Smith \cite{smith1977development} generalized the solution of Proudman and Johnson with both viscous and convection terms in balance by considering the monotonic potential flow $V_0$  not to be a constant when the time is relatively large.\\

When the flow decays so rapidly that viscous force cannot be ignored away from the boundary, the viscous terms term $f_{\eta\eta\eta}$ must be included in the entire flow field. If the potential flow $V_0$ is restricted not to be a constant, then the boundary condition $f_{\eta}(\infty,\tau)$ may be expressed in a time dependent function. Numerical solution of reversed stagnation-point flow for this particular case has been studied in \cite{chio2012unsteady}. 
\\

Now we go through the analysis of this particular case. As with the governing equation of reversed stagnation-point flow, we can write the stream function as
\begin{subequations}
\begin{gather}
\psi = -\sqrt{A\nu}xf(\eta, \tau)\label{psi1}\\
\eta = \sqrt{\frac{A}{\nu}}y\\
\tau = At
\end{gather}
\end{subequations}
where $A$ is a constant proportional to $V_0(\tau)/L$, $V_0(\tau)$ is the external flow velocity removing from the plane and $L$ is the characteristic length. These result in the governing equation (\ref{e7a})
$$
   [f_{\eta\tau}-(f_{\eta})^2+ff_{\eta\eta}-f_{\eta\eta\eta}]=\mathrm{function~of~}\tau\mathrm{~only}.
$$
or the function of $\tau$ may be expressed as
\begin{equation}
f_{\eta\tau}-(f_{\eta})^2+ff_{\eta\eta}-f_{\eta\eta\eta}=-C(\tau),
     \label{e3_1a}
\end{equation}

Under the boundary conditions $f_{\eta}(\infty,\tau) = 1$, the value of $C(\tau)$ should be a constant and equal to $1$.  If the boundary condition $f_{\eta}(\infty,\tau)$ is restricted not to be a constant, following the assumption of Shapiro \cite{shapiro2006analytical}, a particular time-dependence function $C(\tau)$ may be expressed in the form
\begin{equation}
C(\tau)=\frac{c}{\tau^2}
\end{equation}
where $c$ is an arbitrary constant. The partial differential equation can be simplified by a similarity transformation when a new similarity variable is introduced. This converts the original partial differential equation into an ordinary differential equation.
\\\\
When $\tau$ is small the solution may be obtained by the method developed by Blasius \cite{blasius1908grenzschichten} and the solution satisfying the early stages of the diffusion are of the form
  \begin{equation}
f=\sqrt{\tau}\times \mathrm{function~of} \left(\frac{\eta }{\sqrt{\tau}}\right)
  \end{equation}
For small values of $\tau$, therefore, the variable ${\eta }/{\sqrt{\tau}}$ is more appropriate than $\eta$ itself. When we consider equation (\ref{e3_1a}), if a time dependent function is taken into account, the diffusion variable transformation is introduced
\begin{subequations}
\begin{gather}
\varsigma  = \frac{\eta }{\sqrt{\tau}} \\
f(\eta,\tau)=\frac{1}{\sqrt{\tau}}F(\varsigma)
\end{gather}
\end{subequations}
Here $\varsigma$ is the time combined nondimensional variable and $F$ is the nondimensional velocity function;  $F$ is then the sole function of $\varsigma$ and insertion of the similarity transformation yields an ordinary differential equation
  \begin{equation}
   -\frac{1}{2}\varsigma F''-F'-F'^2+FF''-F'''=-c
 \label{e3_6}
  \end{equation}
where the prime denotes the derivative with respect to the variable $
\varsigma$.
\\

Equation (\ref{e3_6}) is a third-order nonlinear ordinary differential equation. A crucial step in obtaining an analytical solution involves rearranging the equation as an autonomous differential equation. In mathematics, an autonomous differential equation is a system of ordinary differential equations which does not explicitly depend on the independent variable.
\\

In order to omit the variable $\varsigma$ in the differential equation, it is generally accepted as a change of variable
  \begin{equation}
  Q= F-\frac{1}{2}\varsigma
 \label{e3_0}
  \end{equation}
and the equation becomes to an autonomous differential equation
  \begin{equation}
QQ''-2Q'-Q'^2-Q'''=-c+\frac{3}{4}
 \label{e3_9}
  \end{equation}
In our analysis, $P=Q'$ is the dependent variable and $Q$ is the independent variable. Equation (\ref{e3_9}) is reversed ranged as
  \begin{equation}
QP'-2P-P^2-P''=-c+\frac{3}{4}
 \label{e3_7}
  \end{equation}
and the chain rule reduces equation (\ref{e3_7}) to a second-order ordinary differential equation
\begin{equation}
QP\frac{dP}{dQ}-2P-P^2-P\frac{d}{dQ}\left(P\frac{dP}{dQ}\right)=-c+\frac{3}{4}
 \label{e3_7a}
  \end{equation}
Equation (\ref{e3_7a}) is analytically solvable that the solution might be expressed as a low order polynomial. It is suggested that
  \begin{equation}
P=a+bQ+dQ^2
 \label{e3_8}
  \end{equation}
and substituting into equation (\ref{e3_7}) and comparing the coefficients in the powers of $Q$ results in a system of linear algebraic equations
\begin{subequations}
\begin{gather}
2a^2d+ab^2+2a-a^2=-c+\frac{3}{4}\\
8abd+b^3+2b+ab=0\\
8ad^2+(7b^2+2)d=0\\
12bd^2-bd=0\\
6d^3-d^2=0
\end{gather}
\end{subequations}
Solving the related algebraic equation, we have
  \begin{equation}
a=-\frac{3}{2}, ~~~~b=0, ~~~~c=\frac{3}{4},~~~~d=\frac{1}{6}
  \end{equation}
Substituting the constant into equation (\ref{e3_8}) yields a first-order differential equation
  \begin{equation}
Q'=-\frac{3}{2}+\frac{1}{6}Q^2
 \label{e3_15}
  \end{equation}
Equation (\ref{e3_15}) is Riccati equation, which is any ordinary differential equation that is quadratic in the unknown function. The standard form of classical Riccati equation is
  \begin{equation}
Q'=RQ^2+SQ+T
  \end{equation}
The solution of Riccati equation can be obtained by a change of dependent variable, where the dependent variable $y$ is converted to $q$ by \cite{zwillinger1998handbook}
  \begin{equation}
Q=-\frac{q'}{q}\frac{1}{R}
  \end{equation}
By identifying $R=\frac{1}{6}$, $S=0$ and $T=-\frac{3}{2}$, the change of variables in equation (\ref{e3_15}) becomes
  \begin{equation}
Q=-\frac{q'}{\frac{1}{6}q}=-\frac{6 q'}{q}
 \label{e3_16}
  \end{equation}
so the equation (\ref{e3_15}) becomes a second-order linear differential equation
  \begin{equation}
q''-\frac{1}{4}=0
  \end{equation}
of which the general solution is
  \begin{equation}
q=A\cosh\frac{\varsigma}{2}+B\sinh\frac{\varsigma}{2}
  \end{equation}
where $A$ and $B$ are arbitrary constants. Applying this solution in equation (\ref{e3_0}) leads to the general solution of equation (\ref{e3_6})
  \begin{equation}
F(\varsigma)=\frac{\varsigma}{2}-\frac{3A\sinh\frac{\varsigma}{2}+3B\cosh\frac{\varsigma}{2}}{A\cosh\frac{\varsigma}{2}+B\sinh\frac{\varsigma}{2}}
 \label{e3_21}
  \end{equation}
Application of the impermeability condition $F(0)=0$ leads to the determination of the constant $B=0$, so the exact solution becomes
  \begin{equation}
F(\varsigma)=\frac{\varsigma}{2}-3\tanh\frac{\varsigma}{2}
 \label{e3_22}
  \end{equation}
Collecting results, the velocity function becomes
  \begin{equation}
f(\varsigma,\tau)=\frac{1}{\sqrt{\tau}}\left(\frac{\varsigma}{2}-3\tanh\frac{\varsigma}{2}\right)
 \label{e3_23}
  \end{equation}
where $\varsigma=\displaystyle \sqrt{\frac{A}{\nu\tau}}y$ is the non-dimensional distance from the plate. In view of (\ref{e3_23}), the flow far away from the boundary becomes
  \begin{equation}
\lim_{\varsigma \rightarrow \infty}f(\eta,\tau)=\frac{1}{\sqrt{\tau}}\left(\frac{\varsigma}{2}-3\right)=\frac{\eta}{2{\tau}}-\frac{3}{\sqrt{\tau}}
 \label{e3_24}
  \end{equation}
where $f'$ tends exponentially to a positive constant as $\varsigma \rightarrow \infty$. The flow field is not able to remain unchanged at sufficient distances far away from the wall at any finite time, the potential flow cannot be assumed as the outer boundary condition for all values of $\tau.$ A continuous change as $V_0(\tau)$ decreases in magnitude for large $\tau$ should be expected outside the boundary.  
\\\\
Our objective is to obtain a particular solution of the unsteady reversed stagnation-point flow. The solution is obtained in the similarity transformation for unsteady viscous flows. The first term of (\ref{e3_24}) shows that the external flow is directed toward the $y-$axis and away from the wall. The appearance of a negative value in the second term in (\ref{e3_24}) describes a uniform velocity directed toward the wall.  The function $\tanh \varsigma$ has a Taylor series expansion with only odd exponents for $\varsigma$, that is
  \begin{equation}
\tanh  \varsigma =  \varsigma- \frac { \varsigma^3} {3} + \frac {2 \varsigma^5} {15} - \frac {17 \varsigma^7} {315} + \cdots
  \end{equation}
Thus, the flow near the boundary becomes
  \begin{equation}
\lim_{\varsigma \rightarrow 0} f(\eta,\tau) = -\frac{ \varsigma}{\sqrt{\tau}}=-\frac{ \eta}{\tau}<0
 \label{e3_25}
  \end{equation}

Surprisingly, the component of velocity normal to a wall is not outward the wall in the region near the reversed stagnation point.  The vorticity created at the wall will be convected outward the wall, which spreads the vorticity towards its source at the boundary. An explanation is that an adverse pressure gradient in the region close to the wall leads to a boundary-layer separation and associated flow reversal, and therefore the flow divides into a wall region of reversed flow and an outer region of forward flow. 
\\

At this part it is particular to emphasize a point which seems to been ignored in the analysis. Near the wall region or in the boundary layer the phenomenon of reversed flow with boundary-layer separation occurred. No trouble arose from the idealization of Proudman and Johnson that the viscous forces are of the same order as the inertial forces near the stagnation point. Since no information concerning the nature of the flow for finite times has yet been included, there is no justification, theoretical or experimental, for supposing that at large distances from the wall ($\eta \rightarrow \infty$) the velocity $v(x,\eta,\tau)$ should pass over smoothly into that for inviscid $V_0$. Once the reversed flow has occurred, the external boundary condition must be affected and that the whole problem becomes conceptually unsound.

\section{NUMERICAL SOLUTION}
\subsection{VELOCITY DISTRIBUTION}
The particular solution (\ref{e3_23}) is noteworthy in that it is completely analytical. Now this solution satisfies the Navier-Stokes equations; however the equation has no solution that satisfies the necessary no-slip condition at the wall in the presence of non-zero term $F'(0)=-1$.
\\\\
In order to satisfy this too, the effect of no-slip condition $F'(0)=0$ must be taken into account. To do this we apply the numerical analysis for the velocity distribution. The similarity equation and the relevant boundary conditions are
\begin{equation}
  \left\{
  \begin{array}{rr}
      -\frac{1}{2}\varsigma F''-F'-F'^2+FF''-F'''=-c \\
     F(0)= F'(0)=0 \\
     F'(\infty) = \frac{1}{2}
  \end{array}
  \right.
  \label{q:e12}
\end{equation}
where $c={3}/{4}$ to satisfy the unsteady viscous flows in the outer region.
\\\\
Equation (\ref{q:e12}) is a third-order nonlinear ordinary differential equation. It is convenient to describe the problem in terms of a system of first-order equations when solving an ODE system numerically \cite{shampine2003solving}. In numerical analysis, the Runge-Kutta methods are an important family of implicit and explicit iterative methods for the approximation of solutions of ordinary differential equations. This method applies a trial step at the midpoint of an interval to cancel out lower-order error terms, besides; Runge-Kutta formulas are the methods of solving initial value problems for ordinary differential equations. Since (\ref{q:e12}) is a boundary-value problem, apparently we have to alter the boundary value conditions into the initial value conditions.
\\\\
For example solving an $n^{th}$-order problem numerically is common practice to reduce the equation to a system of $n$ first-order equations. Then, by defining $y_1 = F,~y_2 = F',~y_3 = F''$, the ODE reduces to the form
    \begin{equation}
\frac{d\mathbf{y}}{d\varsigma} =
\begin{bmatrix}
  y_2  \\
  y_3  \\
  c-\frac{1}{2}\varsigma y_3-y_2-y_2^2 +y_1y_3
\end{bmatrix}
    \label {eq:e22}
\end {equation}
\\
The first task is to reduce the equation above to a system of first-order equations and define in MATLAB a function to return these. The relevant MATLAB expression for equation~(\ref{eq:e22}) would be:

\lstset{language=MATLAB,
         breaklines=true,
         extendedchars=false,
         showstringspaces=false,
         numbers=left,
         numberstyle=\ttfamily\scriptsize,
         frame=trbl,framesep=5pt,framexleftmargin=8mm,
         frameround=tttt,
         keywordstyle=\ttfamily\bf\color{red},
         ndkeywordstyle=\ttfamily\bf\color{brown},
         commentstyle=\color{blue},
         identifierstyle=\ttfamily\color{black}\bfseries,
         stringstyle=\color{red}\ttfamily
}
\begin{lstlisting}[label=MATLAB,caption=System of first-order equations ]
function dy = stagnation(t,y)
    c=3/4;
    dy = zeros(3,1);
    dy(1) = y(2);
    dy(2) = y(3);
    dy(3) = c-1/2*t*y(3)-y(2)-y(2)*y(2)+y(1)*y(3);
end
\end{lstlisting}

The next step is to convert the boundary value into initial value, because $ode45$, an ODE solver in MATLAB, can only solve the initial-value problem. From equation~(\ref{q:e12}), we gauss the value of $F''(0)$ such that $F'(\infty) = \frac{1}{2}$. The commands written in MATLAB would be

\begin{lstlisting}[label=MATLAB,caption=ODE solver]
function main
    [T,Y] = ode45(@stagnation,[0 10],[0 0 -1]);
end
\end{lstlisting}



\begin{figure}[htbp]
\centering
\subfigure[$F''(0)=0.5$]{\includegraphics[width=15cm]{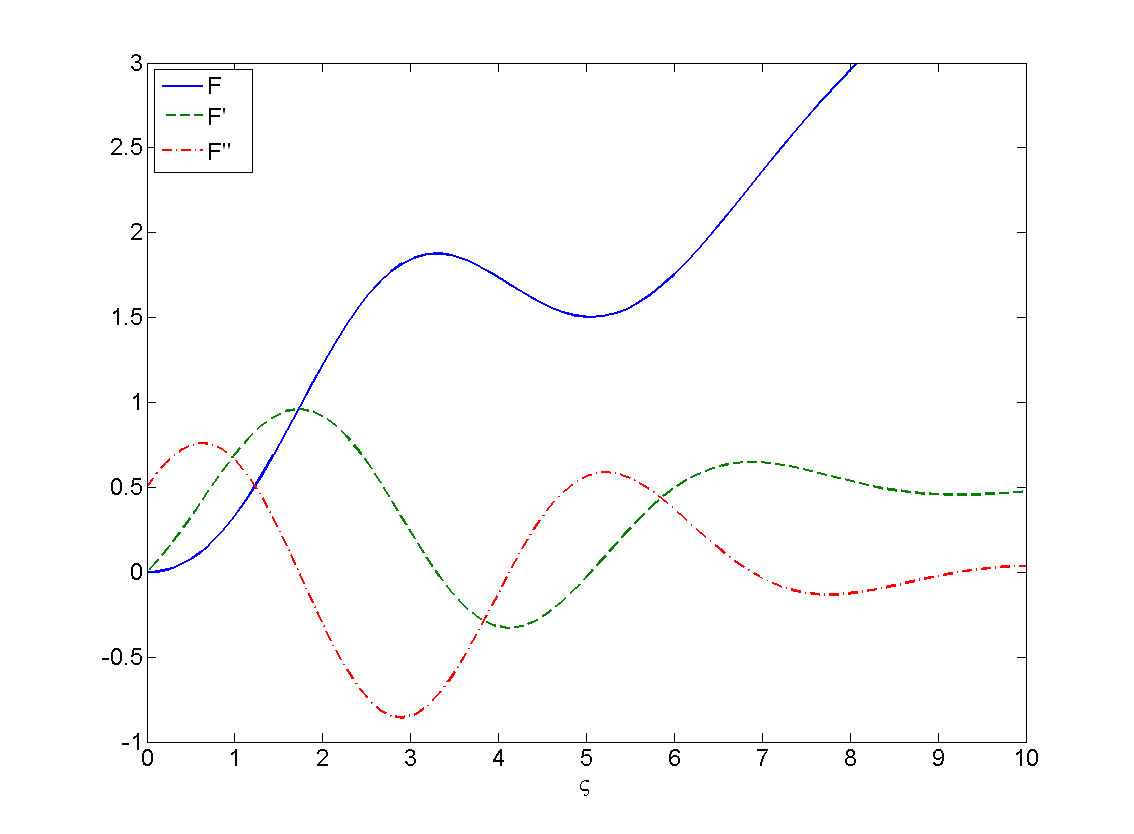}}
\subfigure[$F''(0)=0$]{\includegraphics[width=15cm]{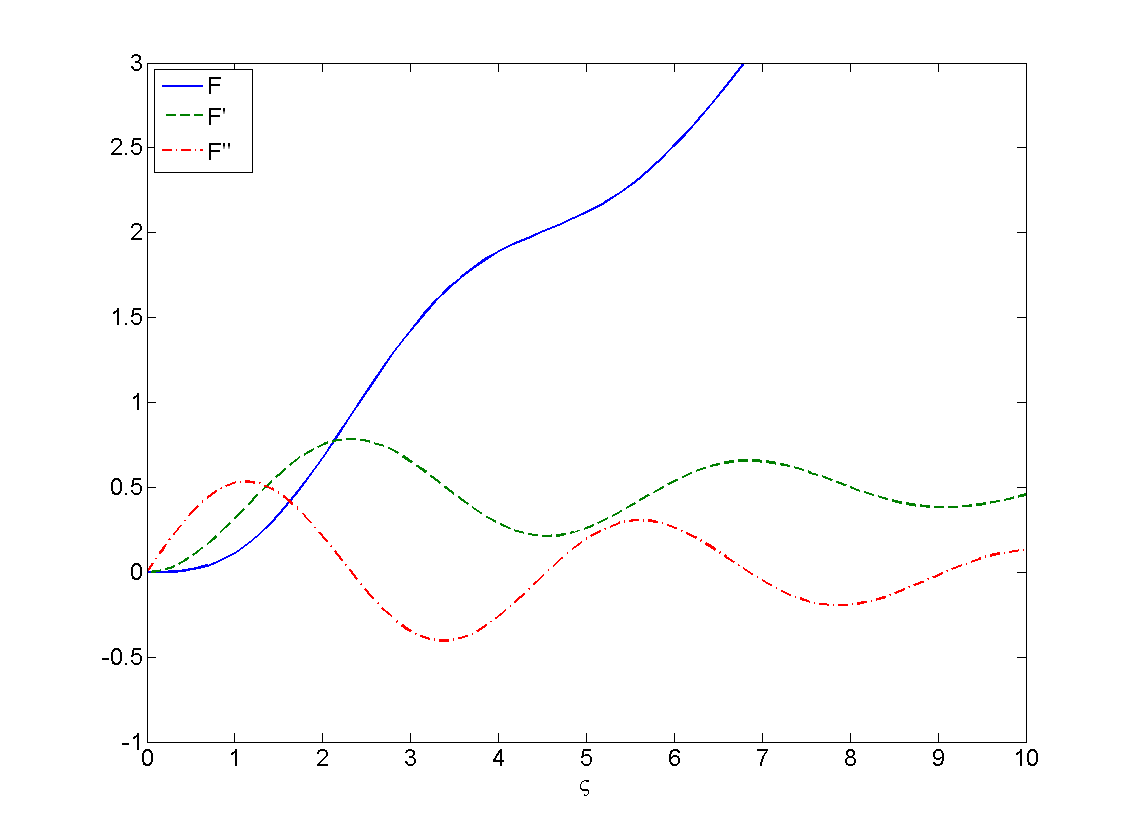}}
\caption{Numerical solutions of viscous reversed stagnation-point flow}
\label{gfpi1}
\end{figure}
\addtocounter{figure}{-1}
\begin{figure}[htbp]
\addtocounter{subfigure}{2}
\centering
\subfigure[$F''(0)=-0.5$]{\includegraphics[width=15cm]{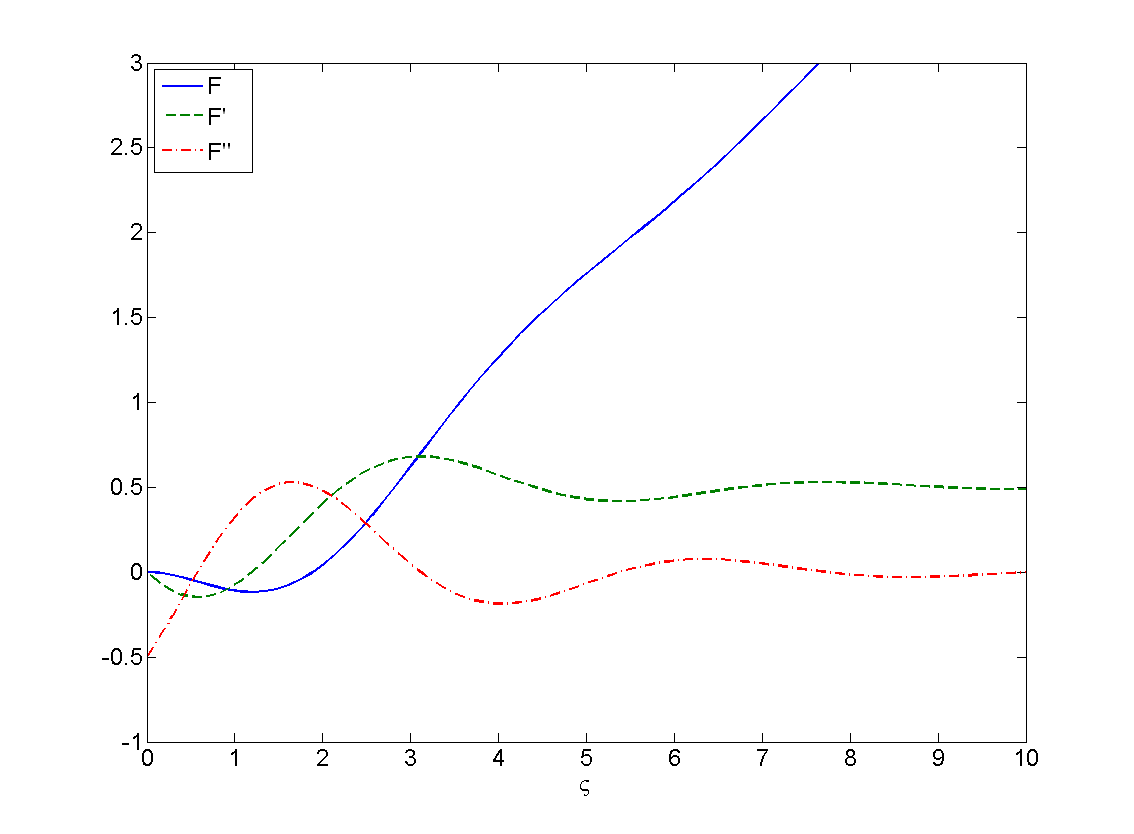}}
\subfigure[$F''(0)=-1$]{\includegraphics[width=15cm]{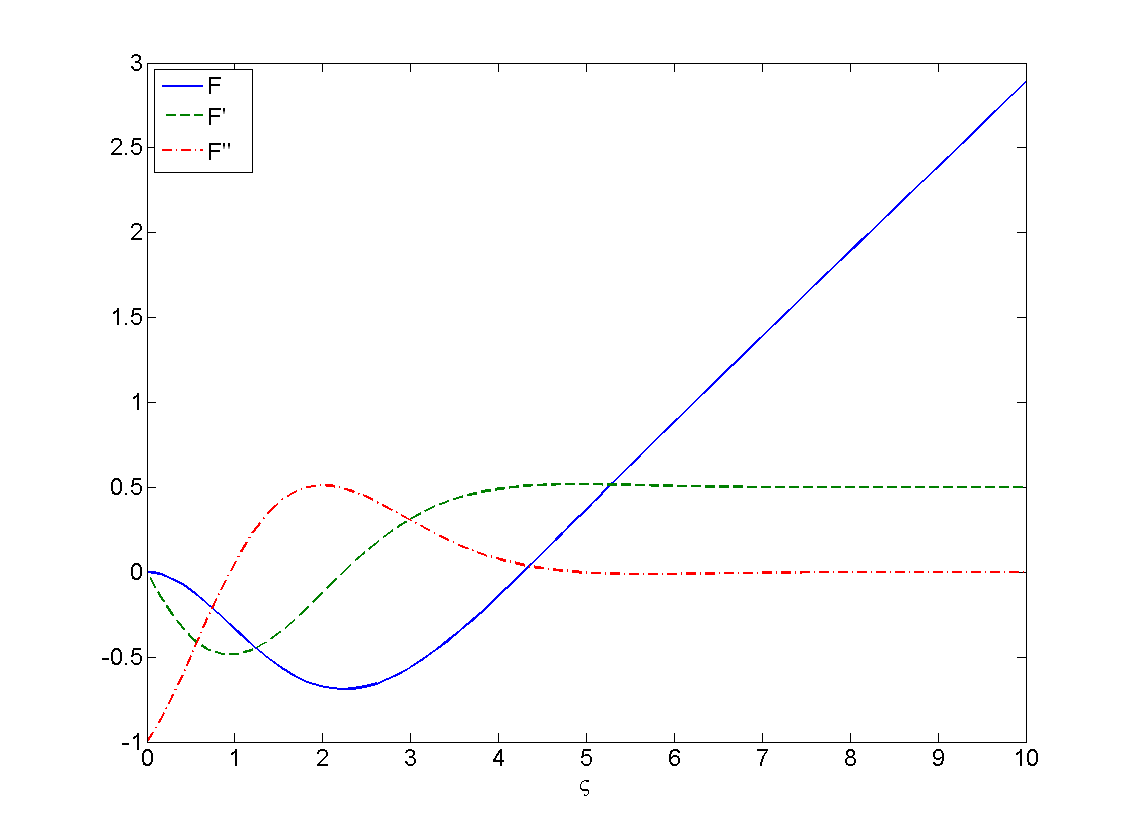}}
\caption{Numerical solutions of viscous reversed stagnation-point flow}
\label{gfpi2}
\end{figure}
\addtocounter{figure}{-1}
\begin{figure}[htbp]
\addtocounter{subfigure}{4}
\centering
\subfigure[$F''(0)=-1.5$]{\includegraphics[width=15cm]{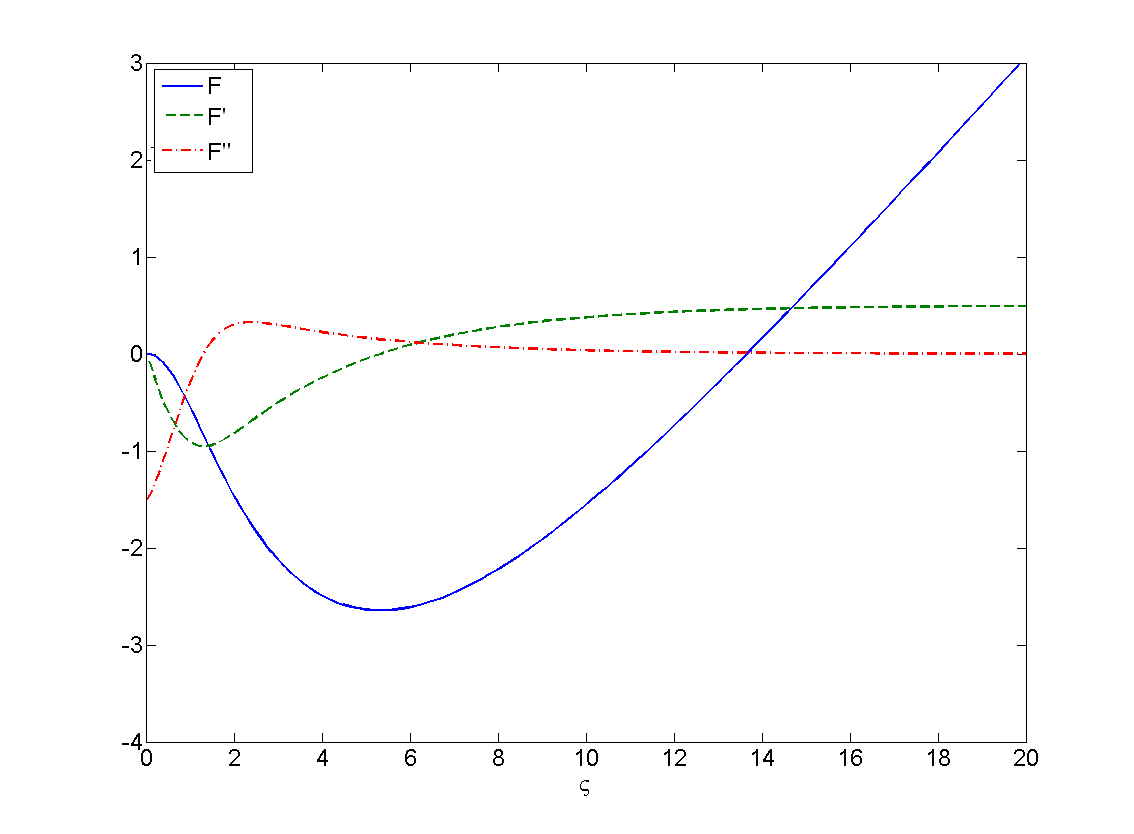}}
\subfigure[$F''(0)=-1.7$]{\includegraphics[width=15cm]{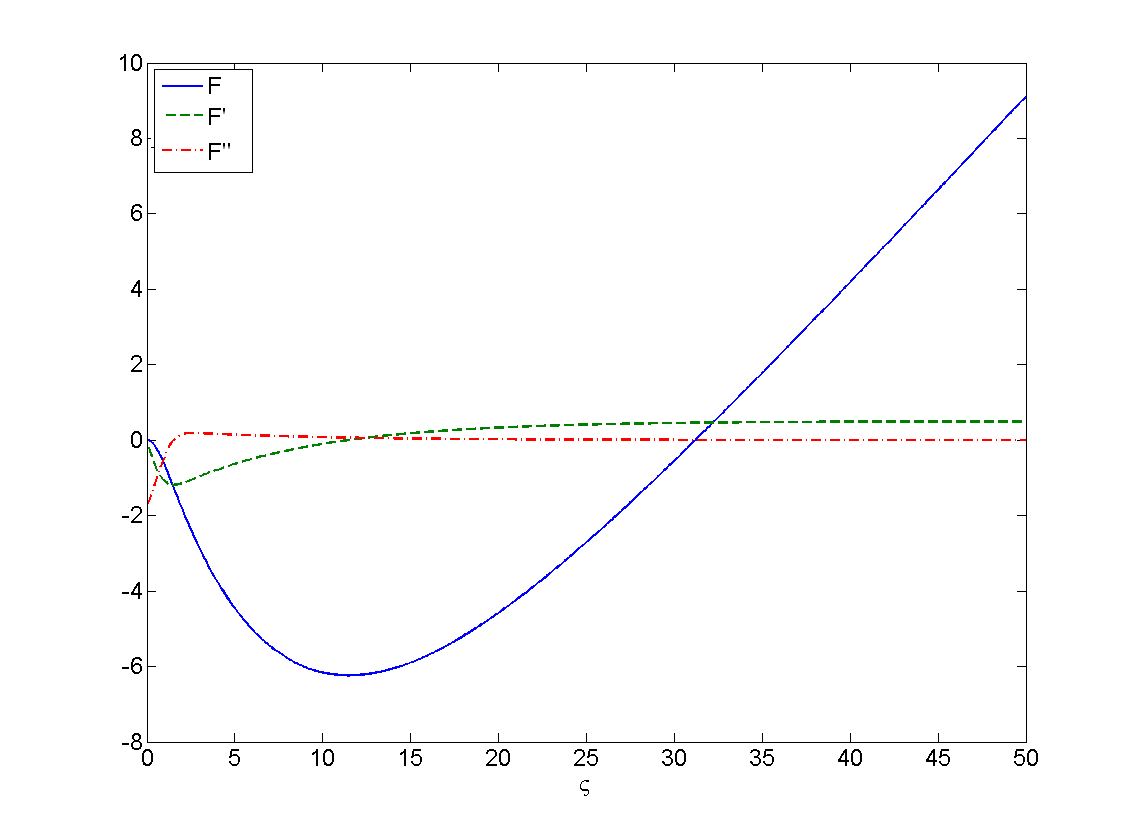}}
\caption{Numerical solutions of viscous reversed stagnation-point flow}
\label{gfpi3}
\end{figure}

The complete solutions of two-dimensional stagnation-point flow with different values of $F''(0)$ are shown from Figures (\ref{gfpi1}) to (\ref{gfpi3}). In these figures the similarity stream function $F$, the velocity profile $F'$ and the shear stress $F''$ are represented. This solution is a similarity solution of the reversed stagnation-point flow over a flat plate, describing an unsteady viscous flow in both outer and inner regions. \\

The result looks interesting from both theoretical and engineering points of view. A single dividing streamline plane separates streamlines approaching the plate from external flow streamlines. The boundary-layer thickness increases as the square root of $\tau$. The boundary layer thickness is the distance from the body at which the velocity is $99\%$ of the velocity obtained from an inviscid solution. When $F''(0)>0$, the values of $F$ and $F'$ are always greater than zero. No separation occurs near the wall region. 
\\
\begin{figure}[htbp]
\centering
\subfigure[$F''(0)=0.5$]{\includegraphics[width=11cm]{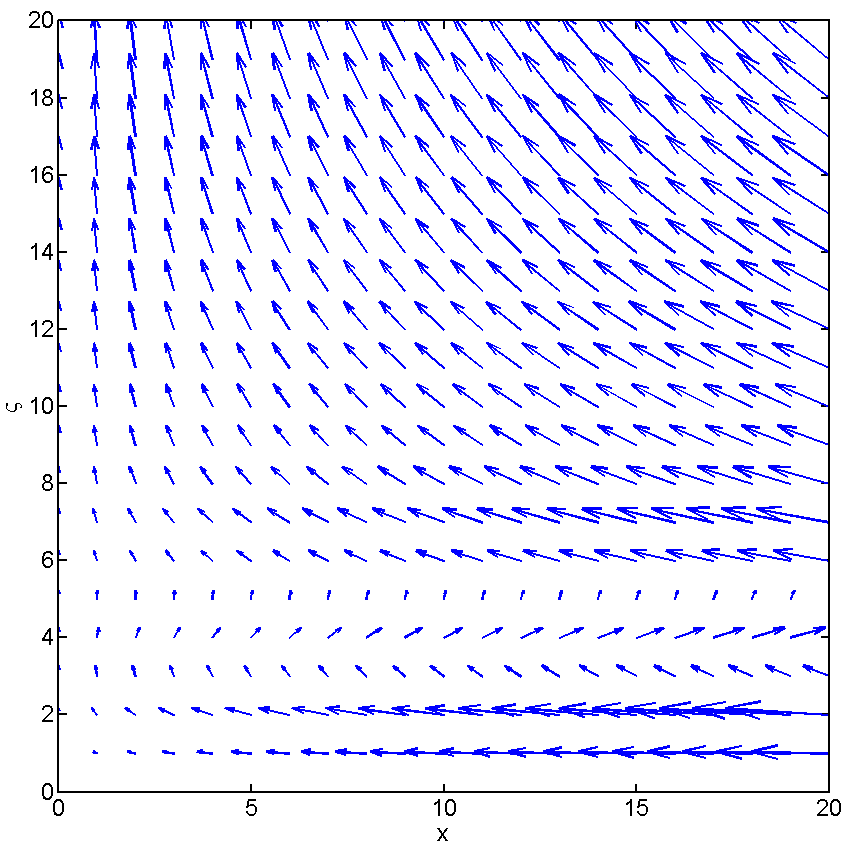}}
\subfigure[$F''(0)=0$]{\includegraphics[width=11cm]{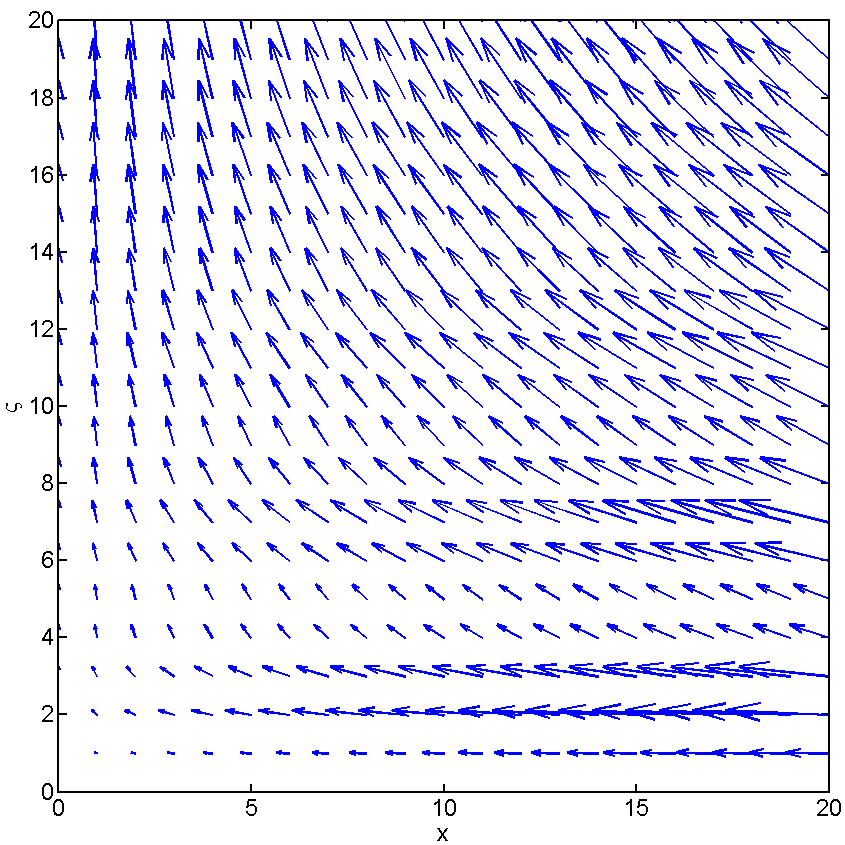}}
\caption{Similarity velocity field as a function of $\varsigma$}
\label{stream}
\end{figure}
\addtocounter{figure}{-1}
\begin{figure}[htbp]
\addtocounter{subfigure}{2}
\centering
\subfigure[$F''(0)=-0.5$]{\includegraphics[width=11cm]{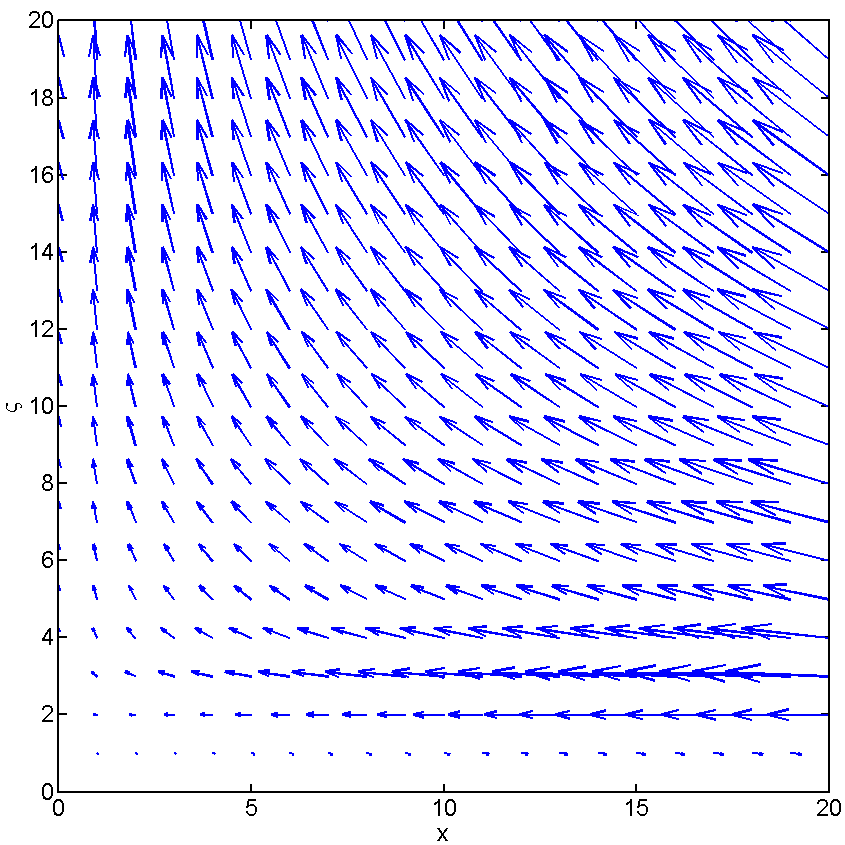}}
\subfigure[$F''(0)=-1$]{\includegraphics[width=11cm]{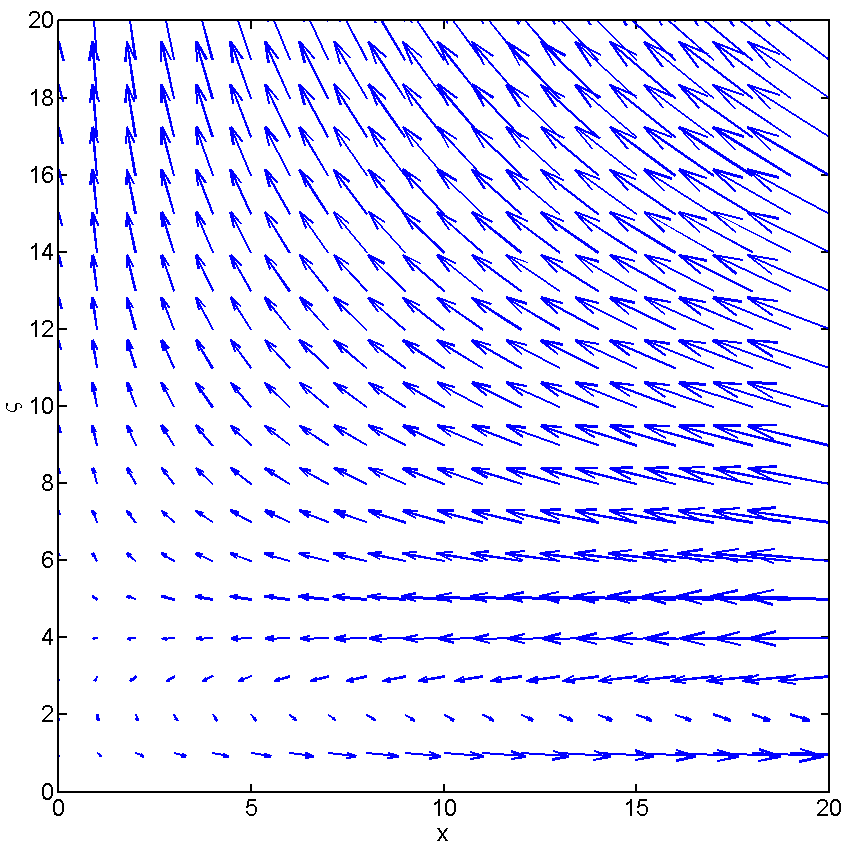}}
\caption{Similarity velocity field as a function of $\varsigma$}
\label{stream1}
\end{figure}
\addtocounter{figure}{-1}
\begin{figure}[htbp]
\addtocounter{subfigure}{4}
\centering
\subfigure[$F''(0)=-1.5$]{\includegraphics[width=11cm]{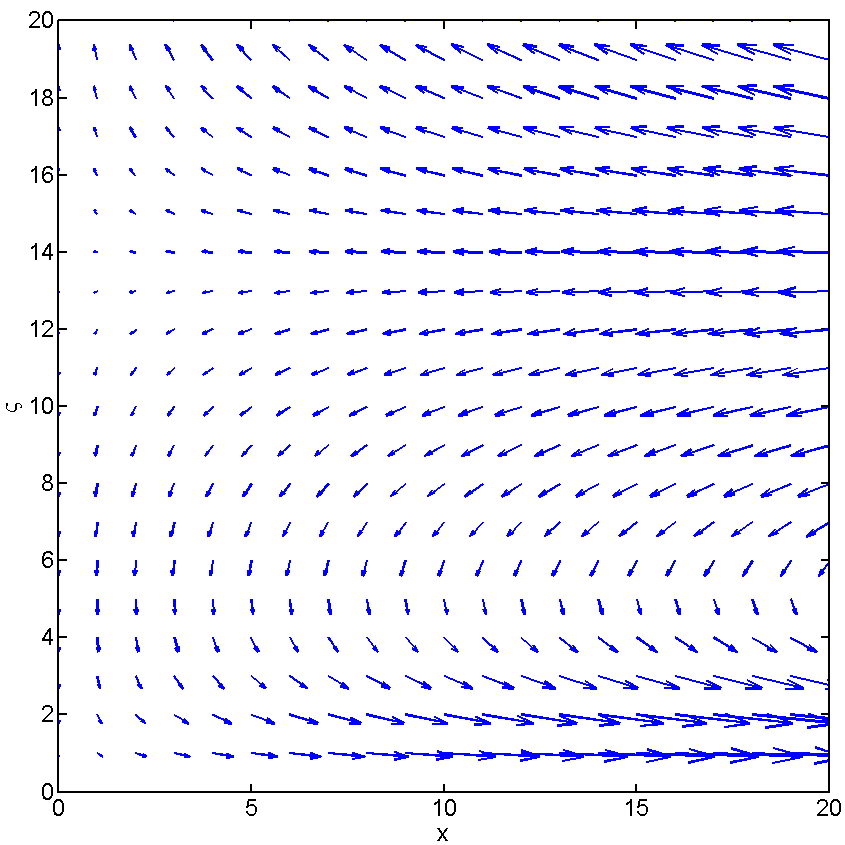}}
\subfigure[$F''(0)=-1.7$]{\includegraphics[width=11cm]{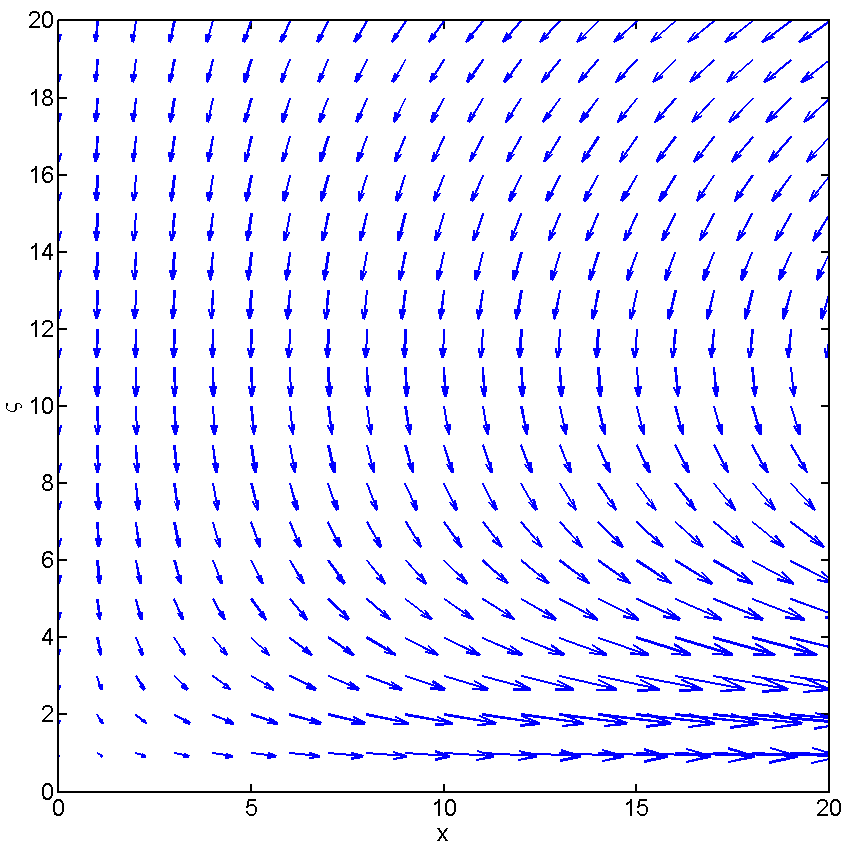}}
\caption{Similarity velocity field as a function of $\varsigma$}
\label{stream2}
\end{figure}

\begin{figure}[htbp]
\centering
\subfigure[]{\includegraphics[width=14cm]{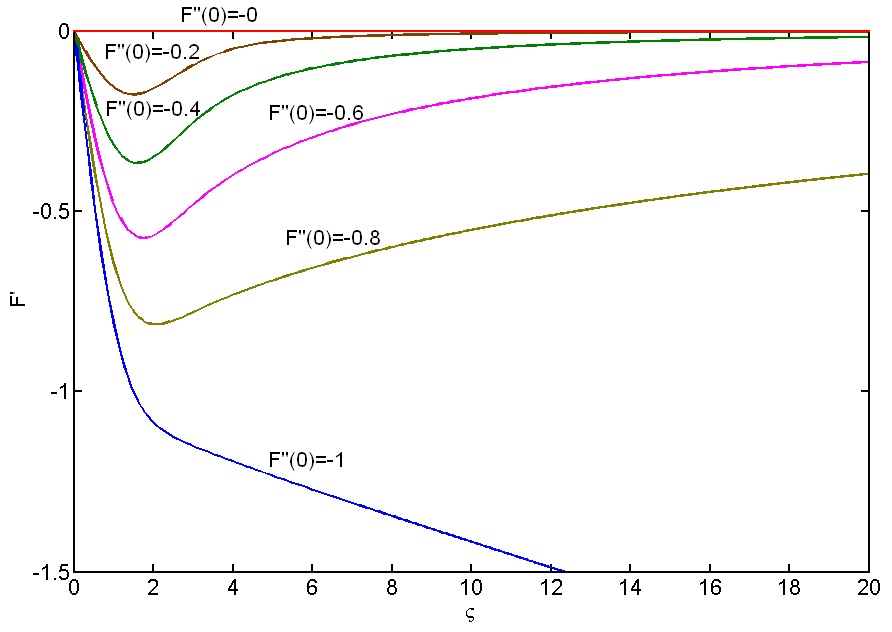}}
\subfigure[]{\includegraphics[width=14cm]{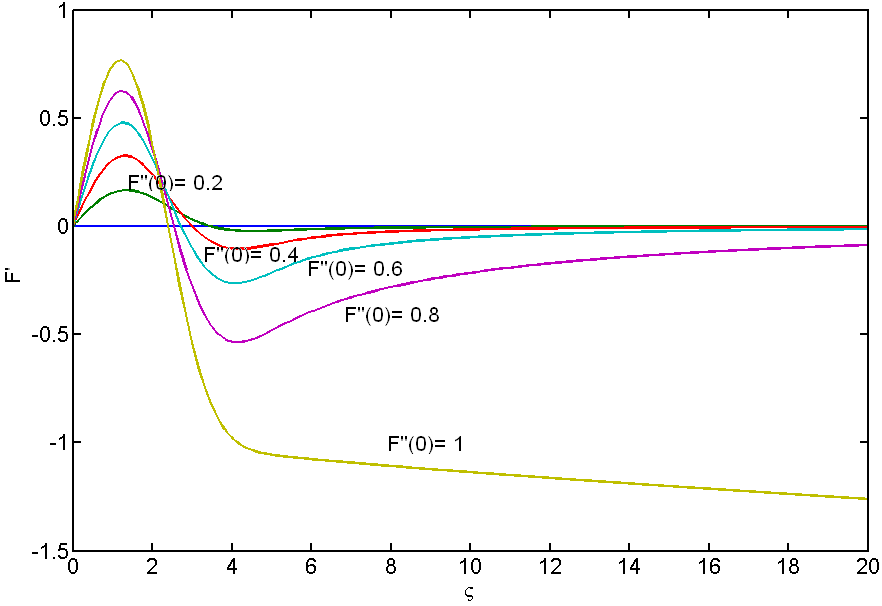}}
\caption{Numerical solutions of viscous reversed stagnation-point flow for $c=0$}
\label{rest}
\end{figure}

The similarity velocity fields are shown in Figures (\ref{stream}) at different values of $F''(0)$. It is reasonable to state that, in general, separation will occur near the wall as $\eta \rightarrow 0$ and the region of reversed flow will move outward away from the wall as $F''(0)<-1$. Moreover, it is noted that given from equation (\ref{e3_24}) the external flow velocity $$f(\eta,\tau)=\frac{\eta}{2{\tau}}-\frac{3}{\sqrt{\tau}}$$ will tend to zero for large times $\tau \rightarrow \infty$.  We have, from equation (\ref{q:e12}) with $V_0$ = 0, the equation
\begin{equation}
  \left\{
  \begin{array}{rr}
      -\frac{1}{2}\varsigma F''-F'-F'^2+FF''-F'''=0 \\
     F(0)= F'(0)=0\\
     F'(\infty)=0 
  \end{array}
  \right.
  \label{q:e12b}
\end{equation}
where $c=0$. Figure (\ref{rest}) shows the numerical solutions at various values of $F''(0)$, indicating that the nonlinear convective terms play a secondary role in fluid motion as $\tau \rightarrow \infty$, the viscous forces may play a significant role to decelerate the velocities to zero. The boundary of this region comes to rest and finally the region of reversed flow does not continue to grow but has finite dimensions. Larger value of $F''(0)$ corresponds to larger dimension of the reversed region. 

\subsection{TEMPERATURE DISTRIBUTION}
Under the assumption that the viscous dissipation is negligible compared to conduction at the wall, $\theta =\Theta(\varsigma)$ is the function of $\varsigma$ only. The energy equation may be written as
  \begin{equation}
\Theta'' + Pr \left(\frac{1}{2}\varsigma -F\right)\Theta'=0
\label{q:e15}
  \end{equation}
subject to the boundary conditions
  \begin{equation}
   \Theta(0)=0, \qquad
   \Theta(\infty)=1
\end{equation}
where $Pr=\displaystyle \frac{\rho c_p\nu}{k}$ is the Prandtl number. Equation(\ref{q:e15}) is a second-order linear ordinary differential equation, and has an exact solution through a transformation. Let
  \begin{equation}
 Z= \frac{d\Theta}{d\varsigma}
     \label {q:e19}
  \end {equation}
Substituting equation (\ref{q:e19}) into equation (\ref{q:e15}) and simplifying gives
$$ \frac {dZ}{d\varsigma}= -Pr \left(\frac{1}{2}\varsigma -F\right)Z$$
A further integration provides
$$Z = Z_0 \ exp \left[\ -Pr\int_{0}^{\varsigma}{\left(\frac{1}{2}s -F\right)ds}\right]\ $$
or
$$\Theta = Z_0 \int_{0}^{\varsigma}{d\varsigma \ exp\left[\ -Pr
\int_{0}^{\varsigma}{\left(\frac{1}{2}s -F\right)ds}\right]\ }+\Theta_0$$
Compare to the boundary conditions, we get
$$\Theta_0=0$$
$$\frac{1}{Z_0}= \int_{0}^{\infty}{d\varsigma \ exp\left[\ -Pr
\int_{0}^{\varsigma}{\left(\frac{1}{2}s -F\right)ds}\right]\ }$$
\newline
An exact solution of equation (\ref{q:e15}) is given as
  \begin{equation}
\Theta(\varsigma) = \frac {\int_{0}^{\varsigma}{d\varsigma \ exp\left[\ -Pr
\int_{0}^{\varsigma}{\left(\frac{1}{2}s -F\right)ds}\right]\ }}
{\int_{0}^{\infty}{d\varsigma \ exp\left[\ -Pr
\int_{0}^{\varsigma}{\left(\frac{1}{2}s -F\right)ds}\right]\  }}
     \label {q:e20}
  \end {equation}


A closed-form solution of the thermal energy equation for forced convection system is obtained. The solution, however, is not anticipated to integrate because equation (\ref{e3_22}) does not satisfy impermeability condition of the wall $F'=0$ and we cannot have an analytical solution of $F$. It is convenient to solve the decoupled momentum and energy equations numerically. Defining $y_4 = \Theta,~y_5 = \Theta'$ and combining the variables in the momentum equation (\ref{eq:e22}), the uncoupled momentum and energy equations reduce to the form

    \begin{equation}
\frac{d\mathbf{y}}{d\varsigma} =
\begin{bmatrix}
  y_2  \\
  y_3  \\
  c-\frac{1}{2}\varsigma y_3-y_2-y_2^2 +y_1y_3\\
  y_5 \\
  Pr\left(y_1-\frac{1}{2}\varsigma\right)y_5
\end{bmatrix}
    \label{eq:e23}
\end{equation}
\\
The relevant MATLAB expression for (\ref{eq:e23}) would be:

\begin{lstlisting}[label=MATLAB,caption=System of first-order equations ]
function dy = stagnation(t,y,Pr)
    c=3/4;
    dy = zeros(5,1);
    dy(1) = y(2);
    dy(2) = y(3);
    dy(3) = c-1/2*t*y(3)-y(2)-y(2)*y(2)+y(1)*y(3);
    dy(4) = y(5);
    dy(5) = Pr*y(5)*(y(1)-1/2*t);
end
\end{lstlisting}

As was previously indicated, the boundary value problem is changed into initial value problem by taking a gauss of $~\Theta''(0)$ such that $~\Theta'(\infty) = 1$. The corresponding commands written in MATLAB would be

\begin{lstlisting}[label=MATLAB,caption=ODE solver when $Pr$ is equal to $1$ ]
function main
    [T,Y] = ode45(@stagnation,[0 10],[0 0 -1.03 0.8]);
end
\end{lstlisting}

\begin{figure}[!htbp]
\begin{center}
\includegraphics[width= 12cm]
{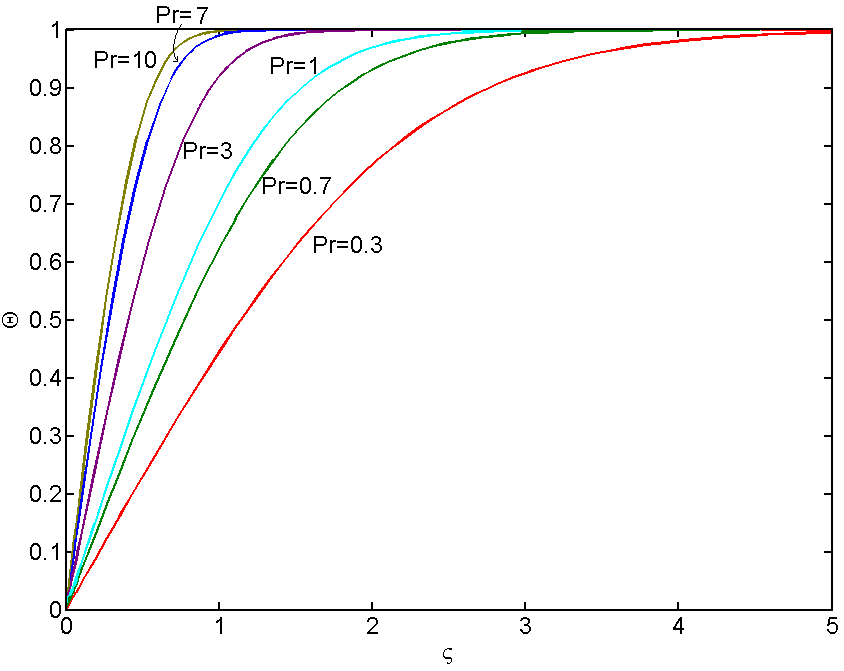}
\caption{Reversed stagnation-point  temperature distributions $\Theta$ for various value of Prandtl number}
\label{tg1}
\end{center}
\end{figure}

The numerical solution for temperature distributions is shown in Figure (\ref{tg1}). It is noticed that the dimensionless wall temperature gradient $\Theta'(0)$ raises with increase of Prandtl number, but the thermal boundary layer thickness decrease with increase of Prandtl number. The thermal boundary layer thickness is the distance from the body at which the temperature is $99\%$ of the temperature obtained from an inviscid solution. The decrease of thickness can be explained by the definition of Prandtl number that is inversely proportional to the thermal diffusivity ${k}/{\rho c_p\nu}$.  If the Prandtl number is greater than $1$, the thermal boundary layer is thinner than the velocity boundary layer. If the Prandtl number is less than $1$, which is the case for air at standard conditions, the thermal boundary layer is thicker than the velocity boundary layer.

\chapter{CONCLUSION AND RECOMMENDATION}
\thispagestyle{empty}

In this study,  nonisothermal stagnation-point flow is studied by applying an unsteady numerical model in Computational Fluid Dynamics. Beyond this, we explored the velocity and temperature profile of the reversed stagnation-point flow. In present studies, investigations on the behaviors of dimensionless velocity in the reversed stagnation-point flow reveal that:
\begin{enumerate}
\item
Compared to the previous research, it is not quite appropriate to say Proudman and Johnson are wrong because of neglecting the viscous term in their analytic result for region sufficient far from the wall. Also, their inviscid result is impressing; because one can expect the flow pattern (see Figure \ref{vort}) from the inviscid field. 
\item
In the region near the reversed stagnation point the numerical simulation agrees remarkably well for smaller values of $\tau$ with the known similarity solution, thus confirming the predictions of the viscous Proudman-Johnson solution. Their idealization that the viscous forces are of the same order as the inertial forces is acceptable near the stagnation point. 
\item
Separation will occur near the wall as $\eta \rightarrow 0$ and the region of reversed flow will move outward away from the wall. For large times $\tau \rightarrow \infty$,  the reversed flow comes to rest. Viscous forces are dominant to decelerate the velocities to zero and ultimately the region of reversed flow does not continue to grow but has finite dimensions. 
\item
For the external flow outside the boundary layer, the hypothesis that the velocity $v(x,\eta,\tau)$ should pass over smoothly into that for inviscid $V_0$ is not valid.  The influence of backflow must be taken into account and a continuous change as $V_0(\tau)$ decreases in magnitude for large $\tau$ should be expected outside the boundary. 
\end{enumerate}

On the other hand, investigations on the behaviors of dimensionless temperature in the nonisothermal reversed stagnation-point flow illustrate that:
\begin{enumerate}
\item
The solution of the thermal energy equation is also provided. The temperature distribution is monotonically increasing.  The nondimensional temperature $\Theta$ drops from its remote value to its wall value in a thin thermal boundary layer adjacent to the wall. It is surprising that, near the backflow region, there is no discernible temperature signature between the dividing streamlines.  
\item
Larger Prandtl number results in thinner boundary layer and higher temperature gradient near the wall. When Prandtl number is small, the heat diffuses very quickly compared to the velocity field. This implies that for liquid metals the thickness of the thermal boundary layer is much bigger than that of the velocity boundary layer. 
\item
The numerical simulation indicates that heat transfers mainly from the cooled wall to the heated external flow by conduction in the beginning. The heated flow passes though the wall, rises and cools down by conduction and convection of heat.  Because of the motion of backflow,  heated flow sinks to the wall where it is prohibited from sinking further and becomes dense near the reversed stagnation point along the edges of the wall. 
\end{enumerate}

With the establishment of this frame work, a similarity method applied to the two-dimensional unsteady reversed stagnation-point has induced new physically significant solutions, and application of the method to other case may be even more fruitful. Recommendations on the study of this type of fluid flow problem are given below:
\begin{enumerate}
\item
The similarity solution is valid only at the reversed stagnation point $x=0$. In order to study the flow for non-zero values of $x$, we must revert the whole problem to the full boundary-equation. 
\item
Three-dimensional simulation is much better than the two-dimensional case that we have been studying so far.  However, more realistic simulation comes with high requirements in memory and CPU time so that the three-dimensional case is generally not simulated. A rapid development of computer hardware and software will further increase the opportunities for numerical simulation.
\item
More execution time would be sufficient in the simulation. Because of the time constraints, only a few cases of simulation are completed. More cases of simulations should be performed to obtain a more reliable data set of this type of fluid flow problem. 
\item
In the result of numerical simulation, one may be observed that there are small vortices generated near the reversed stagnation point when the Reynolds is sufficient high. Some factors of affecting the probability of getting firm results of the investigations on the small vortices near the plate are thought to be:
\begin{enumerate}
\item
Sizes of the time steps;
\item
Sizes of finite volume near the reversed stagnation point;
\item
Magnitudes of the external flow velocity $u_0$
\item
Differences between the wall temperature $T_w$ and the ambient temperature $T_\infty$
\end{enumerate}
\item
The more important practical properties in engineering and technology application, like the velocity of wall is function of time $\tau$ and the temperature of wall is function of time $\tau$, can be investigated and should be performed in the next phase of this study.
\end{enumerate}

\newpage

\chapter*{APPENDIX}
\thispagestyle{empty}
\addcontentsline{toc}{chapter}{APPENDIX}
\section*{MATLAB}
\lstset{language=MATLAB,
         breaklines=true,
         extendedchars=false,
         showstringspaces=false,
         numbers=left,
         numberstyle=\ttfamily\scriptsize,
         frame=trbl,framesep=5pt,framexleftmargin=8mm,
         frameround=tttt,
         keywordstyle=\ttfamily\bf\color{red},
         ndkeywordstyle=\ttfamily\bf\color{brown},
         commentstyle=\color{blue},
         identifierstyle=\ttfamily\color{black}\bfseries,
         stringstyle=\color{red}\ttfamily
}

\begin{lstlisting}[label=MATLAB,caption=Finite-difference formulations for reversed stagnation-point flow]
clear all
deta=0.1;           dtau=0.05;         Pr=1;
IMAX=100;           NMAX=6;
beta=dtau/(deta^2);
IM=IMAX/deta+1;     NM=NMAX/dtau+1;
h=zeros(IM,NM);     g=zeros(IM,NM);    s=zeros(1,IM-2);
% =======================================================
% Setting the initial and boundary conditions
  eta(1,1) = 0.0;
for i= 2:IM
  eta(i,1) = eta(i-1,1) + deta;
  h(i,1)=1-erf(eta(i,1)/2/sqrt(dtau/10));
  g(i,1)=erf(eta(i,1)/2/sqrt(dtau/10/Pr));
end

for n= 1:NM
  h(1,n)=1;           % no slip boundary condition
  g(IM,n)=1;
end
tic
for n=1:NM-1
       s(1,1)=h(1,n)+0.5*beta*(h(3,n)-2*h(2,n)+h(1,n))...
              +dtau*(2*h(1,n)-(h(1,n))^2)...
              -dtau/2*(-h(3,n)+4*h(2,n)-3*h(1,n))*...
              0.5*(1-h(1,n))+0.5*beta*h(1,n);
    for i=2:IM-2
       s(1,i)=h(i,n)+0.5*beta*(h(i+1,n)-2*h(i,n)+h(i-1,n))...
              +dtau*(2*h(i,n)-(h(i,n))^2)...
              -dtau/2*(h(i+1,n)-h(i-1,n))*...
              (sum(1-h(1:i,n))-0.5*(1-h(i,n)));
    end
    s(1,IM-2)=s(1,IM-2)+0.5*beta*h(IM-2,n);

% =======================================================
% Thomas algorithm for a tridiagonal system
% a,b,c: diagonal, superdiagonal,
%        and subdiagonal elements
    a=(1+beta)*ones(1,IM-2);
    b=-0.5*beta*ones(1,IM-2);
    c=-0.5*beta*ones(1,IM-2);
    x=ones(1,IM-2);
    d=ones(1,IM-2);
    d(1,1)=b(1,1)/a(1,1);
    y=ones(1,IM-2);
    y(1,1)=s(1,1)/a(1,1);

    for p=1:(IM-3)
        den=a(1,p+1)-c(1,p+1)*d(1,p);
        d(1,p+1)=b(1,p+1)/den;
        y(1,p+1)=(s(1,p+1)-c(1,p+1)*y(1,p))/den;
    end

    x(1,IM-2)=y(1,IM-2);
    for p=IM-3:-1:1
        x(1,p)=y(1,p)-d(1,p)*x(1,p+1);
    end

    for p=2:IM-1
        h(p,n+1)=x(1,p-1);
    end

    s(1,1)=g(1,n)-dtau*(-g(3,n)+4*g(2,n)-3*g(1,n))...
           *sum(1-h(1,n))+beta/Pr*g(1,n);
    for i=2:IM-2
       s(1,i)=g(i,n)...
              -dtau/2*(g(i+1,n)-g(i-1,n))*...
              (sum(1-h(1:i,n))-0.5*(1-h(i,n)));
    end
    s(1,IM-2)=s(1,IM-2)+beta/Pr*g(IM-2,n);
% =======================================================
% Thomas algorithm for a tridiagonal system
% a,b,c: diagonal, superdiagonal,
%        and subdiagonal elements
    a=(1+2*beta/Pr)*ones(1,IM-2);
    b=-beta/Pr*ones(1,IM-2);
    c=-beta/Pr*ones(1,IM-2);
    x=ones(1,IM-2);
    d=ones(1,IM-2);
    d(1,1)=b(1,1)/a(1,1);
    y=ones(1,IM-2);
    y(1,1)=s(1,1)/a(1,1);
    for p=1:(IM-3)
        den=a(1,p+1)-c(1,p+1)*d(1,p);
        d(1,p+1)=b(1,p+1)/den;
        y(1,p+1)=(s(1,p+1)-c(1,p+1)*y(1,p))/den;
    end
    x(1,IM-2)=y(1,IM-2);
    for p=IM-3:-1:1
        x(1,p)=y(1,p)-d(1,p)*x(1,p+1);
    end
%=======================================================
    for p=2:IM-1
        g(p,n+1)=x(1,p-1);
    end
end
\end{lstlisting}
\clearpage

\section*{OpenFOAM}
\lstset{language=C++,
         breaklines=true,
         extendedchars=false,
         showstringspaces=false,
         numbers=left,
         numberstyle=\ttfamily\scriptsize,
         frame=trbl,framesep=5pt,framexleftmargin=8mm,
         frameround=tttt,
         keywordstyle=\ttfamily\bf\color{violet},
         ndkeywordstyle=\ttfamily\bf\color{brown},
         commentstyle=\color{blue},
         identifierstyle=\ttfamily\color{black}\bfseries,
         stringstyle=\color{red}\ttfamily
}

\begin{lstlisting}[label=C++,caption=myicoFoam solver]
/*-----------------------------------------------------*\
Application
    myicoFoam

Description
    Transient solver for incompressible, laminar flow
    of Newtonian fluids and temperature profile
\*-----------------------------------------------------*/

#include "fvCFD.H"

int main(int argc, char *argv[])
{
    #include "setRootCase.H"
    #include "createTime.H"
    #include "createMesh.H"
    #include "createFields.H"
    #include "initContinuityErrs.H"

    Info<< "\nStarting time loop\n" << endl;

    while (runTime.loop())
    {
        Info<<"Time = "<<runTime.timeName() << nl << endl;

        #include "readPISOControls.H"
        #include "CourantNo.H"

        fvVectorMatrix UEqn
        (
            fvm::ddt(U)
          + fvm::div(phi, U)
          - fvm::laplacian(nu, U)
        );

        solve(UEqn == -fvc::grad(p));

        // --- PISO loop

        for (int corr=0; corr<nCorr; corr++)
        {
            volScalarField rUA = 1.0/UEqn.A();

            U = rUA*UEqn.H();
            phi = (fvc::interpolate(U) & mesh.Sf())
                + fvc::ddtPhiCorr(rUA, U, phi);

            adjustPhi(phi, U, p);

            for (int nonOrth=0; nonOrth<=nNonOrthCorr; nonOrth++)
            {
                fvScalarMatrix pEqn
                (
                    fvm::laplacian(rUA, p)== fvc::div(phi)
                );

                pEqn.setReference(pRefCell, pRefValue);
                pEqn.solve();

                if (nonOrth == nNonOrthCorr)
                {
                    phi -= pEqn.flux();
                }
            }

            #include "continuityErrs.H"

            U -= rUA*fvc::grad(p);
            U.correctBoundaryConditions();
        }

        // ---  Temperature transport

        fvScalarMatrix TEqn
        (
            fvm::ddt(T)
            + fvm::div(phi, T)
            - fvm::laplacian(DT, T)
        );

        TEqn.solve();
        runTime.write();

        Info<< "ExecutionTime = " << runTime.elapsedCpuTime() << " s"
            << "  ClockTime = " << runTime.elapsedClockTime() << " s"
            << nl << endl;
    }

    Info<< "End\n" << endl;
    return 0;
}
// ***************************************************//
\end{lstlisting}
\newpage

\begin{lstlisting}[label=C++,caption=Geometry Analysis]
FoamFile
{
    version     2.0;
    format      ascii;
    class       dictionary;
    object      blockMeshDict;
}
convertToMeters 1;
vertices
(
    (0 0 0)
    (1 0 0)
    (1 1 0)
    (0 1 0)
    (0 0 1)
    (1 0 1)
    (1 1 1)
    (0 1 1)
);
blocks
(
    hex (0 1 2 3 4 5 6 7) (200 400 1)
    simpleGrading (1 5 1)
);
edges ();
patches
(
    patch left_inlet    ((2 6 5 1))
    patch right_inlet   ((0 4 7 3))
    patch outlet        ((3 7 6 2))
    wall fixedWalls     ((1 5 4 0))
    empty frontAndBack  ((0 3 2 1)
                         (4 5 6 7))
);
mergePatchPairs ();
\end{lstlisting}
\newpage
\begin{lstlisting}[label=C++,caption=Fluid Transport Properties]
// * * * * * * * * * * * * * * * * * * * * * * *//
FoamFile
{
    version     2.0;
    format      ascii;
    class       dictionary;
    location    "constant";
    object      transportProperties;
}
nu              nu [ 0 2 -1 0 0 0 0 ] 2.94e-7;
DT              DT [ 0 2 -1 0 0 0 0 ] 1.68e-7;
// ************************************************** //
\end{lstlisting}
\newpage
\begin{lstlisting}[label=C++,caption=Initial Pressure Profile]
// * * * * * * * * * * * * * * * * * * * * * * *//
FoamFile
{
    version     2.0;
    format      ascii;
    class       volScalarField;
    object      p;
}
dimensions      [0 2 -2 0 0 0 0];
internalField   uniform 0;
boundaryField
{
    left_inlet   {type    zeroGradient;}
    right_inlet  {type    zeroGradient;}
    outlet       {type    fixedValue;
                  value   uniform 0;}
    fixedWalls   {type    zeroGradient;}
    frontAndBack {type    empty;}
}
// ************************************************** //
\end{lstlisting}
\newpage
\begin{lstlisting}[label=C++,caption=Initial Velocities Profile]
// * * * * * * * * * * * * * * * * * * * * * * *//
FoamFile
{
    version     2.0;
    format      ascii;
    class       volVectorField;
    object      U;
}
dimensions      [0 1 -1 0 0 0 0];
internalField   uniform (0 0 0);
boundaryField
{
    left_inlet   {type    fixedValue;
                  value   uniform (-1 0 0);}
    right_inlet  {type    fixedValue;
                  value   uniform (1 0 0);}
    outlet       {type    zeroGradient;}
    fixedWalls   {type    fixedValue;
                  value   uniform (0 0 0);}
    frontAndBack {type    empty;}
}
// ************************************************** //
\end{lstlisting}
\newpage
\begin{lstlisting}[label=C++,caption=Initial Temperature Profile]
// * * * * * * * * * * * * * * * * * * * * * * *//
FoamFile
{
    version     2.0;
    format      ascii;
    class       volScalarField;
    object      T;
}
dimensions      [0 0 0 1 0 0 0];
internalField   uniform 373;
boundaryField
{
    left_inlet   {type    fixedValue;
                  value   uniform 373;}
    right_inlet  {type    fixedValue;
                  value   uniform 373;}
    outlet       {type    zeroGradient;}
    fixedWalls   {type    fixedValue;
                  value   uniform 273;}
    frontAndBack {type    empty;}
}
// ************************************************** //
\end{lstlisting}
\newpage
\begin{lstlisting}[label=C++,caption=Simulation Control]
// * * * * * * * * * * * * * * * * * * * * * * *//
FoamFile
{
    version     2.0;
    format      ascii;
    class       dictionary;
    location    "system";
    object      controlDict;
}
application     myicoFoam;
startFrom       startTime;
startTime       0;
stopAt          endTime;
endTime         10;
deltaT          0.001;
writeControl    timeStep;
writeInterval   100;
purgeWrite      0;
writeFormat     ascii;
writePrecision  6;
writeCompression uncompressed;
timeFormat      general;
timePrecision   6;
runTimeModifiable yes;
// ************************************************** //
\end{lstlisting}
\newpage
\begin{lstlisting}[label=C++,caption=System Solver]
// * * * * * * * * * * * * * * * * * * * * * * *//
FoamFile
{
    version     2.0;
    format      ascii;
    class       dictionary;
    location    "system";
    object      fvSchemes;
}
ddtSchemes
{
    default         Euler;
}
gradSchemes
{
    default         Gauss linear;
    grad(p)         Gauss linear;
}
divSchemes
{
    default         none;
    div(phi,U)      Gauss linear;
    div(phi,T)      Gauss upwind;
}
laplacianSchemes
{
    default         none;
    laplacian(nu,U) Gauss linear corrected;
    laplacian((1|A(U)),p) Gauss linear corrected;
    laplacian(DT,T) Gauss linear corrected;
}
interpolationSchemes
{
    default         linear;
    interpolate(HbyA) linear;
}
snGradSchemes
{
    default         corrected;
}
fluxRequired
{
    default         no;
    p               ;
}
// ************************************************** //
\end{lstlisting}
\newpage
\begin{lstlisting}[label=C++,caption=Preconditioner and Tolerance]
// * * * * * * * * * * * * * * * * * * * * * * *//
FoamFile
{
    version     2.0;
    format      ascii;
    class       dictionary;
    location    "system";
    object      fvSolution;
}
solvers
{
    p
    {
        solver          PCG;
        preconditioner  DIC;
        tolerance       1e-06;
        relTol          0;
    }
    T
    {
       solver            BICCG;
        preconditioner   DILU;
        tolerance        1e-7;
        relTol           0;
    }
    U
    {
        solver          PBiCG;
        preconditioner  DILU;
        tolerance       1e-05;
        relTol          0;
    }
}
PISO
{
    nCorrectors     2;
    nNonOrthogonalCorrectors 0;
    pRefCell        0;
    pRefValue       0;
}
// ************************************************** //
\end{lstlisting}
\newpage

\section*{SPECIFICATIONS OF THE SIMULATION COMPUTER}
\begin{center}
\begin{tabular}{|c|c|}
\hline
Computer Model& Lenovo Thinkstation Workstation D20\\
\hline
CPU & Intel \textregistered Xeon \textregistered CPU X5690 @3.47~GHz\\
\hline
RAM & 24.0~GB\\
\hline
Operation System & Ubuntu Linux 10.04\\
                 & Windows 7\\
\hline
Software & OpenFOAM\\
         &MATLAB\\
\hline
\end{tabular}
\end{center}
\newpage


\addcontentsline{toc}{chapter}{BIBLIOGRAPHY}
\renewcommand{\bibname}{\thispagestyle{empty}BIBLIOGRAPHY}
\thispagestyle{empty}
\bibliographystyle{ieeetr}	
\bibliography{myrefs}

\newpage
\end{document}